\documentclass[12pt,onecolumn,twoside]{pnas-arxiv}
\makeatletter
\usepackage{xcolor}
\usepackage{siunitx}
\usepackage{hyperref}
\newcommand{\xRightarrow}[2][]{\ext@arrow 0359\rightarrowfill@{#1}{#2}}
\newcommand{\coli}{\emph{E.~coli }}
\newcommand{\ansatz}{\emph{ansatz} }
\newcommand{\zd}{z^\dagger}
\newcommand{\zmin}{z_\text{min}}

\newcommand{\amin}{a_\text{min}}
\newcommand{\rhomin}{\rho_\text{min}}

\newcommand{\N}{N}%\text{front}}
\newcommand{\OD}{\text{OD}_{600}}
\renewcommand{\eqref}[1]{Eq.~\textbf{\ref{#1}}}
\usepackage{mathptmx}
\usepackage{setspace}

\newcommand\blfootnote[1]{%
  \begingroup
  \renewcommand\thefootnote{}\footnote{#1}%
  \addtocounter{footnote}{-1}%
  \endgroup
}
\usepackage{color}
\definecolor{forestgreen}{rgb}{0.13,0.55,0.13}
\definecolor{orange}{rgb}{0.96,0.5,.14}
\definecolor{brown}{rgb}{0.81,0.54,.27}

\templatetype{pnas-arxiv} % Choose template 
\title{A Traveling-Wave Solution for Bacterial Chemotaxis with Growth}

\author[a]{Avaneesh V. Narla}
\author[b]{Jonas Cremer} 
\author[a,*]{Terence Hwa}

\affil[a]{Department of Physics, University of California, San Diego, 9500 Gilman Drive, La Jolla, CA 92093}
\affil[b]{Biology Department, Stanford University, 318 Campus Drive, Stanford, CA 94305}
\affil[*]{Contact: hwa@ucsd.edu}

\leadauthor{Narla} 


\begin{abstract}
Bacterial cells navigate around their environment by directing their movement along chemical gradients. This process, known as chemotaxis, can promote the rapid expansion of bacterial populations into previously unoccupied territories. However, despite numerous experimental and theoretical studies on this classical topic, chemotaxis-driven population expansion is not understood in quantitative terms. Building on recent experimental progress, we here present a detailed analytical study that provides a quantitative understanding of how chemotaxis and cell growth lead to rapid and stable expansion of bacterial populations. We provide analytical relations that accurately describe the dependence of the expansion speed and density profile of the expanding population on important molecular, cellular, and environmental parameters. In particular, expansion speeds can be boosted by orders of magnitude when the environmental availability of chemicals relative to the cellular limits of chemical sensing is high. As analytical understanding of such complex spatiotemporal dynamic processes is rare, the results derived here provide a mathematical framework for further investigations of the different roles chemotaxis plays in diverse ecological contexts.
\end{abstract}

\begin{document}
\onehalfspacing
\maketitle
\thispagestyle{firststyle}
\ifthenelse{\boolean{shortarticle}}{\ifthenelse{\boolean{singlecolumn}}{\abscontentformatted}{\abscontent}}{}
\blfootnote{A.V.N., J.C. and T.H. designed research; A.V.N. and T.H. performed research; A.V.N. developed computational tools; A.V.N., J.C., and T.H. analyzed data and wrote the paper. The authors declare no conflict of interest.}
% If your first paragraph (i.e. with the \dropcap) contains a list environment (quote, quotation, theorem, definition, enumerate, itemize...), the line after the list may have some extra indentation. If this is the case, add \parshape=0 to the end of the list environment.
Many species of bacteria are motile and respond to environmental changes by directing their movement along gradients of certain chemicals~\cite{sourjik2012responding}. This process, known as chemotaxis, is among the most extensively-investigated topics in molecular biology~\cite{berg1975chemotaxis,adler1975chemotaxis}. Beyond driving striking cell movements, chemotaxis also drives the collective movement of cells leading to emergent patterns and behaviors at the population level. For example, when encountering preferred chemicals referred to as \textit{attractants}, cells consume the attractants and collectively move up self-generated attractant gradients~\cite{erban2004individual}.\par 

A characteristic population-level behavior is the emergence of clear \textit{migrating bands} when the bacteria encounter a region of uniform attractant concentration~\cite{adler1966chemotaxis,adler1966effect,adler1969chemoreceptors}. The migrating bands typically comprise of one or two peaks in population density, which stand in contrast to the predictions of canonical models of front propagation and population expansion~\cite{fisher1937wave,kolmogorov1937etude,van2003front}. %As such, understanding these migrating bands mathematically holds relevance beyond microbiology to describe propagating fronts, biotic or abiotic, that interact with their local environment.\par
The first attempt to understand these migrating bands mathematically was made by Keller and Segel who recovered a traveling wave solution using a pair of reaction-diffusion-convection equations to describe the bacterial population and the concentration of the attractant they consume~\cite{Keller1971traveling}. While being highly influential, the Keller-Segel (KS) Model neglected cell growth, a substantial factor in the expansion process. It further required unrealistic assumptions without which the migrating bands would lose stability~\cite{novick1984gradually}. Subsequent modeling efforts including cell growth managed to recover the stability of the bands, but their predictions did not match with major experimental observations such as the sharply peaked density profiles and their rapid migration speeds~\cite{lapidus1978model,keller1980assessing,tindall2008overview,koster2012surface,fraebel2017environment}.\par
%
%A notable modeling effort to incorporate taxis (while excluding population growth) by Keller and Segel~\cite{Keller1971model} was incredibly influential in the study of bacterial and eukaryotic chemotaxis, i.e., taxis driven by the sensing of gradients in concentrations of chemical attractants. A key result of the Keller-Segel (KS) Model was the emergence of experimentally-observed stable migrating bands in chemotactic bacterial populations~\cite{Keller1971traveling} resulting from self-generated gradient of attractant concentration exerted by the bacteria on the environment, and the navigational role of the attractant gradient on the population. However, upon relaxation of some of the assumptions made in the KS Model, the migrating bands lose stability~\cite{novick1984gradually}, and subsequent efforts to recover the stability did not match experimental observations.\par
%

Recent experimental work by Cremer and Honda et al.~\cite{Cremer2019a} demonstrated that the major features of the migrating bands can be accurately captured by including bacterial growth that is independent of the attractant. They were able to quantitatively predict the observed expansion dynamics of \coli in soft agar for a wide range of experimental conditions through numerical solutions to their Growth-Expansion (GE) model~\cite{Cremer2019a}. Their results established the role of attractants as environmental cues which bacteria exploit independent of their possible nutritional values to promote rapid expansion.\par

To gain an analytical understanding on how, and in what conditions, growth, diffusion, and chemotaxis interact to generate rapid stable traveling waves, we here perform a heuristic traveling-wave analysis of the GE model. We derive analytic relations that describe the dependence of the expansion speed and density profile on important molecular, cellular, and environmental parameters, including the rate of cell growth, the diffusivity and availability of the attractants, the motility and sensitivity of the bacteria, and the limit of attractant sensing. These relations provide the necessary mathematical framework to investigate the consequences of population-level chemotaxis in a wide range of ecological contexts. %To derive a more mechanistic understanding of the expansion, we here present an analytical analysis of the expansion dynamics. Building on the GE model, we performed a traveling wave analysis to obtain analytical relations for the expansion speed and the profile. Our results establish how the expansion dynamics depend on physiological and enviornmental parameters, including....\jc{give exmaples.}\jc{can could better integrate some math language here - traveling wave analysis, ....}.

\section*{Growth-Expansion Model}

In the GE model the evolution of the bacterial density, $\rho$, in space and time ($t$) is given by:
\begin{align}
\frac{\partial \rho}{\partial t}&=D_\rho\nabla^2\rho-\nabla\cdot(\vec{v}\rho)+r\rho\ (1-\rho/\rho_c). \label{eq-rho}
\end{align}
The growth of the population is given by the logistic equation where $r$ is the growth rate and $\rho_c$ is the carrying capacity of the system. The non-directed run-and-tumble movement of bacteria is described by a diffusion-like term with the \textit{motility coefficient} $D_\rho$, while directed movement along the gradient of the attractant concentration $a$ is described by a convection term with the drift velocity $\vec{v}(a,\nabla a)$, where
\begin{align}
    \vec{v}&\equiv\chi_0\vec{\nabla}a/(a+a_m). \label{eq-v}
\end{align}
%\eqref{eq-v} is adapted from the Weber-Fechner law~\cite{gescheider2013psychophysics,fechner1966elements} and 
$\chi_0$ is the \textit{chemotactic coefficient} which describes how cells translate the sensing of the local attractant gradients into directed movement. The value of $\chi_0$ depends on the strain, the internal cell state, the medium cells move through (e.g., liquid medium or soft agar), and the type of attractant being used~\cite{si2012pathway}.
$a_m$ describes the finite sensitivity of the attractant-sensing receptors~\cite{borkovich1992attenuation,berg1975transient} and ensures that $\vec{v}\to 0$ as $a\to 0$. Finally, the dynamics of the attractant are determined by its diffusion and consumption by the bacteria:
\begin{align}
\frac{\partial a}{\partial t}&=D_a\nabla^2 a-\mu\frac{a}{a+a_k}\rho, \label{eq-a}
\end{align}
where $D_a$ is the molecular diffusion coefficient of the attractant, $\mu$ is the rate of uptake of the attractant by the bacteria, and $a_k$ is the Michaelis-Menten constant describing attractant uptake. We note that the GE model defined by Eqs.~\textbf{\ref{eq-rho}-\ref{eq-a}} is a slight simplification of the one studied numerically in~\cite{Cremer2019a}. However, the simplifications do not significantly impact our results, even at the quantitative level. (see \textit{Supplemental Figure} S2 for comparison with the generalised GE model used in~\cite{Cremer2019a}).

%need to put gradient into equation
%The attractant is consumed by cells with an uptake rate $\mu$ and the attractant diffuses with the diffusion coefficient $D_a$. $\chi_0$ is a phenomenological parameter known as the ``chemotactic coefficient''. 

%\jc{this is disrupting flow, the reader doesn't yet know what the paper analysis is about and has to think already about these details. I would bring this later.} \an{But it will stand out to readers that $a_m$ is repeated, and we should address it. Even if they don't notice it, maybe we want them to}. 
%I think this important feature of simplifcation formullation is what I found was too distructive. 
%Another important feature of our simplification is that the parameter $a_m$ is taken to be \emph{both} the lower Weber cutoff for sensitivity to attractant and the Michaelis saturation constant in the rate of attractant uptake. The equality of these two parameters is biologically motivated since for aspartate (one of the two major attractants for \emph{E.~coli} for which most is known molecularly), the same periplasmic binding protein serves as both the receptor for signaling chemotaxis and for uptake~\cite{schellenberg1977resolution,springer1977sensory}. 
%This equality is also equivalent to the case of sampling without replacement or chemoreception by capture~\cite{berg1977physics}. This approximation turns out to be crucial for deriving a closed-form expression for the expansion speed of the expansion band. Numerical results for the system where the two parameters are different will be discussed towards the end. \par 

Without growth ($r=0$), the GE model resembles the original KS model which additionally also assumed negligible attractant diffusion, i.e., $D_a=0$, and infinitesimal sensitivity in sensing, i.e., $a_m=0$. The latter assumption of the KS model is necessary for stable traveling waves~\cite{Keller1971traveling,novick1984gradually} as otherwise the portion of the band with $a<a_m$ is not able to migrate as fast as the rest of the band and falls behind, leading to a gradually-diminishing and slowing band. Many models have attempted to ``replenish'' the bands by including cell growth~\cite{lauffenburger1982effects,lapidus1978model,kennedy1980traveling,lauffenburger1982effects,lauffenburger1984traveling,pedit2002quantitative,hilpert2005lattice} and while they are able to recover stability, they fail to reproduce the fast-moving expansion dynamics as they take growth to depend on the same substrate that the bacteria deplete to generate a gradient and migrate. Thus, fast expansion is only obtained when growth and chemotaxis do not depend on the same substrate.\par

In the absence of  chemotaxis ($\chi_0=0$), the GE model reduces to the Fisher-Kolmogorov–Petrovsky–Piskunov (F-KPP) equation which describes expansion by growth and non-directed motion alone~\cite{fisher1937wave,kolmogorov1937etude,ablowitz1979explicit}. The F-KPP equation has been used canonically to describe range expansion into unoccupied habitats~\cite{skellam1951random,lubina1988spread,hastings2005spatial}, including the expansion dynamics of non-motile bacteria in colony growth and long-range dispersal~\cite{gandhi2016range,warren2019spatiotemporal,hallatschek2010life}. While growth and non-directed motion movement result in a traveling-wave solution with a stable expansion speed given by $c_F=2\sqrt{D_\rho\, r}$ (known as the Fisher speed)~\cite{fisher1937wave,kolmogorov1937etude,tikhomirov1991study,van2003front}, it is not sufficient to explain the high expansion speeds of the bacterial front observed in populations of chemotactic bacteria~\cite{Cremer2019a}. Indeed, as we will see below, the expansion speed for the GE model can lead to expansion speeds orders of magnitude higher than the Fisher speed.\par
 
Remarkably, while the two different reaction-diffusion models (KS and F-KPP) fail to even qualitatively describe the experimental observation of fast-moving stable migrating bands by themselves, when combined together they are able to to quantitatively explain the prominent features of bacterial chemotaxis for a broad range of physiological and environmental conditions~\cite{Cremer2019a}.\par

%not sure if we need so much details here....so I skipped it. 
%and is strongly influenced by parameters related to chemotaxis, such as the chemotactic coefficient $\chi_0$, and the lower-bound of attractant detection, $a_m$.

%To obtain analytical solutions of the GE model and the described expansion dynamics, we first translate the equation into....\jc{say a few words what is done....how to call this - traveling wave analysis?}
The GE model describes a system of nonlinear coupled partial differential equations (Eqs.~\textbf{\ref{eq-rho}-\ref{eq-a}}) which has a degree of 4 and is accompanied by appropriate initial values and boundary conditions. For our system, we specify the initial values to be a localized profile for $\rho$ (any localized profile converges to the same steady state solution) and a uniform attractant concentration denoted by $a_0$. In 1D and with $x$ denoting the spatial coordinate, we look for a stable traveling-wave solution of the form $$\rho(x,t)=\rho(z),\ a(x,t)=a(z);\ \text{with } z=x-ct$$ where $c>0$ is the expansion speed. This converts the system of coupled partial differential equations to two one-dimensional ordinary differential equations as follows:
\begin{align}
-c\frac{d\rho}{dz}&=D_\rho\frac{d^2}{dz^2}\rho-\chi_0\frac{d}{dz}\left(\frac{\rho}{a+a_m} \frac{da}{dz}\right)+r\rho\left(1-\frac{\rho}{\rho_c}\right), \label{eq:simplrho}\\
-c\frac{da}{dz}&=D_a\frac{d^2}{dz^2} a-\mu\frac{a}{a+a_m}\rho. \label{eq:simpla}
\end{align}
In Eqs.~\textbf{\ref{eq:simplrho}-\ref{eq:simpla}}, we have taken an additional simplifying assumption that $a_m=a_k$. For the well-characterized model organism \coli, the uptake and sensitivity of the major attractant aspartate are both $\sim 1~\mu M$~\cite{schellenberg1977resolution,Wong-ng2016,vaknin2007physical,neumann2012chemotactic,Cremer2019a}. Relaxation of this assumption affects the results only weakly as will be discussed below.\par 
Eqs.~\textbf{\ref{eq:simplrho}-\ref{eq:simpla}} are supplemented by boundary conditions that describe limiting values for the bacterial density and attractant concentration far from the front:
\begin{align}
\lim_{z\rightarrow-\infty}\rho\rightarrow\rho_c,\ \lim_{z\rightarrow-\infty}a\rightarrow0;\ \lim_{z\rightarrow\infty}\rho\rightarrow0,\ \lim_{z\rightarrow\infty}a\rightarrow a_0.\label{bc}
\end{align}\par

Fig.~1 shows the numerically obtained steady state profiles emerging from Eqs.~\textbf{\ref{eq:simplrho}-\ref{eq:simpla}} using experimentally established model parameters~\cite{Cremer2019a}. All numerical solutions were obtained using Finite-Element simulations~\cite{Alnaes2012a,LoggMardalEtAl2012a} (see \emph{Materials and Methods}). The density profile (orange line) has a distinct peak at the front which defines the appearance of the ``migrating band'' observed in experiments~\cite{adler1966chemotaxis,adler1966effect,Cremer2019a}. %\jc{is this comment needed?} The formation of a distinct density peak requires a large ratio of $a_0/a_m$.
%and as is shown numerically in Fig.~2B.
and can be divided into three distinct regimes: the \emph{Growth regime} (left of the density trough), the \emph{Chemotaxis Regime} (the rising part of the density profile), and the \emph{Diffusion Regime} (right of the density peak), as indicated in Fig.~2B. Such a distinction reflects the fact that, as we will show below, in each of these regimes, either the Growth, Chemotaxis, or Diffusion term dominates in \eqref{eq:simplrho} respectively.\par %These regimes are individually discussed below.\jc{add clarifying senence why this is done....analysis of dynamics in these regimes and their coupling allows to obtain an analytical understanding of expansion dynamics}.
%In the Growth regime, $a(z)\ll a_m$, and thus $v\approx 0$ and the chemotactic term is very small and negligible. Also, in the Diffusion regime, $a(z) \approx a_0$ and thus $\frac{da}{dz}$ is very small, such that $v\approx 0$ and thus the chemotactic term is very small and again negligible. The Chemotaxis Regime is the regime in which $a_m\lesssim a(z)\ll a_0$ and the chemotactic term is significant, and thus leads to a locally maximal bacterial density.\par Though the profiles are affected by the carrying capacity, $\rho_c$, we will first analyze the case where $\rho_c\rightarrow\infty$ where there is no density-dependent inhibition of growth. The effect of $\rho_c$ will be explored later. As shown in Fig.~3, as long as $\rho_c$ well exceeds the locally maximal value of the density peak in the Chemotaxis Regime, it makes a negligible difference to the propagation of the traveling wave. Our analysis of the Chemotaxis Regime for biologically relevant parameters demonstrates the selection of a single expansion speed that will establish the profiles in the other two regimes.\par

\section*{Chemotaxis Regime}
\textbf{Heuristic derivation of the expansion speed.} We first analyze the most striking feature of the traveling wave, the density bulge. % As we will see, the density bulge determines the 
%The Chemotaxis Regime determines the expansion speed and is thus discussed first.
%
Initially, we consider Eqs.~\textbf{\ref{eq:simplrho}-\ref{eq:simpla}} in the limit that $\rho_c\to\infty$ (this assumption will be relaxed later). We start with the following \emph{ansatz}:
\begin{equation}
\rho(z)=\beta\cdot (a(z)+a_m) \label{ansatz}
\end{equation}
with $\beta$ being a proportionality constant. This reduces \eqref{eq:simpla} to a homogeneous linear differential equation in $a(z)$ with constant coefficients. The solution to such an equation is an exponential function, $a(z)\propto \exp(\lambda z)$, with  $\lambda$ satisfying
\begin{equation}
    -c\lambda = D_a\lambda^2-\mu \beta . \label{beta}
\end{equation}\par
The \emph{ansatz} \eqref{ansatz} also simplifies \eqref{eq:simplrho} considerably, with the penultimate term on the right hand side now proportional to $d^2a/dz^2$. Another consequence of the \emph{ansatz} is that $\frac{d}{dz} \rho(z)=\beta \frac{d}{dz} a(z)$, a relation that will be used often in our calculations. With the \emph{ansatz}, \eqref{eq:simplrho} simplifies to
\begin{equation}
      - c \lambda= (D_\rho - \chi_0) \lambda^2+r\left(1+\frac{a_m}{a(z)}\right). \label{lambda1}
\end{equation}
To proceed further, we consider the case that growth is much smaller than chemotactic drift so that the term proportional to $r$ on the RHS of \eqref{lambda1} may be neglected. This requires that both of the following conditions be true: The first is a condition on the parameters such that 
\begin{equation}
    r\ll \lambda c, \label{rllc}
\end{equation}
which is equivalent to assuming that the timescale of growth is much larger than the timescale of chemotactic drift and thus the two timescales may be separated. As we will show later, this assumption corresponds to a broad, biologically relevant parameter regime and is independent of the growth rate itself (because $\lambda c$ turns out to be proportional to $r$). The second is a condition on the values of attractant concentration $a(z)$, 
\begin{equation}
    a(z)\gg \frac{ r}{\lambda c}a_m. \label{a_condition}
\end{equation}
As we will show below, the quantity on the RHS of \textbf{\ref{a_condition}} is approximately the value of the attractant concentration at the trough of the density profile (i.e, the left boundary of the Chemotaxis Regime).% . This second condition is therefore equivalent to the definition of the Chemotaxis Regime\jc{too complicated to understand - needs to be simplified - maybe better to start with that condition}. 
Thus, for growth small compared to chemotactic drift (i.e., the condition \textbf{\ref{rllc}}), \eqref{lambda1} becomes independent of $a(z)$ in the Chemotaxis Regime. This means \eqref{eq:simplrho} is a linear equation involving $\rho(z)$, $a(z)$, and their derivatives, and it (self-consistently) admits the \ansatz \eqref{ansatz} as a solution. With the last term in \eqref{lambda1} neglected, the solution to $\lambda$ is readily obtained, i.e.,
\begin{equation}
\lambda=\frac{c}{\chi_0-D_\rho}, \label{c_lambda}
\end{equation}
where the solution $\lambda=0$ is rejected as it does not solve \eqref{beta}.  In this regime, the solution to the attractant concentration can be explicitly written as \begin{equation}
    a(z) = a_m\, \exp[\lambda\cdot (z-z_m)] \label{a_C}
\end{equation}
where $z_m$ is defined by $a(z_m) = a_m$. 
\par 
To obtain a relation for the expansion speed $c$ and its dependence on the model parameters, we note that Eqs.~\textbf{\ref{beta}} and \textbf{\ref{c_lambda}} are by themselves insufficient, since there are three unknown quantities: $c$, $\lambda$ and $\beta$. To obtain a defined solution, we thus invoke the boundary conditions at $z=+\infty$ well outside the Chemotaxis Regime (\eqref{bc}) . This is done by integrating \eqref{eq:simplrho} and \eqref{eq:simpla} from a position $z=\zd$ in the Chemotaxis Regime to $z=+\infty$. For \eqref{eq:simplrho} with $\rho_c\to +\infty$, we obtain
\begin{equation}
    c\rho(\zd) = -D_\rho \frac{d\rho}{dz}(\zd)  + \chi_0  \frac{\rho(\zd)}{a(\zd)+a_m}\frac{da}{dz}(\zd) + r \N(\zd), \label{eq:int-rho}
\end{equation}
where $\N(\zd) \equiv \int_{\zd}^\infty \rho(z)dz$ is the total bacterial population to the right of $\zd$. Note that \eqref{eq:int-rho} is exact and independent of our \emph{ansatz}. For $\zd$ located in the Chemotaxis Regime, we plug in our \emph{ansatz} \eqref{ansatz} and \eqref{a_C}  to \eqref{eq:int-rho}, yielding
\begin{equation}
    c\beta (a(\zd)+a_m) = -(D_\rho-\chi_0) \beta \lambda a(\zd) +r\N(\zd). \label{Nzd}
\end{equation}
Note that while the term with growth rate $r$ was negligible in \eqref{lambda1}, it
%While we assumed that growth is much smaller than chemotactic drift in the differential form of \eqref{eq:simpla}, net growth 
cannot be neglected in the integral form as it involves contributions by $\rho(z)$ outside of the Chemotaxis Regime. %Thus, the last term is kept in \eqref{eq:int-rho} and \eqref{Nzd}. 
Using \eqref{c_lambda}, \eqref{Nzd} simplifies to
\begin{equation}
    c\beta a_m = r \beta (a(\zd)+a_m)/\lambda  +r\N(\zd). \label{Nzd1}
\end{equation}
\par 
Now, while \eqref{Nzd1} provides us another equation for $c,\beta$ and $\lambda$, we have a new unknown, $N(\zd)$. But, another relation for $\N(\zd)$ is obtained by integrating both sides of \eqref{eq:simpla} from $\zd$ to $+\infty$, yielding
\begin{equation}
-c(a_0-a(\zd))=-D_a\lambda a(\zd)-\mu [\N(\zd) - \Delta \N(\zd)], \label{Nzd2}
\end{equation}
where $\Delta \N(\zd) \equiv \int_{\zd}^\infty {a_m \rho(z)}/{(a(z)+a_m)}\ dz$. We show in \textit{Supplemental Text} S5 that $\Delta \N(\zd) \sim {\cal O}({a_m\beta}/{\lambda}) \ll \N(\zd)$ for $r\ll\lambda c$. Neglecting $\Delta \N(\zd)$ in \eqref{Nzd2} and using \eqref{beta},  we obtain
\begin{equation}
c a_0 \approx \mu\beta a(\zd)/\lambda +\mu \N(\zd). \label{Nc}
\end{equation}

Eqs.~\textbf{\ref{Nzd1}} and \textbf{\ref{Nc}} allow us to eliminate $\N(\zd)+\beta a(\zd)/\lambda$ and explicitly obtain the proportionality constant of the \emph{ansatz} \eqref{ansatz},
\begin{equation}
    \beta=\frac{ra_0}{\mu a_m}\frac{1}{\left(1-\frac{r}{\lambda c}\right)}\approx \frac{ra_0}{\mu a_m}\label{beta_form}.
\end{equation} 
The explicit value of $\beta$ now allows us to solve for $\lambda$ and $c$ using Eqs.~\textbf{\ref{beta}} and \textbf{\ref{c_lambda}}:
\begin{align}
\lambda &\approx \sqrt{\frac{r\, a_0/a_m}{\chi_0-D_\rho+D_a}} , \label{lambda} \\
c &\approx(\chi_0-D_\rho)\sqrt{\frac{r\, a_0/a_m}{\chi_0-D_\rho+D_a}}. \label{c_main}
\end{align}
%with an error of the order of $r/\lambda c\ln(r/\lambda c)$ \an{this matches numerical error, though numerical error could also be due to limits in resolution}. \par 
%
From Eqs.~\textbf{\ref{lambda}-\ref{c_main}}, we find that the condition $r\ll \lambda c$ amounts to the following condition of the parameters:
\begin{equation}
%r\ll \lambda c\implies 
\frac{a_0}{a_m}\gg 1+ \frac{D_a}{\chi_0-D_\rho} \label{rlc_ao}.
\end{equation}
Thus, the requirement for our \emph{ansatz} to hold translates to an equivalent condition on the chemotactic model parameters that is independent of the growth rate $r$. As detailed below, this parameter regime is typical for the study of migrating bands, with $(\chi_0-D_\rho)$ a few fold below $D_a$ for bacteria in soft agar, and comparable to $D_a$ in liquid medium, while $a_m$ is several orders of magnitude smaller than $a_0$.
%The characteristics of $c$ will be explored in Sec.~2b\jc{remove link}.
\par

\vspace{6pt}\noindent
\textbf{Parameter dependences of the expansion speed.} To validate our heuristic approach we compared the derived relation for the expansion speed, \eqref{c_main}, with numerical simulations, obtaining an excellent match for a broad range of model parameters. We show the dependences on growth rate, uptake rate, background attractant concentration and the attractant diffusion coefficient in Fig.~2.

Firstly, there is a square root dependence on the growth rate $r$, as validated by numerical results in Fig.~2A. This demonstrates that the well-known square-root dependence of $c_F$, the Fisher speed, on growth rate is preserved in the GE model. The expansion speed is further increased by the square root of the relative background attractant concentration, $\sqrt{a_0/a_m}$ (Fig.~2B). However, the expansion speed $c$ does not depend on the specific rate of attractant uptake $\mu$ (Fig.~2C) nor the inoculum population size (as the steady state bulge size is an emergent property, independent of the initial population size). %The dependences on $\mu$, $a_0$ and the initial population size are in contrast to those of the KS solution; we obtain the same dependence of $c$ on $\mu$ and $N_0$ ($c=\mu N_0/a_0$) for the GE model as that given by the KS solution. However, unlike in the KS model where the population of the migrating band is fixed by the initial condition (due to the lack of growth), in the GE model, %changes in $\mu$ is ``absorbed'' by $N_0$ so that $c$ is unchanged.
%
%we find that the effects of $\mu$ and $N_0$ are `canceled' out due to the emergent proportionality constant, $\beta$, of our \ansatz.\par% (see \eqref{N0}).\par 
%
%$N_0$ has a factor of $\beta$, an emergent proportionality constant in our system that determines the height of $\rho(z)$, and $\beta\propto1/\mu$ (see \eqref{beta_form}). Thus, the effects of $\mu$ and $N_0$ are `canceled' out.
The independence on $\mu$ is particularly counter-intuitive since it is the uptake of attractant that establishes the attractant gradient which in turn drives the chemotactic movement. The independence on $\mu$ is in contrast to the KS model, which predicts that $c=\mu N_{KS}/a_0$ (where $N_{KS}$ is the inoculum population size), but is in agreement with experimental results~\cite{Cremer2019a,adler1966chemotaxis}. We will show below that our solution for $c$ can be similarly expressed in terms of $\mu$ and $N_0$, the size of the density bulge. But unlike the KS solution, $N_0$ is here an emergent quantity that turns out to be inversely proportional to $\mu$. Thus, the dependence on $\mu$ `cancels' out, making the expansion speed independent of $\mu$.

The most nontrivial aspect of \eqref{c_main} is perhaps the predicted dependence of the expansion speed $c$ on the attractant diffusion coefficient $D_a$ (Fig.~2D) which was not considered in most previous models~\cite{Keller1971traveling,keller1974mathematical,novick1984gradually,rosen1974propagation}. Although this dependence itself is not so strong, it significantly affects the dependence of $c$ on the cellular motility characteristics as we discuss next.\par %and reveals how non-trivial scaling behaviors emerge in this system due to the interplay of the three diffusion-like parameters $D_\rho, \chi_0$, and $D_a$. 

%In fact, there is a non-monotonic relation between $c$ and $a_0/a_m$ for a finite carrying capacity derived below.\par

To see how the expansion speed depends on the cellular motility parameters $D_\rho$ and $\chi_0$ we first note that $D_\rho$ and $\chi_0$ result from the run-and-tumble dynamics and are thus both proportional to $v_0^2\tau$%(assuming that the direction of motion does not change much over the scale of $\tau$)
, where $v_0$ is the run velocity, and $\tau$ is the average duration of runs. The  ratio $\chi_0/D_\rho$ results from the properties of the flagella motor, the ligand/chemotactic receptor interaction, and the chemotactic signaling network~\cite{si2012pathway}.
%\jc{correct? before only sensing was written here so it was unclear} \an{I don't understand the distinction. I thought that only sensing was taking place, and no signaling} %Consequently, if the internal state of the cell or the medium is changed, $D_\rho$ and $\chi_0$ are affected in the same way. 
To better describe the differences, we here define the chemotactic sensitivity, $\phi\equiv(\chi_0-D_\rho)/D_\rho$, a dimensionless parameter such that a large value of $\phi$ represents a strong chemotactic response to a ligand. Notably, $D_\rho$ can vary across a broad range depending on the environment, with $D_\rho\sim 50~ \si{\mu m^2/s}$ for \coli swimming in soft agar~\cite{Cremer2019a}, and $D_\rho\sim 1000~\si{\mu m^2/s}$ in liquid media~\cite{ford1991measurement}. In contrast, $\phi$ is not expected to be affected by environmental obstacles but by the chemotactic properties of the cell and the type of attractant, and is found to vary from from 1 to 5~\cite{si2012pathway}. We can thus keep $\phi$ and $D_\rho$ as independent parameters and write the expansion speed, \eqref{c_main}, as
%Thus, $D_\rho$ is an effective motility coefficient, while $\phi$ reflects the microscopic chemical and biological properties of the system. Increasing $D_\rho$ while keeping $\phi$ constant reflects greater motility (either due to changes in the medium or the internal state of the bacteria), and increasing $\phi$ while keeping $D_\rho$ constant reflects a stronger sensitivity of the bacterial species to the attractant. 
\begin{equation}
c \approx D_\rho\phi \sqrt{\frac{r\, a_0/a_m}{D_\rho\phi+D_a}}. \label{cphi}
%c \approx D_\rho\sqrt{\frac{\phi r\, a_0/a_m}{D_\rho+D_a/\phi}}. \label{cphi}
\end{equation}
%$D_\rho$ is typically around 50$$ \si{\mu m^2/s} for bacteria such as \coli in soft agar, and 1000 $$ \si{\mu m^2/s} for fast swimming bacteria in liquid media.

%this is too complicated, why to stress this here?: and the error in our calculations is of the order of the error observed due to the finite spatiotemporal resolution.

%\jc{the following is still not quite right, we need to distinguish referring to panels a and b in Fig.~4. Also, in line with the discussion about the role of $D_a$ wouldn't it be much easier to show two different cases in each panel with different $D_a$ values?} \an{That's what we had initially, but the fact is that $D_a$ doesn't change much between agar and liquid media, so the more reasonable change should be in $D_\rho$ and $\phi$}

The predicted comparison with numerical solutions confirms the dependence on the cellular parameters: Notably, for high cellular motility, $D_\rho\phi\gg D_a $, \eqref{cphi} gives $c\approx \sqrt{D_\rho\phi r a_0/a_m}$, as seen in Fig.~3A-B (the solid dark blue lines show the analytical prediction for $\phi=5$). The thick cyan lines show a square root fit. On the other hand, in the range $D_\rho\phi\ll D_a$, $c\propto D_\rho\phi\sqrt{ r a_0/a_m}$ and thus, has a linear dependence on the motility parameter and the chemotactic sensitivity (thick yellow lines).% \an{I need to change the colors in the plots to make it uniform}.

%The dependence of the expansion speed on the relative value of $D_a$ reveals a crucial role of the molecular diffusion of the attractant which has historically been assumed to be of a much smaller scale than the motility-induced bacterial diffusion and chemotaxis~\cite{Keller1971traveling,ahmed2010microfluidics,horstmann20031970,keller1980assessing}. Large $D_a$ can be understood to result in a ``smoothening'' of the attractant gradient, thereby slowing down chemotaxis. Quantitatively, the molecular diffusivity ($D_a \approx 800 $ \si{\mu m^2/s}) well exceeds the chemotactic coefficient and the effective cell diffusitivy of \coli in soft agar ($\chi_0 \approx 300 $ \si{\mu m^2/s} and $D_\rho \approx 50 $ \si{\mu m^2/s} according to Cremer and Honda et al.~\cite{Cremer2019a}), thus making the dependence of $c$ on $\chi_0$ approximately linear. Note also that with these parameter values, $(\chi_0-D_\rho)/D_a \approx 0.3$, and the condition \eqref{condition} is satisfied for $a_0 \gg 3 a_m \approx 3 $ \si{\mu M}. \par
The dependence of the expansion speed on the value of $D_a$ (Fig.~2D) and its relation to $D_\rho$ (Fig.~3) reveals a crucial role of the molecular diffusion of the attractant, which has historically been assumed to be of a much smaller scale than the motility-induced bacterial diffusion and chemotaxis~\cite{Keller1971traveling,keller1974mathematical,novick1984gradually,rosen1974propagation,ahmed2010microfluidics,horstmann20031970,keller1980assessing}. Large $D_a$ can be understood to result in a ``smoothening'' of the attractant gradient, thereby slowing down chemotaxis. In fact, for extremely large values of $D_a$, we note that the bacterial population is unable to establish a gradient in the attractant concentration and our analysis fails to hold as seen in the self-consistency condition \eqref{a_condition}.
%This is clearly seen in the self-consistency condition \eqref{a_condition} which requires that the background attractant concentration relative to the limit of sensitivity be greater than the spread of the attractant relative to the spread of the bacteria due to chemotaxis. Otherwise, diffusion of the attractant is too great for a significant attractant gradient to be generated by the consumption of motile bacteria. 
Quantitatively, the molecular diffusivity ($D_a \approx 800~ \si{\mu m^2/s}$) well exceeds the chemotactic coefficient and the effective cell diffusivity of \coli in soft agar ($D_\rho \approx 50~\si{\mu m^2/s}$)~\cite{Cremer2019a}. Hence, the condition \eqref{a_condition} is satisfied for $a_0 > 4 a_m \approx 4~\si{\mu M}$; thus explaining the deviation seen at small $a_0/a_m$ for $D_\rho=50~\si{\mu m^2/s}$ (see red circles in Fig.~2B).\par% In contrast the molecular diffusivity is comparable to the effective cell diffusivity of \coli in liquid media ($D_\rho \approx 1000 $ \si{\mu m^2/s})~\cite{ford1991measurement}, thus making the dependence of $c$ on $\phi$ approximately linear. With these parameter values, $D_\rho\phi/D_a$ ranges from $\sim 1/3$ to $\sim 7$. \par

We also verified the dependence of the expansion speed on $\phi$ itself for $\phi> 1$ (Fig.~3B). For $\phi<1$, the numerical values do not match the analytical values as they are beyond the regime of self-consistency discussed above. In this case, the traveling-wave solutions transition to the pulled wave dynamics of the F-KPP equation, with a lower bound on the expansion speed given by the Fisher Speed ($c_F=2\sqrt{D_\rho r}$); see \textit{Supplemental Figure }S3. \par
%The form of $c$ in "eq. 23" shows that $c$ is positively dependent on both $\phi$ and the growth rate $r$. Biologically, $\phi$ and $r$ are properties of the cell that depend on the anture of the substrates: $\phi$ depends on the attracant molecule and $r$ depends on the nutrient source. It is not necessary easy to find a substrate that is both a good nutrient and a good substrate. 
%underscores the benefit of using a substrate as an attractant as opposed to as a nutrient. For example, both aspartate and serine are strong attractants but poor nutrients, and are thus most potent when used in combination with a good nutrient source such as glucose, which is a mediocre attractant~\cite{adler1966effect}. Thus, separating the role of substances as being either nutrients or attractant not only relaxes mathematical constraints that emerge if growth is dependent on the attractant, but it also relaxes the biological constraint so that good attractants need not be good nutrients. These results provide an important support for the central thesis of the work by Cremer and Honda et al., that chemotactic cells gain fitness by expanding in nutrient-replete conditions as a `foresighted' navigation strategy~\cite{Cremer2019a} and thus reiterates that nutrients utilized for growth and attractant should be distinguished in mathematical models.\par
\vspace{6pt}\noindent 
\textbf{Effect of carrying capacity.} Next, we consider the effect of a finite carrying capacity $\rho_c$ and the corresponding effect on expansion. To do so, we follow a similar approach as above; see \textit{Supplemental Text S6} for details of the calculations performed. Incorporating the effect of $\rho_c$ lead us to the following form for the expansion speed,
\begin{equation}
c=\left. c_\infty \middle/ \sqrt{1+\frac{ r a_0}{\mu\rho_c}\frac{D_\rho\phi\gamma}{\left(D_\rho\phi+D_a\right)}\frac{a_0}{a_m}}\right. ,
\label{c-vs-a0}
\end{equation}
where $c_\infty$ is the expansion speed for infinitely large carrying capacities, $\rho_c\rightarrow\infty$ as given by \eqref{c_main}, and $\gamma$ is a dimensionless function determined by the shape of the density bulge. While we are unable to determine the exact functional form of $\gamma$, we find an excellent agreement between the numerical results and analytical solution for the best-fit value of $\gamma$ (found to be $\gamma=0.26$ for $D_\rho = 50~\si{\mu m^2/s}$ and $\gamma=0.36$ for $1000~\si{\mu m^2/s})$ as seen in Fig.~4A.\par% and Sxyz. From the fits obtained for the relation of $c$ to all the other model parameters, we can numerically confirm that $\gamma$ is only be a function of $\phi,\ D_a$ and $D_\rho$.\par

An intriguing prediction of \eqref{c-vs-a0} is a peak in the relation between $c$ and $a_0$ whose existence is numerically confirmed (Fig.~4A). Thus, too much attractant actually reduces the expansion speed, i.e., the expansion speed of the population cannot be arbitrarily increased merely by increasing the ambient attractant concentration, but is limited ultimately by the physiological and molecular parameters. %The maximum expansion speed is obtained for $a_0 \approx 0.4$ mM, and the numerical results are closely described by our analytical expression. 
\par

To understand this non-monotonic dependence, we note that in \eqref{c-vs-a0}, the effect of $\rho_c$ is insignificant for $\rho_c \gg r a_0^2/(\mu a_m) = \beta\cdot a_0$, i.e., if $\rho_c$ is large compared to the highest density expected from the \emph{ansatz} \eqref{ansatz} when $a(z)\to a_0$. For sufficiently large $a_0$ such that $\rho_c < \beta a_0$, the quantity 
%$ \mu \rho_c/r \ll a_0^2/a_m$, where 
$\mu \rho_c/r$ (which describes the amount of attractant taken up by bacteria at the peak density, where $\rho(z)\approx\rho_c$, in one doubling time) becomes small, and the population is unable to take up the attractant fast enough to generate a substantial gradient in $a(z)$. The lack of a substantial gradient in turn leads to mitigated expansion speeds. % and for large enough $a_0$, leads to a decrease in expansion speeds. %As an alternate perspective, the dependence of $c$ on the uptake rate $\mu$ can be understood as a consequence of the inability of the population to compensate for changes in $\mu$ by adjusting the height of the density peak (as it did when $\rho_c\to\infty$) when the latter is capped by the carrying capacity. 
We note that the existence of a peak in expansion speed for varying background attractant concentrations was observed experimentally and reported already over 30 years ago~\cite{wolfe1989migration,Cremer2019a}, but was believed to be due to receptor saturation. Our analytical solution in \eqref{c-vs-a0}, validated by simulation (Fig.~4A), provides an excellent quantitative explanation of this phenomenon even in the absence of receptor saturation. We note that for small $\rho_c$,\eqref{c-vs-a0} simplifies to $c\propto \sqrt{\mu/a_0}$. Thus, for small carrying capacity, $c$ increases with $\mu$ and decreases with $a_0$, qualitatively similar to the relation found by Keller and Segel ($c\propto \mu/a_0$). \par
The attractant concentration for the maximum expansion speed is found to be
\begin{equation}
%\frac{a_0^{\text{max}}}{a_0}=\sqrt{\frac{\rho_c}{\beta a_0\gamma}\left(1+
%\frac{D_a}{\chi_0-D_\rho}\right)}.
\frac{a_0^{\text{max}}}{a_m}=\sqrt{\frac{\mu\rho_c}{ra_m\gamma}\left(1+
\frac{D_a}{D_\rho\phi}\right)}
\label{a0-max}
\end{equation}
and is validated numerically in Fig.~4B.
The corresponding maximum expansion speed is $c_\text{max}=c_\infty(a_0=a_0^{\text{max}})/\sqrt{2}$, and the corresponding carrying capacity is proportional to $(a_0^{\text{max}})^2$. Thus, for the population to maximize its expansion speed at high attractant concentrations, a very high carrying capacity is required. As the carrying capacity is typically no more than a few OD for aerobically grown cells, the attractant concentration for the maximum expansion speed, $a_0^{\text{max}}$, is not expected to be above $\sim 0.1~ \text{mM}$; see \eqref{a0-max} and Fig.~4A. This result provides a further explanation for the origin of slow expansion speeds typically obtained for populations growing on substrates that serve as both the attractant and the nutrient~\cite{Cremer2019a}: To support substantial cell growth, the nutrient concentration needs to be substantial, i.e., $5\sim 10~\text{mM}$. But if the nutrient is also the attractant, then the expansion speed for such high attractant concentrations would be substantially less than the maximal expansion speed (see Fig.~4A). This effect likely underscores why it is so advantageous for the nutrient and the attractant to be decoupled as shown experimentally by Cremer and Honda et al.%Its value can be read off by using the value of $a_0^\text{max}/a_m$ (from Fig.~5F) in Fig.~3C.
%The values of $a_0^{\text{max}}$ and $c_\text{max}$ are plotted in Fig.~5B and 5C against the carrying capacity $\rho_c$ for the reference parameters given in Ref.~\cite{Cremer2019a}, for three values of $\gamma$. [some comments after seeing the plot.]
%The relation between $a_0^{\text{max}}$ and $c_\text{max}$ and the other model parameters is plotted in Figs.~5C for the reference parameters given in Ref.~\cite{Cremer2019a}, $\gamma=0.26$ for $D_\rho=50 $ \si{\mu m^2/s} and $\gamma=xyz$ for $D_\rho=1000 $ \si{\mu m^2/s}. Thus, numerical results closely match the analytical expression \eqref{a0-max}, thus strongly confirming our result, \eqref{c-vs-a0}.
 
% \par We also want to remark that due to chemotaxis, the local maximum in the density profile can exceed $\rho_c$, thus becoming the global maximum. In our numerical simulations, this was found to occur for $a_0\geq 0.65$ mM, which is higher than $a_0=0.4$ mM, the value at which the maximum expansion speed for varying $a_0$ was obtained (the locally maximal value of $\rho$ obtained for $a_0=0.4$ mM was $\approx 0.6\rho_c$). Thus, we may conclude that the effective reduction in growth due to the carrying capacity is responsible for the reduction of expansion speeds at higher values of $a_0$.

\vspace{6pt}\noindent 
\textbf{Case of $a_k\neq a_m$:} If we relax the assumption that $a_k=a_m$ and take as our \emph{ansatz} $\rho(z)=\beta(a(z)+a_k)$, we note an additional term in \eqref{lambda1} that is of the order
\begin{equation}
     \frac{(a_m-a_k)a_m a(z)}{(a(z)+a_k)(a(z)+a_m)^2} \label{residual}  
\end{equation}
relative to the dominant chemotactic drift term. It is due to this term that our \emph{ansatz} \eqref{ansatz} fails to hold if $a_k\neq a_m$. A similar term is found in \eqref{Nzd2}. While trivially negligible if $a_k=a_m$, the terms are also negligible for $a(z)\gg a_k,a_m$ and as $a(z)\to0$. Thus, we expect our analysis of the Chemotaxis Regime (and the Growth Regime which we perform below) to also be applicable for the case that $a_k\neq a_m$ as long as $a(z)\gg a_k, a_m$. However, when $a(z)\sim a_m\sim (a_m-a_k)$, our \emph{ansatz} won't hold and the value of $a(z)$ where $\rho(z)$ switches from being relatively constant as in the Growth Regime to rising exponentially as in the Chemotaxis Regime is undetermined by our current analysis. %Thus, the correct \emph{ansatz} would be $\rho(z)=\beta (a(z)+\eta a_m)$ where $\eta$ is in the range $(\min(a_m,a_k)/a_m,\max(a_m,a_k)/a_m)$.
We expect the transition to be at $\eta a_m$, between $a_k$ and $a_m$, as both of these values are crucial in determining the transition in $\rho(z)$. The coupled nature of $\rho(z)$ and $a(z)$ make it difficult to determine $\eta$ exactly. Such an assumption leads to a similar expression for expansion speed, but where $\eta a_m$ replaces $a_m$ in the final form. We find an excellent agreement with numerical results for $a_k\neq a_m$ for just one fitting parameter, $\eta$, which we find to be approximately 2/3 for $a_k=0.1~\si{\mu M}=10 a_m$, and $\eta\approx 3$ for $a_k=10~\si{\mu M}=0.1 a_m$. The range of exponential speeds for different values of $a_k$ while keeping $a_m$ fixed at 1\si{\mu M} is shown in Fig.~5A, 5B. Notably, $c$ is seen to decrease only two-fold for a 2000-fold increase in $a_k$, from 50 nM to 100 $\mu$M for standard parameters (Fig.~5B), while if both $a_k$ and $a_m$ increase 2000-fold, $c$ would decrease 45-fold (see Fig.~2B).

\section*{Diffusion Regime and the Density Peak}
Next, we describe the dynamics of the propagating density profile at its asymptotic front. This is the Diffusion Regime which lies to the right of the density peak (Fig.~1), where the exponential increase of the concentration of the attractant observed in the Chemotaxis Regime is curtailed by the right boundary condition, i.e., $a(z\to\infty) \to a_0$. Here, the drift velocity becomes $v \propto \frac{d}{dz}a(z)/a_0\rightarrow0$, and thus negligible as $z\rightarrow\infty$. The equation for $\rho(z)$ is no longer affected by the attractant, and %Also, as $\rho(z)/\rho_c \ll 1$ at the asymptotic front, the effect of the carrying capacity may be ignored.%We thus again have the F-KPP equation, \eqref{FK}, with the solution given by \eqref{lambda_F}.
the dynamics are thus described by the F-KPP equation. The solution is
\begin{equation}
    \rho(z) = \rho_0 \exp(-\lambda^\pm_D z) \quad \text{with }
    \lambda_D^\pm=\frac{c_D\pm \sqrt{c_D^2-4rD_\rho}}{2D_\rho}, \label{rho_D}
\end{equation}
 where $\rho_0$ is a proportionality constant (see below) and $c_D$ is the speed of propagation of the asymptotic front. \par 
For the front to be a part of the stationary solution that propagates at the same speed as the Chemotaxis Regime, $c$ (\eqref{c_main}), we must have $c_D = c$, which well exceeds the F-KPP speed, $c_F = 2\sqrt{rD_\rho}$. 
 %It is well known for the F-KPP equation that if a uniformly translating front solution with $c_D>c_F$ exists for which the leading asymptotic behavior is given by $\rho(z)\propto\exp(-\lambda_D^+z)$ where $\lambda_D^+>\lambda_F=c_D/(2D_\rho)$, then this front solution is the asymptotic pushed-front solution emerging from compact initial conditions~\cite{van2003front}. This is because the linearized front be at least as steep as $\frac{r}{D_\rho}$, which is only true for $\lambda_D^+$.
 It is well known for the F-KPP equation that if the dynamical system admits a uniformly translating front solution with $c_D>c_F$, then the front solution corresponding to the traveling speed $c_D$ is the stable solution~\cite{van2003front}. And for the case that the front is asymptotic, the initial conditions are compact, and the right boundary condition is the unstable state, $\rho(z\to\infty)=0$), the steeper front solution is selected for~\cite{van2003front} (see \textit{Supplemental Text} S7A for a brief description). 
 Thus, our dynamical system selects for a solution with the leading asymptotic behavior given by 
 \begin{equation}
\lambda_D \equiv \lambda_D^+ =\frac{c_D+ \sqrt{c_D^2-4rD_\rho}}{2D_\rho}\approx c_D/D_\rho \label{lambda_D}
 \end{equation}
for the Diffusion Regime. \par 
 
We then turn to the form of $a(z)$ in the Diffusion Regime. As $a(z)\to a_0 \gg a_m$ in this Regime, \eqref{eq:simpla} becomes
\begin{equation}
    -c_D\frac{da}{dz}=D_a\frac{d^2a}{dz^2}-\mu\rho_0\exp(-\lambda_D z) \label{a_diffusion}
\end{equation}
This is a non-homogeneous linear differential equation in $a(z)$ with the solution %\an{we should keep only one of the two equations below. One is exact, while the other gets rid of $\lambda_D$. I am fine with either}
\begin{equation}
    a(z)=a_0-\frac{\mu\exp(-\lambda_D z)}{\lambda_D(c_D-D_a\lambda_D)}-a_1\exp(-c_Dz/D_a)\label{a_D}
\end{equation}
where $a_1$ is an undetermined constant of integration. The leading behavior is determined by whichever exponential term decaying more slowly: For $\lambda_D > c_D/D_a$ (or $D_\rho<D_a$), 
\begin{equation}
    a_0-a(z) \propto \exp(-c_Dz/D_a),
\end{equation}while for $\lambda_D < c_D/D_a$ (or $D_\rho>D_a$), \begin{equation}
a_0-a(z) \propto\exp(-\lambda_D z).
\end{equation}\par

\section*{Growth Regime and the Density Trough} 
Next, we turn to the Growth Regime which is the region with exponential density profile trailing the density bulge (Fig.~1B).
In this Regime, the increase in $\rho(z)$ as $z\to -\infty$ drives the attractant concentration to zero according to \eqref{eq:simpla}, i.e.,  $a(z)\to0,\ da(z)/dz\to0$ as $z\to -\infty$. Consequently $v(z)\to0$ and 
\begin{equation}
\left| \frac{d}{dz}\left(v(z)\rho(z)\right)\right|\ll c\cdot \left| \frac{d\rho}{dz}\right| \label{v_G}
\end{equation}
in the Growth Regime, sufficiently to the left of the density trough. In the next section, we will quantitatively define the condition where the $v$ term is negligible compared to $c$. Here we briefly describe characteristics of the solution when this condition holds. \par 
Eliminating the term associated with chemotactic drift removes the dependence of $\rho(z)$ on $a(z)$ in \eqref{eq:simplrho}, with the only remaining processes determining $\rho(z)$ being growth and diffusion. Thus, we recover the F-KPP equation, with the solution $\rho(z) \propto \exp[-\lambda^\pm_G z]$, where
\begin{equation}
\lambda^\pm_G = \frac{c_G}{2D_\rho}\pm\frac{\sqrt{c_G^2-4D_\rho r}}{2D_\rho},
\end{equation}
$c_G$ being the traveling velocity of the Growth Regime. As in the Diffusion Regime, here $c_G$ must be the same as $c$, the speed of the Chemotaxis Regime, in order for \eqref{eq:simplrho} to admit a stationary solution. Since $c \gg c_F=2\sqrt{rD_\rho}$, the two solutions are $\lambda_G^+ \approx r/c \ll \lambda_F$ and $\lambda_G^- \approx c/D_\rho \gg \lambda_F$ for $\chi_0\gg D_\rho$. It is well established for the F-KPP equation that for a solution to move stably at a speed exceeding $c_F$, its front must be shallower than $\lambda_F$; see \cite{ebert2000front} and \textit{Supplemental Text} S7B. Hence $\lambda^+_G$ is selected. 
%In the next section, we will show that the latter is indeed the case, with the Growth Regime seeded by a steady ``leakage'' of cells from the Chemotaxis Regime. Thus, the density in the Growth Regime is given by
%Below we will show that the front of the wave (given by the transition to the Chemotaxis Regime) does in fact have a slope shallower than $\lambda_F$, and the Growth Regime is effectively ``seeded'' by a steady ``leakage'' of cells from the Chemotaxis Regime.\par 
Thus, the form of density sufficiently to the left in the Growth Regime must be given by
\begin{equation}
\rho_G(z) = \rho_1 \exp[-\lambda_G\cdot z], \quad \text{with } \lambda_G \equiv \lambda_G^+ \approx r/c, \label{rho_G}
\end{equation}
$\rho_1$ being a proportionality constant that sets the z-scale as will be specified below.

%\section*{Transition between Growth and Chemotaxis Regimes}

To understand how the front of the Growth Regime is ``set'', we focus on the transition region between the Growth and Chemotaxis Regimes (located close to the density trough). %, which determines $\rho(\zbar)=\rho(x_0,t_0)$, the initial seeded population at $(x_0,t_0)$. For simplicity, we will take the limit of large $\rho_c$ since at the front, $\rho(z)/\rho_c\to 0$.\par 
A magnified view of this transition region is shown in Fig.~5A, with the location of the density minimum defined to be at $\zmin$.\par  %(as we will show below, there must necessarily be exactly one local minimum to the left of $z_m$, the position where $a(z)=a_m$). 
%In the Chemotaxis Regime, we had neglected growth in the differential form of the equation, \eqref{eq:simplrho} and thus found that the \emph{ansatz} was valid for the condition that $a(z)\gg (r/\lambda c) a_m$. 
%
Previously, we have shown that for $z>z_m$ (defined by $a(z_m)=a_m$, Fig.~5A) in the Chemotaxis Regime, cell density is given by the \ansatz \eqref{ansatz}, with
the attractant concentration $a(z)$ given by \eqref{a_C}.
We showed that the validity of this \ansatz required $a(z)\gg (r/\lambda c) a_m$, i.e., \eqref{lambda1}. 
However, even with $r\ll \lambda c$, this condition will eventually breakdown for $a(z)\ll a_m$, for $z < z_m$, including possibly the vicinity of $\zmin$; see Fig.~5A. Thus, in order to address the density profile in the transition region, we cannot rely on the \ansatz \eqref{ansatz} anymore. \par 
Here we extend our \ansatz to a new form which we will show to be valid for both the Chemotaxis and Growth Regimes, including all of the transition region:
%, and must first make a new statement on the form of $\rho(z)$ and its relation to $a(z)$ for $z<z_m$. On the other hand, in the Growth Regime, the bacterial density can be described with the exponential function, $\rho(z)\propto\exp(-rz/c)$. However, this is only the solution when $a(z)\ll (r/\lambda c)a_m$. Thus, the transition region, where $a(z)\sim(r/\lambda c)a_m$, requires a different description.\par
%
%For the transition region, we posit a modified version of the \ansatz by combining the Chemotaxis and Growth Regimes to state that
\begin{equation}
    \rho(z)=\beta\, [a(z)+a_m]\cdot\exp[-\lambda_G\cdot (z-z_m)]. \label{modified_ansatz}
\end{equation}
%Further, we claim that the form of the attractant concentration from the Chemotaxis Regime is preserved in the transition region, and we have that
%\begin{align}
%    a(z)=a_m\exp(\lambda (z-z_m)) \label{modified_attr}
%\end{align}
%
Clearly for $a(z)\ll a_m$, \eqref{modified_ansatz} recovers the form of density established for the Growth Regime, i.e., \eqref{rho_G}, with $\rho_1 = \beta a_m e^{\lambda_G z_m}$. For $a(z)\gg a_m$ where $a(z)$ is given by \eqref{a_C} in the Chemotaxis Regime, \eqref{modified_ansatz} becomes
\begin{equation}
\rho(z) \approx \beta a(z) \cdot e^{-\lambda_G (z-z_m)} = \beta a_m \cdot e^{(\lambda-\lambda_G)\cdot (z-z_m)} \approx \beta a(z), \nonumber 
\end{equation}
where the last approximation results from $\lambda_G \ll \lambda$ for our parameter regime $r\ll \lambda c$. Furthermore, we can verify that the new \ansatz \eqref{modified_ansatz} satisfies \eqref{eq:simplrho} for intermediate range of $a(z)$, leaving behind a linear equation for $a(z)$ that is the same as that obtained in the Chemotaxis Regime, with the same solution \eqref{a_C}; see \textit{Supplemental Text} 7B. Our new \ansatz thus leads to the following form for the cell density 
\begin{equation}
\rho(z) = \beta a_m \left[1+e^{\lambda\cdot (z-z_m)}\right] \cdot e^{-\lambda_G (z-z_m)}, \label{rho_CG} 
\end{equation}
which we claim to be valid for the entire regime $-\infty < z< z_m$ (for $r\ll \lambda c$), including the vicinity of the density trough located at $\zmin$. \par 
We can now use the expression given by \eqref{rho_CG} to work out characteristics of the solution in the transition region. By setting $\left.\frac{d}{dz}\rho\right|_{z=\zmin} = 0$, we obtain (for $r\ll \lambda c$):
\begin{align}
    &\zmin = z_m - \lambda^{-1} \ln\left(\frac{\lambda c}{r}\right),  \label{zmin}\\
%&\rhomin = \beta a_m \cdot \left[ 1 + \frac{r}{c\lambda}\cdot \left(1 + \ln\left( 1 + \frac{\lambda c}{r}\right)\right) \right] \approx \beta a_m, \label{rhomin}\\
    &\rhomin \equiv \rho(\zmin) = \beta a_m \cdot \left( 1 + \frac{r}{\lambda c} \right) e^{-(\zmin-z_m)\cdot r/c} \approx \beta a_m, \label{rhomin}\\
    & \amin \equiv a(\zmin) = a_m\cdot \exp[\lambda\cdot (\zmin-z_m)] = \frac{r}{\lambda c} a_m. \label{amin}
\end{align}
These results are validated numerically for a range of parameters; see Fig.~5B-5D. \par  
We can determine the left boundary of the transition region, $z'_m$, by finding the range of $z<z'_m$ where \eqref{rho_CG} is described by the simple exponential form \eqref{rho_G} (dashed green line, Fig.~5A). This can be estimated by setting the asymptotic form 
\begin{equation}
\rho_G(z)\equiv \lim_{z\to-\infty}\rho(z) = \beta a_m e^{-\lambda_G\cdot (z-z_m)} \label{rho_G2}
\end{equation}
to $\rho_G(z'_m) = \rhomin$. 
Using \eqref{rhomin} for $\rhomin$, we find
\begin{equation}
    z'_m = \zmin - \lambda^{-1}\ln\left(\frac{\lambda c}{r}\right). \label{zm2}
\end{equation}
In other words, \eqref{rho_G2} can be written as 
$\rho_G(z) = \rhomin e^{-\lambda_G (z-z'_m)}$.
 Note that because $\lambda_G (\zmin-z'_m) \ll 1$ according to \eqref{zm2} for $r\ll \lambda c$,  $\rho_G(z) \approx \rhomin$ for $z'_m < z < \zmin$, i.e., the density function on the left side of $\zmin$ is constant with relative variation of the order of $r/\lambda c$.  %\par 
[We can verify the self-consistency of the new \ansatz \eqref{rho_CG} by using it to compute the drift velocity $dv(z)/dz$ and hence evaluate the spatial domain where the condition \textbf{\ref{v_G}} is satisfied. We find that \textbf{\ref{v_G}} is satisfied for 
%\begin{equation}
$e^{\lambda\cdot (\zmin-z)} \gg 1$,%\nonumber
%\end{equation}
or $z< \zmin - \lambda^{-1}\ln(\lambda c/r)$, which is the same as the condition \textbf{\ref{zm2}}.]\par 
To summarize, the transition region between the Chemotaxis and Growth Regimes range from $z'_m < z < z_m$ where the distance from $\zmin$ to $z_m$ and $z'_m$ are given by \eqref{zmin} and \eqref{zm2}, respectively. %The density $\rho(z)$ is highly skewed, with $\rho(z'_m)\approx \rhomin$ and $\rho(z_m) \approx 2\rhomin$. 
The total width of the transition zone is 
\begin{equation}
    w\equiv z_m - z'_m = \frac{2}{\lambda}\ln(\lambda c/r). \label{width}
\end{equation}
Note that the time it takes for the wave-front to migrate across the transition region is $\tau = w/c$. Thus, the key condition for our results, $r\ll \lambda c$ corresponds simply to $r\tau \ll 1$, i.e., a separation of time scale between expansion and population growth. This is a condition which we expect to hold for most expanding populations.% and is independent of the growth rate itself (see \eqref{rlc_ao}).
\par
\section*{The Growth-Leakage Balance}
We can finally use the explicit solution for $\rho(z)$ to connect the dynamics in the Chemotaxis and Growth Regimes. We consider the total bacterial population to the right of a position $x=z+ct$, which is co-moving with the population: $\tilde{N}(z;t)\equiv\int_{z+ct}^{+\infty} dx' \rho(x',t)$. The change in $\tilde{N}(z;t)$ over time is given formally by
%\begin{align}
%\frac{dN}{dt} &=\frac{d}{dt}\int_{x-ct}^\infty dx' \rho(x',t)\\
%&=-c\rho(x,t)+\int_{x-ct}^\infty dx' \frac{\partial \rho(x',t)}{\partial t}\\
    %&=-c\rho(x,t)+\int_{x-ct}^\infty dx' \left(D_\rho\frac{\partial^2 \rho(x,t)}{\partial x^2}-\frac{\partial}{\partial x}(v(x,t)\rho(x,t))+r\rho(x,t)\right)\\
%    &=\underbrace{-(c-v(x,t))\rho(x,t)-D_\rho\frac{\partial \rho(x,t)}{\partial x}}_{\text{Loss of cells due to chemotaxis and diffusion}}+\underbrace{rN(x,t)}_{\text{Recovery due to Growth}} \label{loss_reco}
%\end{align}
\begin{equation}
     \frac{d\tilde{N}}{dt} = -\tilde{J}(z;t) + r\cdot \tilde{N}(z;t), \label{growth-leakage balance}
\end{equation}
where 
\begin{equation}
    \tilde{J}(z;t) = (c-v(z+ct,t))\rho(z+ct,t) + D_\rho\left.\frac{\partial \rho}{\partial x}\right|_{z+ct}\nonumber
\end{equation}
obtained from taking time derivative of $\tilde{N}$ using \eqref{eq-rho}, is the ``leakage flux'' which includes the loss of cells across the position $x=z+ct$ in the lab frame due to chemotaxis and diffusion, and the last term in \eqref{growth-leakage balance} describes the growth of the cells in the region $x> z+ ct$. \par
In the absence of growth $r=0$, Novick-Cohen and Segel \cite{novick1984gradually} showed that incorporating the lower Weber cut-off to the KS Model led to the loss of cells from the front, and subsequently the slowdown of the migrating wave-front. We see from \eqref{growth-leakage balance} that the incorporation of growth, even at very low rates, allows the migrating wave-front to ``replenish'' itself and thereby maintain stability.\par
In the stationary state ($\frac{d}{dt}\tilde{N}=0$), quantities in the moving frame have no time dependence, i.e., $\tilde{N}(z;t) = N(z)$. Therefore, 
\begin{equation}
    rN(z) = J(z) \equiv (c-v(z))\rho(z) + D_\rho\frac{d \rho}{dz},
    \nonumber 
\end{equation}
which is just \eqref{eq:int-rho} with $v(z)$ given by $\rho(z)$ and $a(z)$ that solve the stationary equations, \eqref{eq:simplrho} and \eqref{eq:simpla}. Earlier, we solved \eqref{eq:int-rho} using the \ansatz \eqref{ansatz} that holds only in the Chemotaxis Regime with $z>z_m$. We can repeat the calculation using \eqref{rho_CG} and \eqref{a_C} derived from our new \ansatz \eqref{modified_ansatz}. We find the leakage flux to be very weakly $z$-dependent in the vicinity of the density trough, i.e.,
\begin{equation}
    J(z) = J_0 \cdot \left[1+{\cal O}(r\cdot (\zmin-z)/c)\right] \quad \text{for } z'_m < z < z_m. \nonumber 
\end{equation}
where
\begin{equation}
    J_0\equiv J(\zmin) = c\rhomin\cdot \left[1-\frac{r}{\lambda c} \frac{\chi_0}{\chi_0-D_\rho}\right] \approx c\rhomin. \label{J_min}
\end{equation}
Since $|z-\zmin| < \lambda^{-1}\ln(\lambda c/r)$ according to \eqref{zmin} and \eqref{zm2}, we conclude that $J(z)$ is within the order $r/(\lambda c)\ln(\lambda c/r) \ll 1$ around $J_0$. 
 Consequently, $N(z)$ is nearly $z$-independent also, reflecting the sharply-peaked structure of the density front.  %Note that this result does not require knowing the precise shape of the front.  \par 
For convenience, we define $N_0 \equiv N(\zmin)$ as the size of the population in the density bulge. The above results then lead to an important biological relation
\begin{equation}
rN_0 = J_0, \label{seeding}
\end{equation}
with the bulge size given by
\begin{equation}
    N_0 = J_0/r \approx c\rhomin/r. \label{N0}
\end{equation}
\eqref{seeding} describes a balance of the growth of the cells in the front and their leakage behind the front, as depicted in Fig.~6. At a given instance (time $t_0$), the wave-front is shown as the dashed red line in the lab frame. The front region, comprised of $N_0$ cells, grow at a rate $rN_0$. This growth is balanced by cells leaving the front (i.e., across the black dashed line indicating $x_0=z_m + c t_0$), with flux $J_0 = -c\rhomin$. At some time $\delta t$ later, the front has traversed a distance $\delta x = c \cdot \delta t$. The total amount of cells leaving the front during this time is $\delta N = J_0 \delta t$. The corresponding density of the cells left behind the propagating front is $\delta N/\delta x \approx \rhomin$ (shown as the purple region in Fig.~6A). 
The cells left behind will grow at the rate $r$. For $\delta t$ much smaller than the doubling time, the density behind the front will not have grown much and thus remain at $\sim \rhomin$ (Fig.~6A). We have shown that this is the case for the time it takes for the front to traverse the width of the trough region (\eqref{width}). After a time $\Delta t$ large compared to the doubling time, the population size at the back will become $\rho(x_0, t) = \rho(x_0,t_0)\, e^{r\Delta t}=\rho(x_0,t_0)\, e^{r(t-t_0)}$ (Fig.~6B). Given that $t_0 = (x_0-\zmin)/c$, we have 
\begin{equation}
\rho(x_0, t) \approx \rhomin \exp\left[-\frac{r}{c}(x_0-ct)\right]. \label{trail}
\end{equation}
Thus, the trailing exponential density profile \eqref{trail}, while looking like a moving front, is merely a result of the exponential growth of a stationary population, which is \emph{seeded} by the traveling wave-front at density $\rhomin$ and speed $c$. 
\par 
Finally, we note that the picture depicted in Fig.~6A can be used directly to predict the value or $\rhomin$ without going through detailed calculation: Since the bacteria are concentrated in the density bulge, the removal of the attractant is almost entirely due to uptake by cells in the density bulge. %for whom attractant availability is saturated. 
This gives us the mass-conservation condition\footnote{This relation can also be obtained systematically from our solution by using $\rhomin\approx \beta a_m$ (from \eqref{rhomin}) and the expression for $\beta$ from \eqref{beta_form} in \eqref{N0}. Since the result for $\beta$ was invoked, it involves the approximation made following \eqref{Nzd2}. This reflects the fact that in arriving at \eqref{N0.2}, we assumed that attractant uptake is always saturating.}
\begin{equation}
\mu N_0 \approx c a_0. \label{N0.2}
\end{equation}
The growth-leakage balance $rN_0 = J_0$ then gives $J_0 = c a_0 r/\mu$. The consideration described in Fig.~6A then immediately gives the result that the density left behind the front bulge, which would be $\rhomin$, is given by $J_0/c = a_0 r/\mu$. Thus, we obtain a surprisingly simple result,
\begin{equation}
   \rhomin \approx a_0 r/\mu \label{trough}
\end{equation}
independent of the other details of the system. \par
We can also use the expression for $\rhomin$ thus obtained to calculate the consumption of attractant around the density trough. Using $\rho(z) = \rhomin$ and $a(z)$ from \eqref{a_C}, \eqref{eq:simpla} becomes
\begin{equation}
    -c\lambda =D_a\lambda^2-\frac{\mu\rhomin}{a_m}=D_a\lambda^2-r \frac{a_0}{a_m}.
\end{equation}
This relation together with the proportionality between $\lambda$ and $c$, \eqref{c_lambda}, immediately gives the central result on the expansion speed, \eqref{c_main}. This simple line of consideration reveals the underlying origin of the dependence of the expansion speed on $a_0/a_m$: The growth-leakage balance relates the ambient concentration $a_0$ to the trough density $\rhomin$ (\eqref{trough}), and the balance between attractant uptake $\mu\rhomin$ and drift/diffusion at the trough relates $c$ and $\lambda$ to $\rhomin$ and $a_m$.% In the simplest case where $D_a=0$, uptake at the trough, of rate $\mu \rhomin/\lambda=ra_0/\lambda$, is balanced by clearance of $a$ at concentration $a_m$, of rate $ca_m$, to yield $\lambda c = ra_0/a_m$. Together with $\lambda = c/(\chi_0-D_\rho)$, we immediately obtain $c=\sqrt{(\chi_0-D_\rho)ra_0/a_m}$.

%ansion speed also reFor exp$D_a\ll \chi_0-D_\rho$, we thus have that
%\begin{equation}
%    \frac{r}{c\lambda}=\frac{a_m}{a_0}\label{rca0_simple}
%\end{equation}
%While we may obtain a more exact expression for our model from \textbf{\ref{lambda}-\ref{c_main}}, \eqref{rca0_simple} is again independent of other details of the system. \eqref{rca0_simple} also provides an explanation for the square root dependence of $c$ on $a_0/a_m$ in our model as from \eqref{lambda1}, $c\lambda\propto c^2$ and thus from \eqref{rca0_simple}, $c^2\propto a_0/a_m.$
\section*{Discussion}
To reveal the underlying dynamics governing chemotaxis-driven population expansion, we analyzed the experimentally verified GE model mathematically~\cite{Cremer2019a}. Following an extensive traveling-wave analysis, we were able to describe the density and attractant profiles throughout the Chemotaxis and Growth Regimes (Fig.~6, \eqref{modified_ansatz} and \eqref{a_C}). %\jc{this is a quite lengthy and detailed summary, I think it should be much shorter with not much repetition from before, just phrasing for which tings analytical relations has been found and then the discussion should start}\jc{I think it becomes also much easier to read if we remove the references to the KS model .... all important references should have been covered in main section not the discussion part}\jc{also probably best to reduce the number of references to figures and equations, just refering to the main eq on exp speed} 
We determined the expansion speed (\eqref{c_main}), and through it, the value of the slope $\lambda$ which specifies the width of the migrating band (\eqref{lambda}). Our results, which are in excellent agreement with numerical simulations for a broad range of model parameters tested (Figs.~2-6), recover many key experimentally-observed relations of the expansion speed to biological and environmental parameters~\cite{Cremer2019a} that previous models based on the KS model had failed to capture~\cite{tindall2008overview, keller1980assessing}. Notably, while our model agrees with the KS model near the density bulge, with the same relation between expansion speed and the size of the peak ($c=\mu N_0/a_0$, \eqref{N0.2}), the size of the peak itself is not a constant as in the KS model, but an emergent quantity. Consequently, expansion speed depends on many of the model parameters.\par 

Firstly, the expansion speed depends on the ratio of the initial attractant concentration to the lower limit of attractant sensitivity (i.e., $c\propto \sqrt{a_0/a_m}$) for large carrying capacity. % This provides an evolutionary path to increasing expansion speed by increasing detection sensitivity (i.e., reducing $a_m$).
For finite carrying capacity our analysis predicts the non-monotonic dependence of expansion speed with initial attractant concentration, providing an explanation for this long-known experimental
observation~\cite{wolfe1989migration}: For lower attractant concentrations, increasing concentration increases the size of the bulge hence promotes faster expansion. %and for large carrying capacity expansion speed increases as the square-root of $a_0$,
But for higher concentrations, the carrying capacity limits the size of the bulge and expansion speed decreases with increasing attractant concentration as it takes longer for the bulge to consume the attractant and establish a gradient (\eqref{c-vs-a0} and Fig.~4A). The same effect is likely responsible for the slow expansion speeds observed when the nutrient and the attractant are the same substance~\cite{Cremer2019a}, since to provide sufficient boost to cell density, a high concentration of nutrient is desired, while if the nutrient is also the attractant, a high concentration of the latter is detrimental to expansion. Thus, this provides a population-level justification for the physiological observation of the separation of the role of a substrate as a nutrient from its role as an attractant~\cite{Cremer2019a}.  \par% This underscores the result of Cremer and Honda et al. that rapid population expansion requires the decoupling of the roles of the attractant and the nutrient to be oriented towards chemotaxis and growth respectively (to relax biological and mathematical constraints). \par% Thus, rapid expansion speeds require the attractant and nutrient substrates to be different as growth requires nutrient-replete conditions (high concentration), while chemotaxis requires that gradients occur sufficiently fast (low concentration).\par

Secondly, our results reveal a dependence of the expansion speed on the diffusion of the attractant ($D_a$, Fig.~2D). The effect of multiple diffusion coefficient-like parameters ($D_\rho,\ \chi_0$, and $D_a$) is one of the reasons the GE model is difficult to analyze. In Cremer and Honda et al., a scaling theory was developed to describe the dependence of the expansion speed on the chemotaxis coefficient $\chi_0$~\cite{Cremer2019a}. Assuming that $\chi_0$ was the main relevant factor, the scaling theory predicted that $c\propto \chi_0$. Our analysis here reveal that $c\propto ({\chi_0-D_\rho})$ holds for large $D_a$ but $c\propto \sqrt{\chi_0-D_\rho}$ for small $D_a$; see Fig.~3.  \par

The analytical understanding attained in this work quantitatively supports the role of chemotaxis in range expansion found by Cremer and Honda et al.~\cite{Cremer2019a}. Particularly, bacterial chemotaxis does not necessarily occur to fulfill an immediate nutritional need, nor does it necessarily reflect an attempt to avoid starvation. For example, cells move chemotactically towards attractants they cannot metabolize and also swim in nutrient-replete conditions~\cite{adler1969chemoreceptors,adler1966chemotaxis,Cremer2019a}. Instead, chemotaxis could be hard-wired to promote the expansion of bacterial populations into unoccupied territories well before nutrients run out in the existing environment; low levels of attractants thus act as aroma-like cues that establish the direction of expansion and enhance the speed of population movement~\cite{Cremer2019a}. Subsequently, cells left behind by the migrating band fully occupy the region behind the front by growing at the rate determined by nutrient availability. This allows the population to expand rapidly into unoccupied territories while still colonizing the territories it has traversed, without one compromising the other.\par

Our results also expand upon the general theory of front propagation into unstable states and reveal a novel mechanism for speed selection. While many studies of front propagation involve modification of the non-linear growth/reaction term in the original F-KPP equation~\cite{van2003front,dee1988bistable,van1989front}, our model considers a drift term which is a functional of an environmental variable, the attractant concentration. Though the canonical results pertaining to the F-KPP equation are not expected to hold in such a two-variable system, the dynamics in the Growth and Diffusion Regimes in our system are effectively described by the F-KPP equation. While the expansion of an F-KPP wave-front ``pushed'' by the bulk (as in the Diffusion Regime) at rates much higher than the stable Fisher speed has long been known~\cite{van2003front,ablowitz1979explicit}, our results demonstrate how F-KPP wave-fronts can also be ``seeded'' by a transition regime at the front (as in the Growth Regime) to attain large expansion speeds. Alternatively put, we can think of chemotaxis in the leading density bulge as a ``trick'' the population uses to propagate faster than predicted by F-KPP equation based on growth and diffusion alone. \par  
Our analysis assumes a separation of time scales between growth (slow) and chemotactic migration (fast), i.e, $1/r\gg 1/\lambda c$, indicating that cell growth is negligible over the time scale the population migrates across the width of the density bulge given by $1/\lambda$. This condition is fulfilled for a broad parameter regime (\textbf{\ref{rlc_ao}}) and particularly holds for chemotactic bacteria. However, we note that relaxing this assumption in future work would be helpful to understand the regime where the chemotactic bias is small, i.e., when $\chi_0\to D_\rho$ where \textbf{\ref{rlc_ao}} breaks down. Numerically, we find that as $\chi_0$ is reduced to the order of $D_\rho$ or smaller, the expansion speed approaches the stable Fisher speed $c_F$ (\textit{Supplemental Figure }S3), which is the expected speed for a ``pulled'' wave solution determined by the asymptotic front~\cite{van2003front}. A solution to the GE model that includes the small-$\chi_0$ regime would provide an analytical connection to the F-KPP equation and thereby provide insight on the transition from the ``pushed'' and ``seeded'' dynamics observed when $r\ll \lambda c$ to the well-established ``pulled wave'' dynamics~\cite{collet1987stability,van2003front,gandhi2016range,erm2020evolution}. % Thus beyond providing a quantitative picture of the transition between the F-KPP and GE models, it will also provide an understanding of the interaction between growth and chemotaxis.
%A full solution to the GE model, including results when \eqref{rlc_ao} is not satisfied, will be needed to fully connect to the F-KPP dynamics and understand the transition from pushed to pulled waves in our system (SI xyz)~\cite{collet1987stability,van2003front,gandhi2016range,erm2020evolution}. This solution will have to tackle growth and chemotaxis simultaneously, and thus beyond providing a quantitative explanation, will also provide an understanding of how growth and chemotaxis-driven flux interact.% Such a solution would also help bring down the error in our solutions \eqref{c_main} and \eqref{lambda}, which is currently of the order of $(r/\lambda c)\ln(r/\lambda c)$.
\par

%\jc{maybe good final sentence}
Finally, we note that the biological features underlying chemotaxis-driven population expansion, including sensing, directed movement, and the modification of environmental conditions, should be generic to many motile organisms. The traveling-wave solutions of the GE model presented here may thus be employed to understand the growth-expansion dynamics of different organisms in diverse ecological contexts.\par 

\par

\matmethods{To generate all of the numerical results, finite element simulations of the system of equations were performed using FeniCs, a computing platform for solving partial differential equations (PDEs). A 1D mesh of resolution 15-50 $\mu$m was used to simulate a moving window of 30 mm (or 120mm for very fast fronts). Finite elements of $\mathcal{P}_3\Lambda^0$ type were used.\par
The initial bacterial density was specified with $\rho(x,t)=(\tanh((1-x^2))+1)\times 0.029/2$ in order to initiate a sufficiently localized initial population with a differentiable functional form. The initial attractant concentration was specified to be constant everywhere. Neumann boundary conditions of zero flux were specified on both ends of the simulation domain. A difference equation was then solved to approximate the differential equation in time using a small time step (typically between 2 and 25 seconds) The resulting solutions were recorded and used for the subsequent iteration of the difference equation.\par 
%As expected, more accurate solutions were obtained for higher-resolution simulations, for both time and space resolution. In particular, lowering the saturation constants for the different reaction and convection terms (i.e., increasing the sensitivity) required substantial increases in the spatial and temporal resolutions. \par
In order to obtain high spatial and temporal resolutions simultaneously, a moving window technique was utilized. In the moving window technique, only a 30mm (or 120mm for very fast fronts) interval was simulated at a time. But when the front of the wave had gone beyond a certain threshold in the simulation domain, the simulation domain was was translated to the right and the attractant concentrations and bacterial densities were extrapolated for the sections of the new simulation domain for which the values weren't previously known. This technique holds very well as long as a threshold sufficiently far from the right end of the domain is chosen (this is also desirable to ignore edge effects) such that the linear extrapolation is correct within numerical resolution.\par
To analyze the simulations and extract the expansion speeds, the position of the maximum drift velocity was recorded for each timestep. A linear fit over time was then employed for the position to obtain the expansion speed. The fit was also curated manually to ensure that the expansion speed was calculated using a period of steady and constant expansion.}
\showmatmethods{} % Display the Materials and Methods section
\acknow{The authors would like to thank Massimo Vergassola, Lev Tsimring, Roman Stocker, Johannes Keegstra, and Francesco Carrara for helpful discussions, and Ying Tang for advice with numerical simulations. This research was supported by Simons Foundation (Grant No. 542387) and the National Science Foundation (MCB 2029574).}% This work was performed on the unceded ancestral homeland of the Kumeyaay people.}

\showacknow{} % Display the acknowledgments section

% Bibliography
\section*{Bibliography}
\bibliography{pnas-sample}
\newpage

\begin{figure}[b!]
\centering
\includegraphics[scale=0.6]{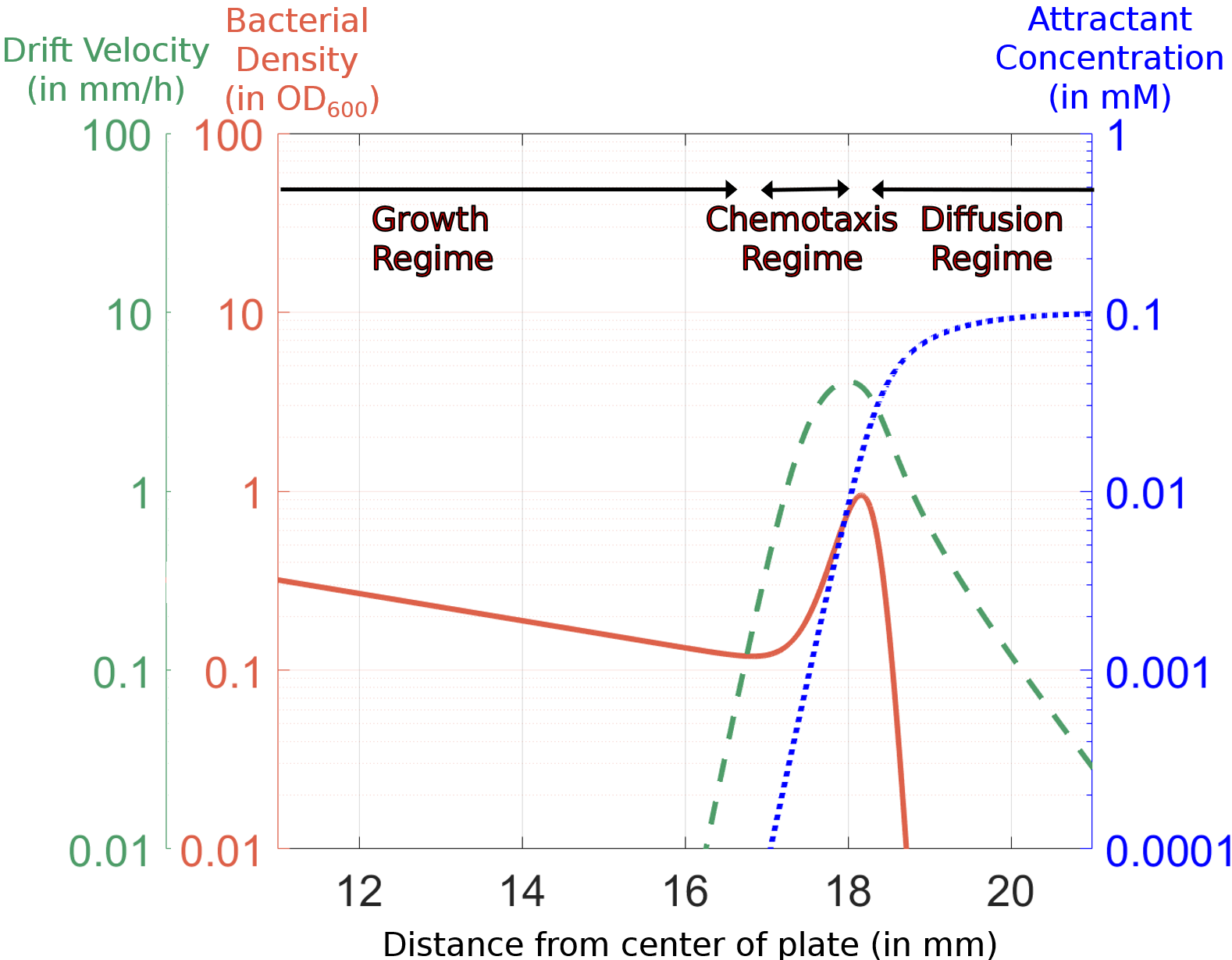}
\caption{%A. A schematic of the GE model as introduced by Cremer and Honda et al. The wave front is shown at two different times. First, it is shown at an earlier time in the top half of the panel where the front is propagating to the right with a given expansion speed. The solid green line is a plot of the bacterial density for different distances from the inoculation site. The front of the wave is shaded cyan. Then, the same front is shown with the solid green line after a doubling time in the bottom half of the panel (the earlier front is represented by a dashed green line). A hypothetical wave front that would result with growth and convection but without diffusion is shown in the orange line. Due to diffusion, the increased proportion of bacteria in the front are left behind to give the resulting wave. Thus, diffusion, growth and chemotaxis act together to result in a stable traveling-wave solution in the GE Model. 
Profiles of bacterial density (solid red line, in $\OD$), drift velocity (dashed green line, in mm/hr) and attractant concentration (dotted blue line, in mM) for a steadily expanding population 14.5 hours after the inoculation. Arrows indicate the different regimes used in the analytical consideration. %C. The height of the density peak (defined as the difference between the locally maximal density in the front, $\rho_{\text{max}}$, and the locally minimal density in the transition regime between the Growth and the Chemotaxis Regimes, $\rho_{\text{min}}$) relative to the locally minimal density $\rho_{\text{min}}$. We plot the relative height for three sets of values of $D_\rho$, $\chi_0$ and $D_a$. It must be noted that the peak height is fairly sensitive to the spatiotemporal resolution of the simulation. 
%The GE model (Eqs.~[\ref{eq:simplrho}-\ref{eq:simpla}]) introduces a finite carrying capacity to account for the limited availability of nutrients. A variant explicitly considering nutrients is showing very similar results (Supplementary Text XY and SI Figure XY). 
Model parameters used are adapted from those determined in Ref.~\cite{Cremer2019a} and are provided in \textit{Supplemental Table }S1 (this simulation used the low motility parameters).}
\end{figure}

\begin{figure}[t!]
\includegraphics[scale=0.85]{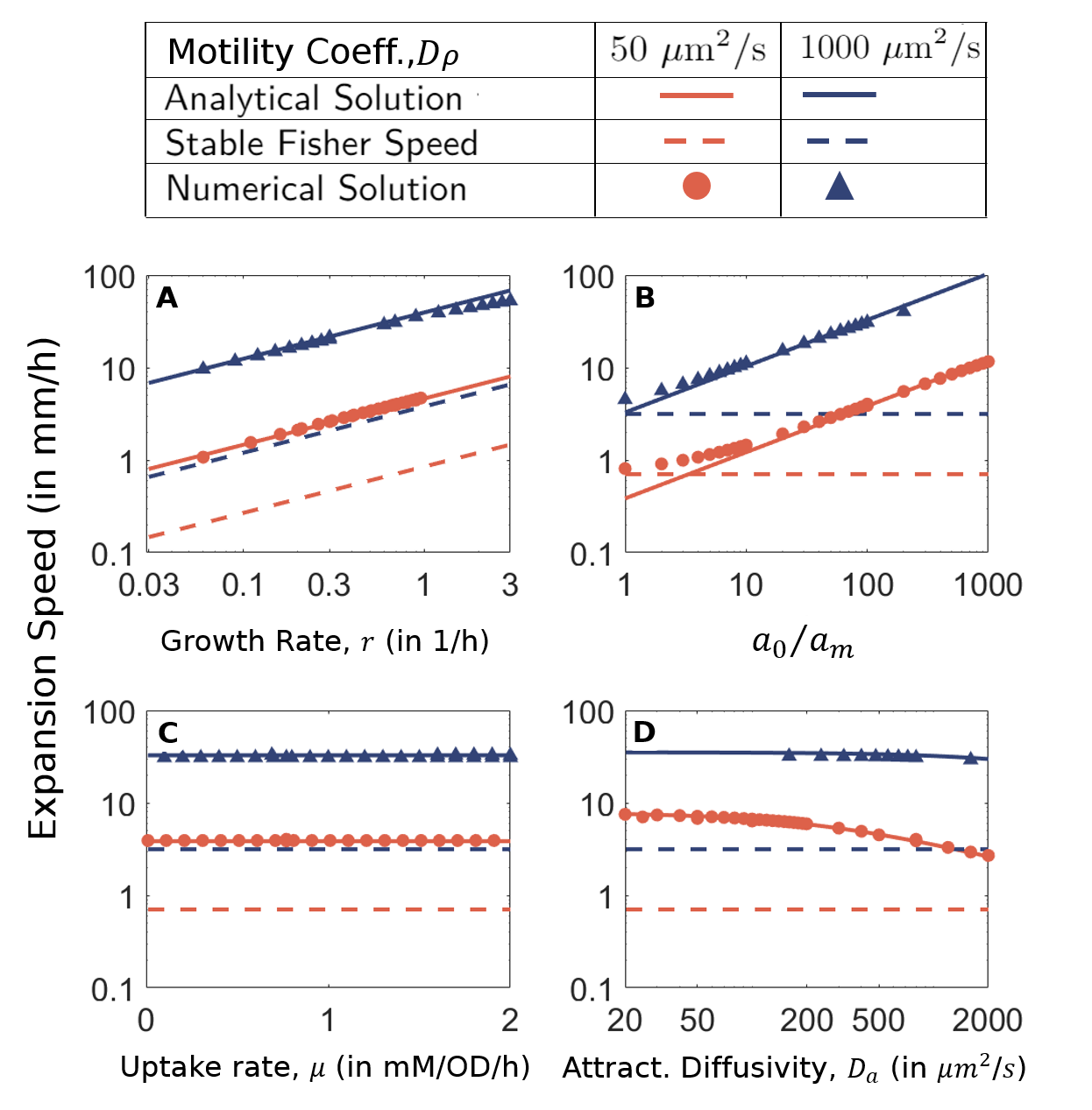}
\caption{Dependence on growth-rate $r$ (\textbf{A}), uptake-rate $\mu$ (\textbf{B}), relative attractant levels $a_0/a_m$ (\textbf{C}), and attractant diffusion $D_a$ (\textbf{D}). Analytical relation for the expansion speed \eqref{cphi} is shown by solid lines ($D_\rho={50,1000}~\si{\mu m^2/s}$ in red and blue, respectively). The corresponding  Fisher speeds, $c_F=\sqrt{r~ D_\rho}$, are denoted by corresponding dashed lines. Numerical solutions of the GE model (Eqs.~\textbf{\ref{eq:simplrho}-\ref{eq:simpla}}) are shown by corresponding symbols. Unless specified, all parameter values are the default values given in \textit{Supplemental Table }S1.}
\end{figure}

\begin{figure}[b!]
\centering
\includegraphics[scale=0.56]{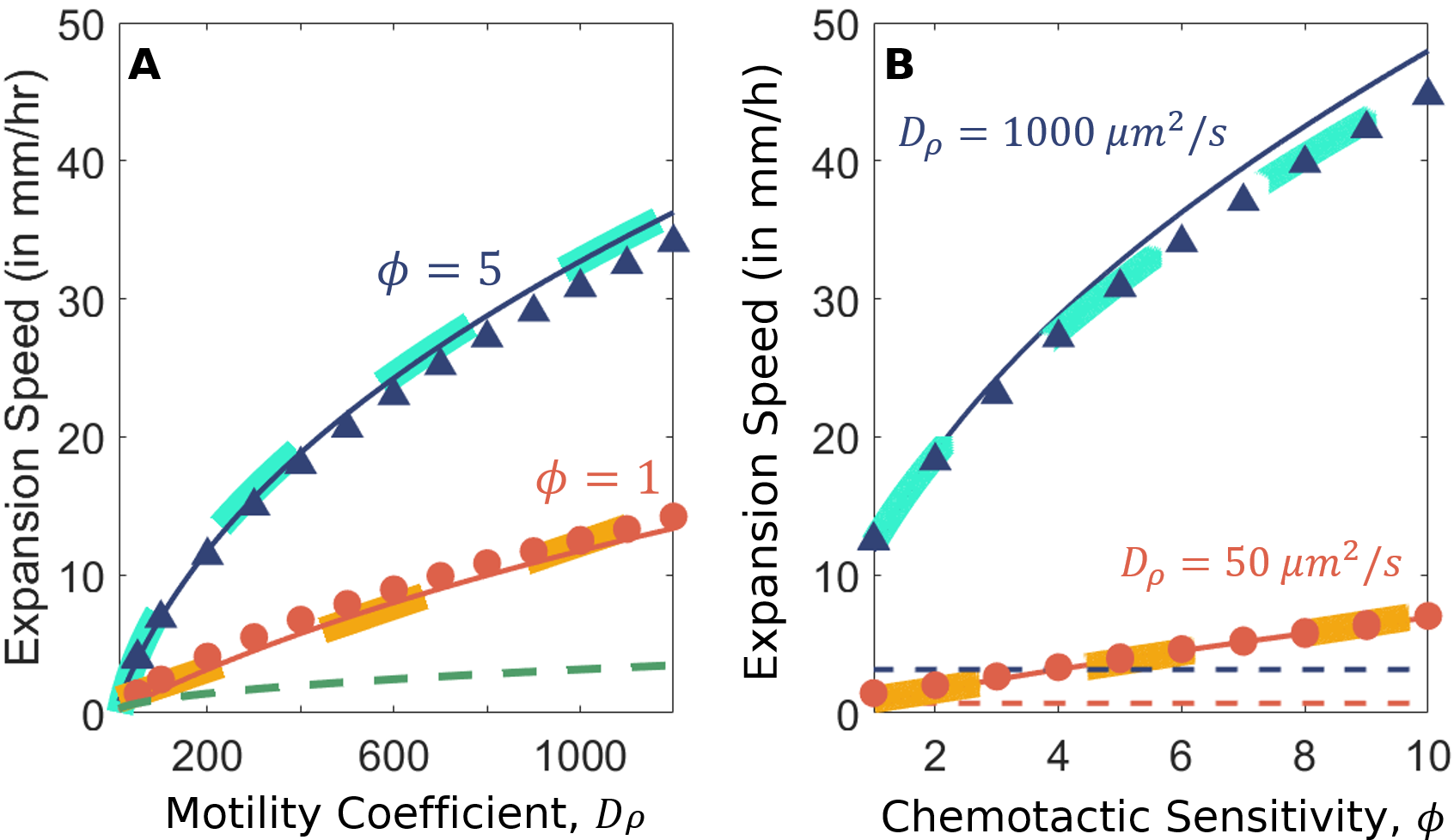}
\caption{Dependence of expansion speed on motility parameters. \textbf{A.} Dependence on cellular motility $D_\rho$.
Numerical solutions for $\phi=1$ and $\phi=5$ are shown by red circles and dark blue circles, respectively. Analytical solutions following \eqref{cphi} are shown by corresponding solid red and blue lines. The green dashed line represents the stable Fisher speed, $c_F=2\sqrt{D_\rho r}$, the minimum expansion speed of our system. \textbf{B.} Dependence on the chemotactic sensitivity, $\phi$. Numerical solutions for $D_\rho = 50~\si{\mu m^2/s}$ and $D_\rho = 1000~\si{\mu m^2/s}$ are shown by red and dark blue circles, respectively. Analytic solution following \eqref{cphi} are shown by the corresponding solid lines. Thick yellow and cyan dashed lines are best fits for the respective values of $\phi$ and $D_\rho$ to demonstrate that $c\propto D_\rho\phi$ for $D_\rho\phi\lesssim D_a$ and that $c\propto \sqrt{D_\rho\phi}$ if $D_\rho\phi$ is large compared to $D_a$. Unless specified, all parameter values are the default values given in \textit{Supplemental Table }S1.}
\end{figure}

\begin{figure}[t!]
\centering
\includegraphics[scale=1]{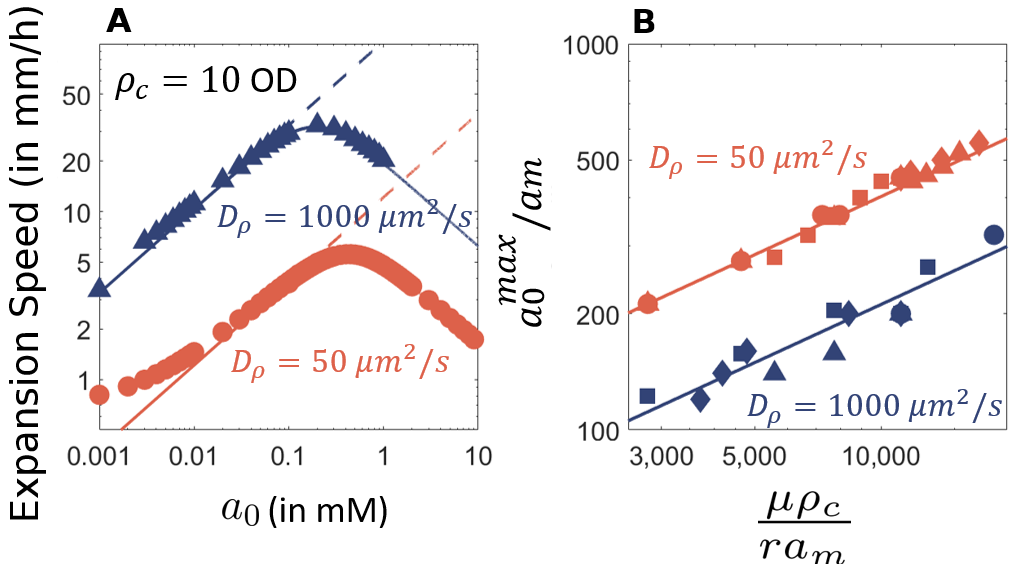}
\caption{Effect of Carrying Capacity.
\textbf{A.} Dependence of expansion speed on the ambient attractant concentration when the carrying capacity is finite ( $\rho_c=10~\OD$). Markers (red circles and blue triangles) indicate numerical values, solid lines indicate analytical predictions as per \eqref{c-vs-a0}, and dashed lines indicate analytical predictions with $\rho_c\to\infty$. All results in red are for $D_\rho = 50~\si{\mu m^2/s}$ and all results in blue are for $D_\rho = 1000~\si{\mu m^2/s} $. \textbf{B.} The ambient attractant concentration resulting in maximum expansion speed  $a_0^{\text{max}}$ is shown depending on the dimensionless parameter $\mu\rho_c/(r a_m)$. The analytical solution, \eqref{c-vs-a0}, is shown as corresponding solid lines. Dashed lines show the solutions ($c_\infty$) without a limiting carrying capacity ($\rho_c\rightarrow\infty$; as shown in Fig.~3). Different symbols in (B) denote which model parameter was varied from its default value (square if $\mu$, circle if $\rho_c$, triangle if $r$, and diamond if $a_m$) for $D_\rho = 50~\si{\mu m^2/s}$ (red) and $D_\rho = 1000~\si{\mu m^2/s}$ respectively. For details please refer to \textit{Supplemental Methods} and to \textit{Supplemental Table }S2 for range of values used for each parameter. Parameters have the default values from \textit{Supplemental Table }S1 unless specified.}
\end{figure}

\begin{figure}[t!]
\centering
\includegraphics[scale=0.55]{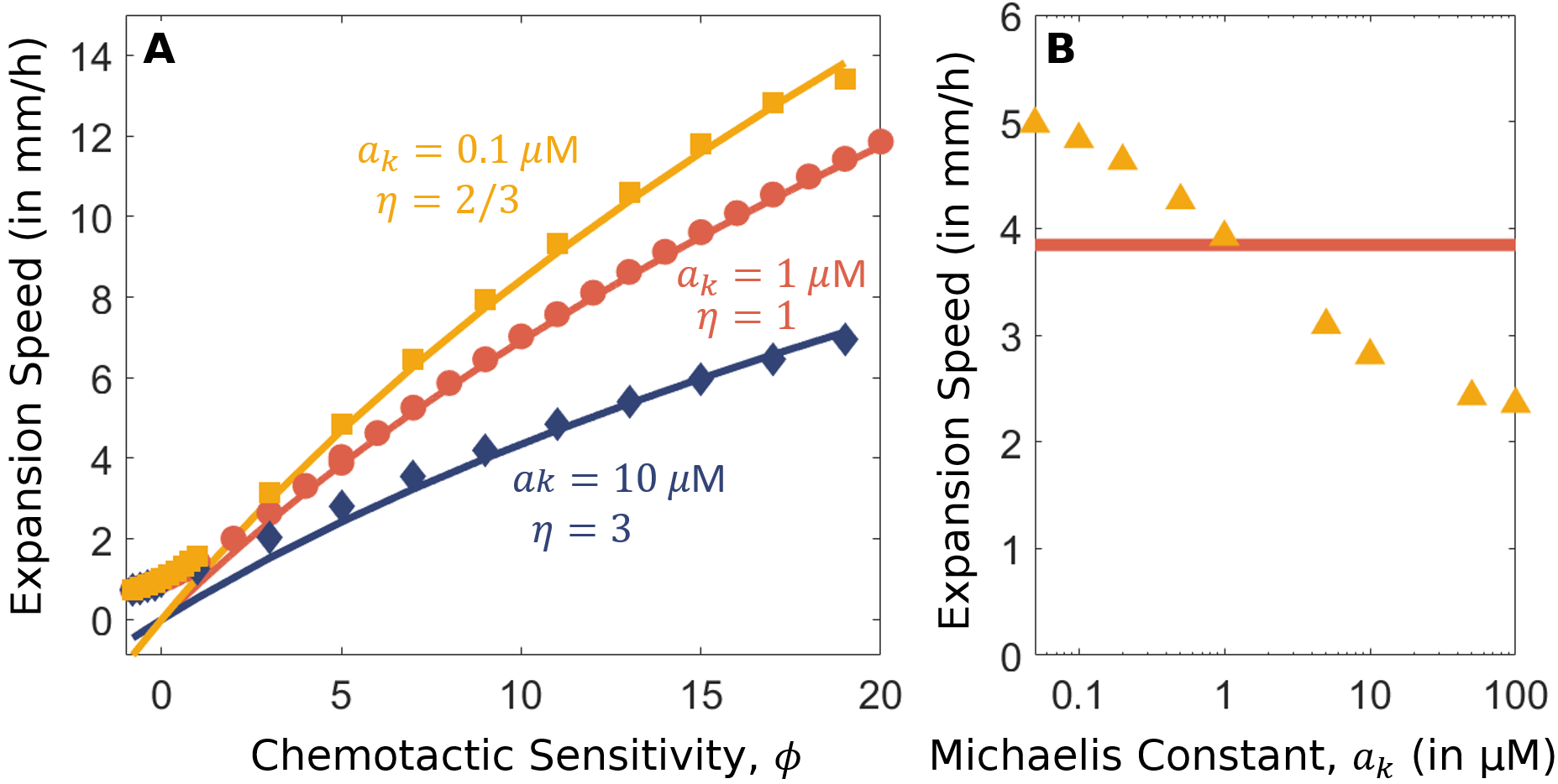}
\caption{Effect of varying Michaelis Constant, $a_k$.
\textbf{A.} Dependence of expansion speed on the chemotactic sensitivity, $\phi$, for different values of $a_k$ and $D_\rho=50~\si{\mu m^2/s}$. Solid lines indicate analytical solutions for corresponding best fit values of $\eta$, and markers denote the numerical solutions. Results for $a_k=0.1~\si{\mu M},\ 1~\si{\mu M} and 10~\si{\mu M}$ are shown in yellow, red and blue respectively. \textbf{B.} Dependence of the expansion speed on model parameter $a_k$. The numerical solutions obtained for $D_\rho=50~\si{\mu m^2/s}$, $\phi=5$ are represented by yellow triangles, and the analytic solution found in \eqref{c_main} for $a_k=a_m=10^{-3}~\si{ mM}$ is shown by the red line. Parameters have the default values from \textit{Supplemental Table }S1 unless specified.}
\end{figure}

\begin{figure}[t!]
    \centering
    \includegraphics[scale=0.5]{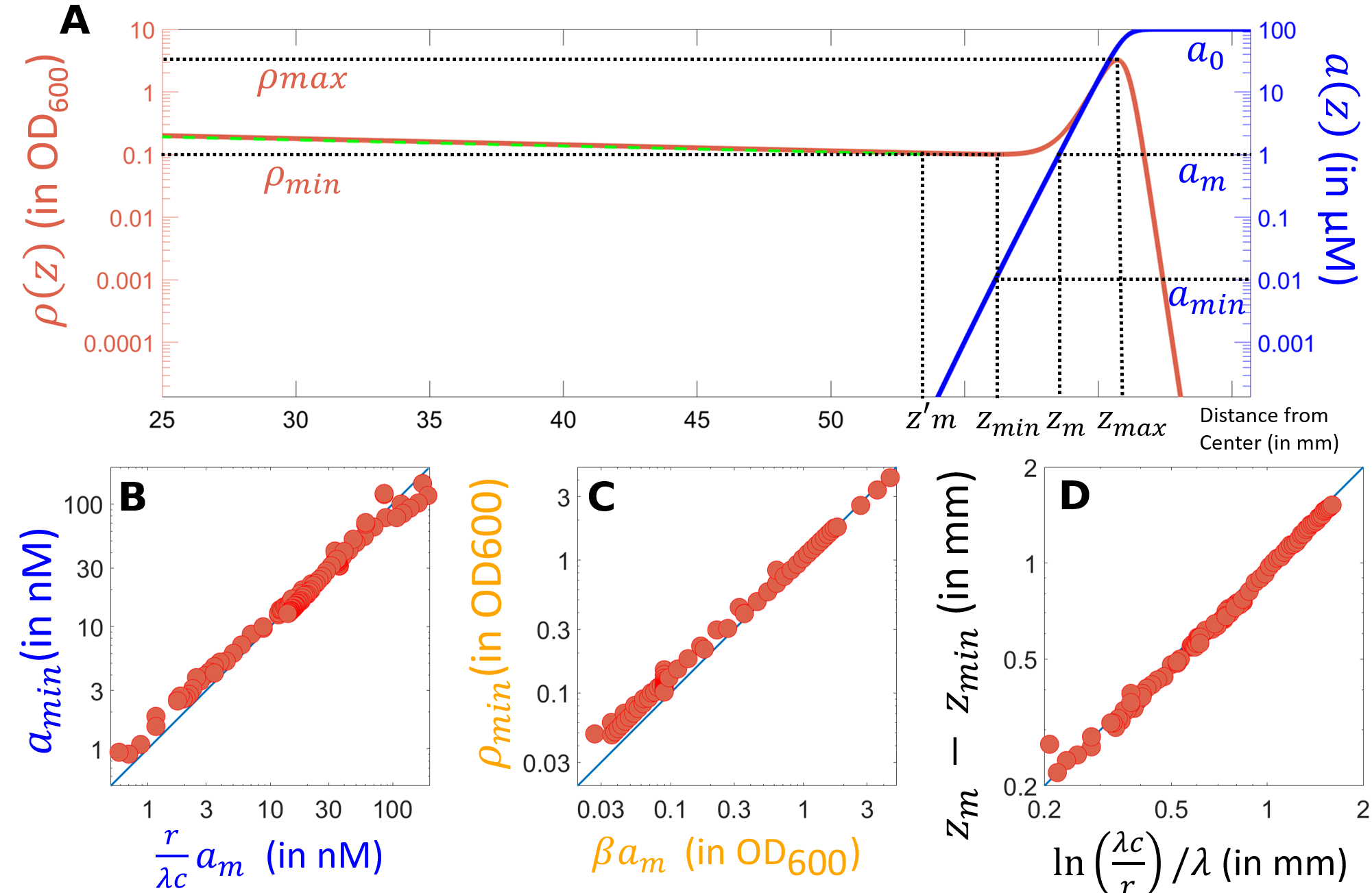}
    \caption{Transition from the Chemotaxis to the Growth Regime. \textbf{A}. steady expansion profiles of $\rho(z)$ (solid red line) and $a(z)$ (solid blue line) for the standard parameters (\textit{Supplemental Table }S1; $D_\rho=50\ \mu m^2/s,\ \chi_0=300 \mu m^2/s$). The profile of $\rho(z)$ as predicted by the \ansatz \eqref{ansatz} is shown using the dashed green line. Dashed horizontal lines indicate distinct values of  $a$ and $\rho$ as indicated. \textbf{B-D}. Numerically obtained values  of $a(\zmin)$, $\rho(\zmin) $, and $z_m-\zmin$ for a broad variation of parameters; seven model parameters in Eqs.~\textbf{\ref{eq:simplrho}-\ref{eq:simpla}} (other than the carrying capacity, which was $>1000$ for all results here) were varied across many decades (see \textit{Supplemental Methods} for details of what was done and \textit{Supplemental Table} S3 for the range of values investigated).  Blue lines show $y=x$ to demonstrate agreement with the predicted values of $a(\zmin)$, $\rho(\zmin)$, and $z_m-\zmin$.}
    \label{growth_figure}
\end{figure}

\begin{figure}[tb!]
    \centering
    \includegraphics[scale=0.64]{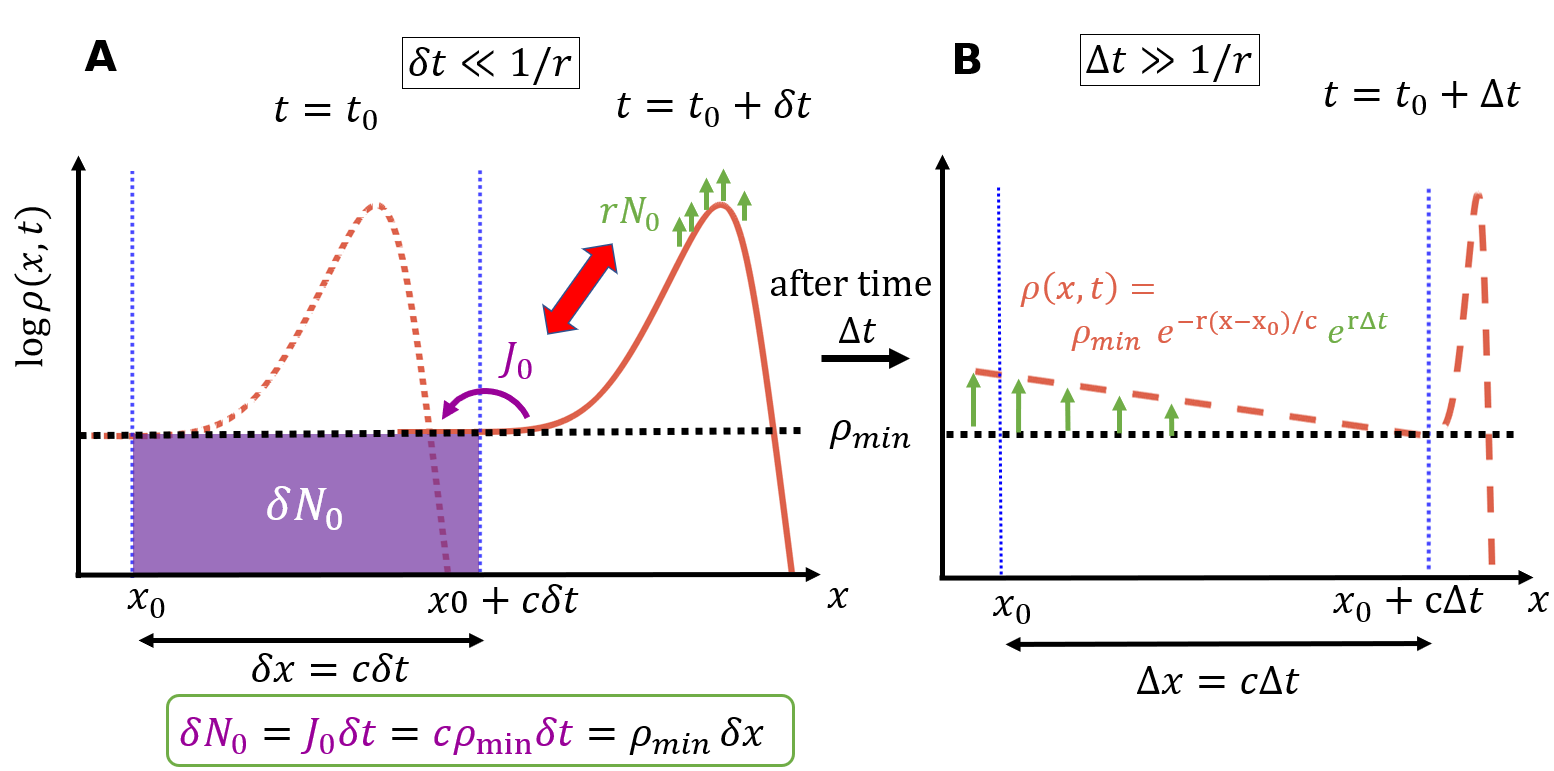}
    \caption{Schematic of the dynamics of the transition between Chemotaxis and Growth Regimes. \textbf{A}. In a short time $\delta t$, the density bulge shown near $x_0$ (dotted red line) moves forward to be near $x_0+c\delta t$ (solid red line). In that time, the density bulge grows by an amount $rN_0\delta t$ and is diminished by ``leakage'' given by an amount $J_0\delta t$. During steady expansion, these values match as stated in our \textit{ansatz} (\eqref{ansatz} and \eqref{modified_ansatz}). The ``leaked'' cells are deposited behind the density bulge where the bacterial density is roughly constant for a distance $\delta x$ (thus, $\rho(x_0,t_0+\delta t)\approx \rhomin$, and the total deposition over time $\delta t$, given by $\delta N_0$ is also equal to $J_0\delta t$. \textbf{B}. After a long time $\Delta t$, the density bulge moves to be near a position $x_0+c\Delta t$ (dashed red line). Cells behind the density bulge grow at a rate $r$ and the density thus accumulates as $\rho(x_0,t)=\rhomin \exp(r\Delta t)$}
    \label{fig:my_label}
\end{figure}
\end{document}

\maketitle

%% Adds the main heading for the SI text. Comment out this line if you do not have any supporting information text.
\onehalfspacing
\tableofcontents
\newpage
\section*{Supplemental Methods}
\addcontentsline{toc}{section}{Supplemental Methods}
All of the numerical results shown in the main text were generated using the finite element method (FEM). Numerical simulations of the time evolution of the system of partial differential equations (PDEs) were performed using FeniCs, a computing platform for solving PDEs~\cite{Kirby2004a,KirbyKnepleyEtAl2005a,KirbyLogg2006a,KirbyLoggEtAl2006a,KirbyScott2007a,KirbyLogg2007a,Logg2007a,OlgaardLoggEtAl2008a,KirbyLogg2008a,Logg2009a,RognesKirbyEtAl2009a,AlnaesLoggEtAl2009a,AlnaesMardal2010a,LoggWells2010a,OlgaardWells2010b,HoffmanJanssonEtAl2012a,JanssonJanssonEtAl2012a,paper17,AlnaesEtAl2012,AlnaesBlechta2015a,LoggMardalEtAl2012a,LoggWellsEtAl2012a,LoggOlgaardEtAl2012a,Kirby2012a,AlnaesMardal2012b,AlnaesLoggEtAl2012a,Alnaes2012a,HoffmanJanssonEtAl2012a,HoffmanJanssonEtAl2012b}. A 1D mesh of resolution 15 $\mu$m was used to simulate a moving window of 30 mm (explained below). Finite elements of $\mathcal{P}_3\Lambda^0$ type were used. The shape function space for $\mathcal{P}_3\Lambda^0$ consists of all differential $0-$forms with polynomial coefficients of degree at most 3, and has dimension 4. The degrees of freedom are given on line segments by moments of the trace weighted by a full polynomial space:
$$u\rightarrow\int_f (\text{tr}_f u)\wedge   q,\ q\in \mathcal{P}_2\Lambda^1(f).$$
The initial bacterial density was specified with $\rho(x,t)=(\tanh((1-x^2))+1)\times 0.029/2$ in order to initiate a sufficiently localized initial population with a differentiable functional form to avoid singularities. The initial attractant concentration was specified to be constant everywhere (with a value of $a_0$ that was an important model parameter). Neumann boundary conditions of zero flux were specified on both ends of the simulation domain. A difference equation was then solved to approximate the differential equation in time using a small time step (typically between 2 and 25 seconds) The resulting solutions were recorded and used for the subsequent iteration of the difference equation.\par 
As expected, more accurate solutions (with a smaller error in the goal functional) were obtained for higher-resolution simulations, for both spatial and temporal resolution. In particular, lowering the saturation constants for the different reaction and convection terms (i.e., increasing the sensitivity) required substantial increases in the spatial and temporal resolutions. In order to obtain high spatial and temporal resolutions simultaneously, a moving window technique was utilized.\par
In the moving window technique, only a 30mm window was simulated at a time. When the front of the wave had gone beyond a certain threshold (chosen to be 60-75\% for our system) in the simulation domain, the simulation domain for the subsequent iteration was then translated to the right by the distance that the front had moved in the last timestep. The attractant concentrations and bacterial densities were extrapolated for the sections of the new simulation domain for which the values weren't previously known (which are just the boundary values and are near constant at steady state for our model formulation). This technique holds very well as long as a threshold sufficiently far from the right end of the domain is chosen (this is also desirable to ignore edge effects) such that the linear extrapolation is correct within numerical resolution. Further, this method requires a smaller time interval (especially for fast-expanding solutions) to ensure that the simulation window isn't translated too much in each timestep.~\par
To analyze the simulations and extract the expansion speeds, the position of the maximum drift velocity was recorded for each timestep. A linear fit over time was then employed for the position to obtain the expansion speed. The fit was also curated manually to ensure that the expansion speed was calculated using a period of steady and constant expansion speed.\par
\textbf{Variation of parameters for Fig.~4B}: For Fig.~4B, the value of $a_0$ that maximizes $c$ for different values of $r$, $a_m$, $\rho_c$, and $\mu$ was sought. To do so, the values of $r$, $a_m$, $\rho_c$, and $\mu$ were varied over the ranges given in Table S2 and while the other values were fixed as given in Table S1 (with the exception of $\rho_c$ that was set at 10 OD unless it was being varied.) Once all the data was generated, we found the value of $a_0$ between $10^{-3}-10$ mM that led to the greatest value of $c$ and plotted the corresponding value of $a_0/a_m$ against the corresponding value of $\mu\rho_c/(ra_m)$ while denoting the parameter varied with a marker as specified in the legend of Fig.~4B.\par
\textbf{Variation of parameters for Fig.~6B-D}: For Figs.~6B-D, all of the data generated for this work was collated and the empirically determined values of $\amin,\ \rhomin,$ and $z_m-\zmin$ was plotted. This involved over 200 data points in which the following 7 model parameters were varied: $D_\rho,\ \chi_0,\ a_m,\ r,\ D_a,\ \mu$, and $a_0$. The ranges of values over which these parameters were varied is given in Table S3. Only results with $\rho_c>1000$, $\phi>1$, $a_k=a_m=1\ \mu M $, and $D_a>10\si{\mu m^2/s}$ were considered. 
\newpage
\section*{Supplemental Figures}
\addcontentsline{toc}{section}{Supplemental Figures}
\begin{figure}[htb!]
\centering
\includegraphics[scale=0.6]{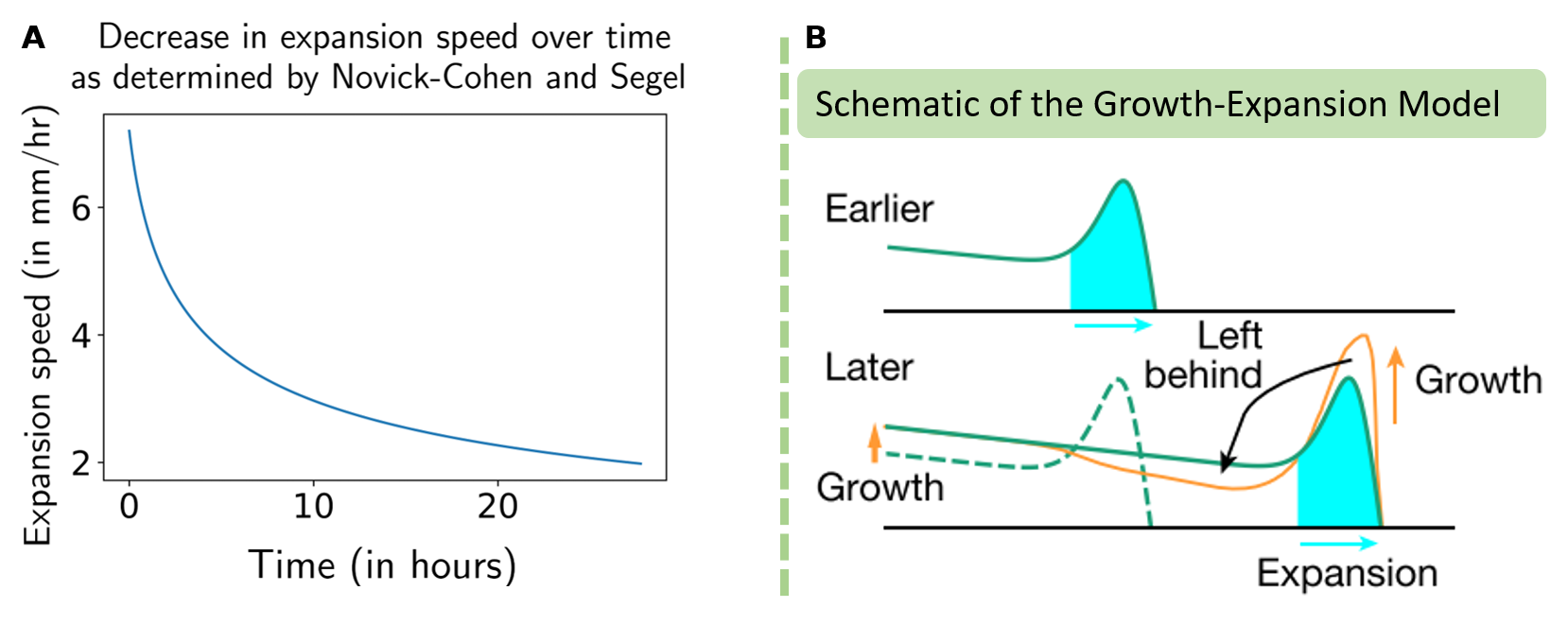}
\caption{The crucial role of growth for traveling-wave solutions. A. Decrease in expansion speed over time upon inclusion of a lower bound in the sensitivity to attractant concentration in the KS model. This is based on the result obtained by Novick-Cohen and Segel~\cite{novick1984gradually} for $D_\rho=50 $ \si{\mu m^2/s} and other parameter values specified in Table S1. B. A schematic of the GE model as introduced by Cremer and Honda et al. The wave front is shown at two different times. First, it is shown at an earlier time in the top half of the panel where the front is propagating to the right with a given expansion speed. The solid green line is a plot of the bacterial density for different distances from the inoculation site. The front of the wave is shaded cyan. Then, the same front is shown with the solid green line after a doubling time in the bottom half of the panel (the earlier front is represented by a dashed green line). A hypothetical wave front that would result with growth and convection but without diffusion is shown in the orange line. Due to diffusion, the increased proportion of bacteria in the front are left behind to give the resulting wave. Thus, diffusion, growth and chemotaxis act together to result in a stable traveling-wave solution in the GE Model. All parameter values are the default values specified in Table S1.}
\end{figure}

\begin{figure}
    \centering
    \includegraphics[scale=0.5]{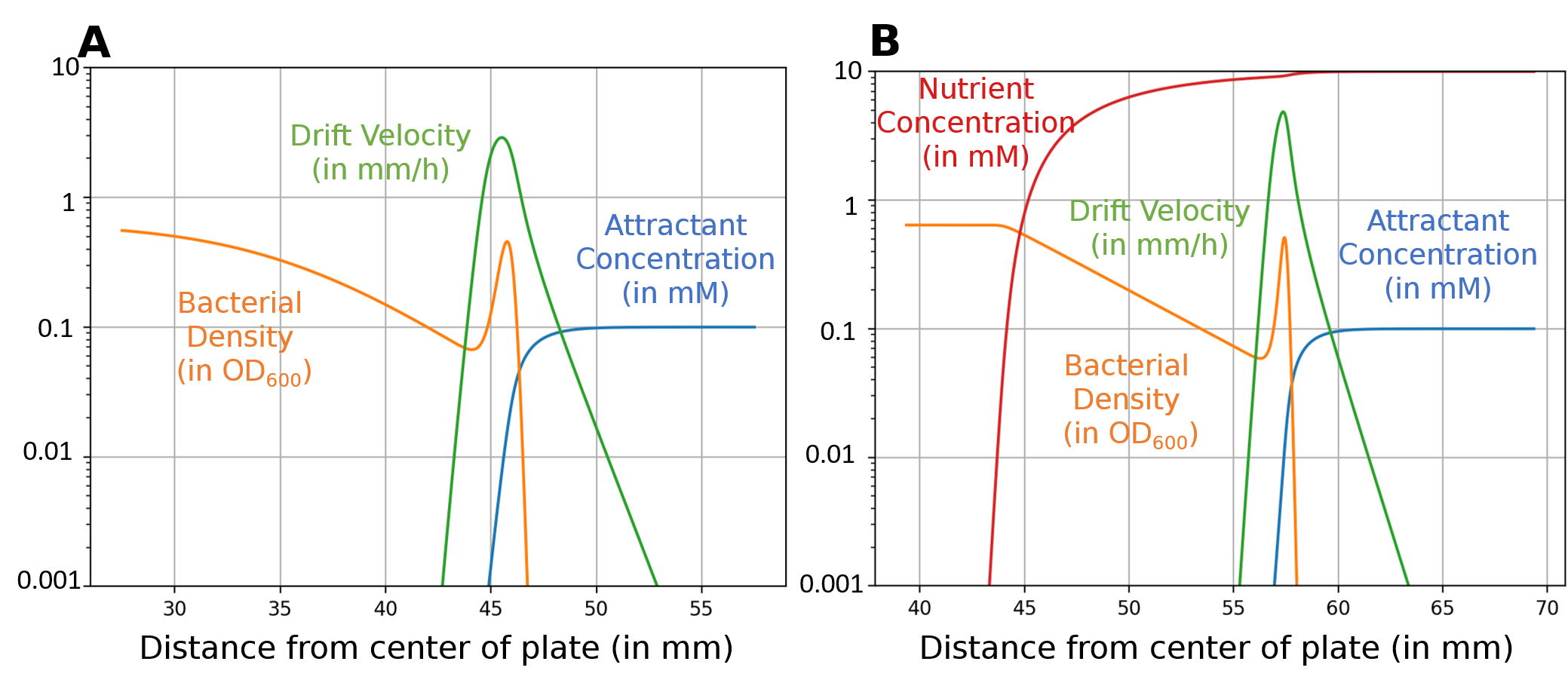}
    \caption{Comparison of density profiles obtained numerically for the simplified GE model (\textbf{A}) analyzed in this study (using parameters specified in Table S1 for low motility, and $\rho_c=0.64$) with the general GE model formulated by Cremer and Honda et al. (\textbf{B}) using the experimentally-determined model parameters found in~\cite{Cremer2019a}. We note that $\rho_c=0.64$ is the carrying capacity corresponding to the initial nutrient concentration (10 mM) used in ~\cite{Cremer2019a}. Both simulations were performed using the Finite Element Method as detailed in \textit{Supplemental Methods}. The corresponding expansion speeds are 2.79 mm/h for the version of the GE model analyzed in this study and 3.45 mm/h for the general GE model formulated by Cremer and Honda et al.}
    \label{fig:cremer_compare}
\end{figure}

% \begin{figure}
%     \centering
% %    \includegraphics{}
%     \caption{Expansion speed does not depend on initial inoculum size}
%     \label{fig:my_label}
% \end{figure}
\begin{figure}
    \centering
    \includegraphics[scale=1]{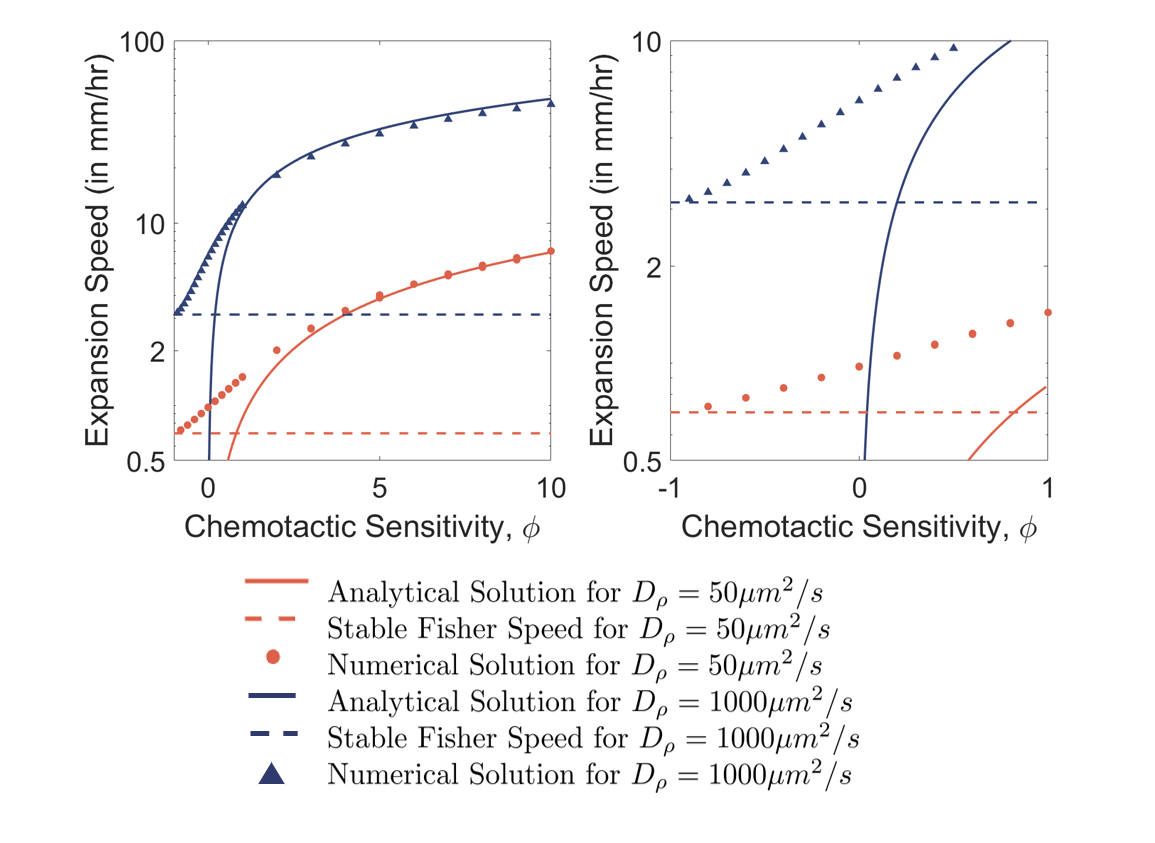}
    \caption{Dependence on the chemotactic sensitivity, $\phi$. Numerical solutions for $D_\rho = 50 \si{\mu m^2/s}$ and $D_\rho = 1000  \si{\mu m^2/s}$ are shown by red and dark blue circles, respectively. Analytic solution following Eq. 23 are shown by the corresponding solid lines.  Parameter values not specified in the legend are provided in Table S1.}
    \label{fig:inset}
\end{figure}

% \begin{figure}
%     \centering
% %    \includegraphics{}
%     \caption{Carrying Capacity}
%     \label{fig:carrying_capacity}
% \end{figure}

\newpage
\section*{Supplemental Table}
\addcontentsline{toc}{section}{Supplemental Table}
\label{param}
\begin{table}[h!]
\centering
\begin{tabular}{|c|c|c|c|}
\hline
\textbf{Parameter} & \textbf{Description} & \textbf{Value Used for low motility} & \textbf{Value Used for high motility} \\ \hline
   $D_\rho$   & Bacterial Motility Parameter    &     50 \si{\mu m^2/s}  & 1000 \si{\mu m^2/s}     \\ \hline
    $D_a$      & Attractant Molecular Diffusion Coefficient &    800 \si{\mu m^2/s} &    800 \si{\mu m^2/s}       \\ \hline
        $\phi$      &  Chemotactic Sensitivity Parameter &   5 & 5       \\ \hline
     $\chi_0$     & Chemotactic Motility Parameter &    300 \si{\mu m^2/s} & 6000 \si{\mu m^2/s}      \\ \hline
     $a_m$     &  Lower Weber Cut-off &   $10^{-3}$ \si{mM} &$10^{-3}$ \si{mM}       \\ \hline
     $a_k$     &  Michaelis Constant for Attractant Uptake by Bacteria &   $10^{-3}$ \si{mM} &   $10^{-3}$ \si{mM}       \\ \hline
    $a_0$      & Background Attractant Concentration &    0.1 \si{mM} &    0.1 \si{mM}        \\ \hline
    $\mu$      &    Rate of Attractant Uptake by Bacteria & 0.77 \si{mM/OD/h} & 0.77 \si{mM/OD/h}        \\ \hline
     $r$     & Rate of growth of bacteria &   0.69/h &   0.69/h        \\ \hline
     $\rho_c$     &  Carrying Capacity &    $10^9$ OD &    $10^9$ OD \\\hline
\end{tabular}
\caption{Standard parameters used for numerical simulations. These parameters were always used unless otherwise explicitly specified.}
\end{table}

\begin{table}[h!]
\centering
\begin{tabular}{|c|c|}
\hline
\textbf{Parameter varied} & \textbf{Range of Values Used}                   \\ \hline
$r$              & 0.03-7.7 /h                            \\ \hline
$a_m$            & $10^{-4}-3\times10^{-2}$ mM            \\ \hline
$\rho_c$         & 0.9-100 OD                             \\ \hline
$\mu$            & 0.25-2.3 \si{mM/OD/h} \\ \hline
\end{tabular}
\caption{The range over which individual parameters were varied to determine the optimal value of $a_0$ for each set of parameters for Fig.~4B. Unless varied, the values were the default values given in Table S1 (except for $\rho_c$ for which the default value was 10 OD) }
\end{table}

\begin{table}[h!]
\centering
\begin{tabular}{|c|c|}
\hline
\textbf{Parameter varied} & \textbf{Range of Values Used}                    \\ \hline
$D_\rho$         & $10^{-3}-1000\ \mu m^2/s$                \\ \hline
$D_a$            & $110-800\ \mu m^2/s$                     \\ \hline
$\chi_0$         & $100-11000\ \mu m^2/s$                   \\ \hline
$r$              & 0.44-11.1 /h                            \\ \hline
$\rho_c$         & 0.9-100 OD                              \\ \hline
$\mu$            & $0.01-1.91\ \si{mM/OD/h}$ \\ \hline
$a_0$            & $0.004-10$ mM                           \\ \hline
\end{tabular}
\caption{The range over which individual parameters were varied to determine the numerical values of $\amin,\ \rhomin,$ and $z_m-\zmin$ for Fig.~6B-D. Unless varied, the values were the default values given in Table S1.}
\end{table}
\newpage
\SItext
\addcontentsline{toc}{section}{Supplemental Text}
\section{Historical Population Models for Chemotaxis}
\label{hist}
Chemotaxis, defined as the biased movement of sensitive organisms along gradients of sensible chemicals (known as chemoattractants or chemorepellants in the case of movement up the gradient and down the gradient respectively), can be described mathematically by stochastic models for the position and direction-dependent velocity of each individual~\cite{stroock1974some}. The mathematical models consider the movement of individuals, independent of each other, that have the following characteristics:
\begin{enumerate}
    \item The movement of each individual is piece-wise linear (each piece is often called a `run'),
    \item Each linear `run' stops probabilistically,
    \item After stopping, the individual chooses a new direction randomly by a `tumble' process.
\end{enumerate}
This is called the run-and-tumble mechanism of chemotaxis. In the stochastic models, the speed of a linear `run', the probability of stopping, and the probability of a direction being chosen after tumbling can depend on the time, the position, and the direction of the individual~\cite{alt1980biased}. These assumptions reflect the observed flagellar motion of many bacteria in liquids and gels~\cite{berg1974chemotaxis}, but can also be appropriate to describe the movement of other cells migrating on surfaces~\cite{alt1980biased}.\par
The stochastic mathematical models used to describe the motion of individual cells are based on quantitative experimental observations of the statistics for the turning frequency and the turn angle distributions~\cite{berg1974chemotaxis}. If these distributions are biased in the direction of the chemical gradient, it leads to a biased random walk for each individual. From these stochastic models and reasonable biological assumptions, an effective coarse-grained theory of population-level behavior can be obtained. The ontological components of the population-level theory are the local cell density and the concentration of the relevant chemical species.\par 
A set of deterministic partial differential evolution equations to approximate the density and the mean direction of the population of moving individuals can be obtained rigorously mathematically~\cite{alt1980biased}. This was first done by Patlak~\cite{patlak1953random} for a general persistent random walk using Taylor expansions, and then rediscovered by Keller and Segel in the context of chemotaxis through multiple derivations~\cite{keller1974mathematical,segel1977theoretical}. For movements with uniform mean run speed affected by a single attractant, the equation takes the form of a reaction-diffusion equation as follows
\begin{align}
\underbrace{\frac{\partial \rho}{\partial t}}_{\text{Rate of change of cell density}}&=\underbrace{\vec{\nabla}\cdot\left(\,D_\rho(a)\ \vec{\nabla}\rho\,\right)}_{\text{Non-chemotactic ``Diffusion''}}-\underbrace{\,\vec{\nabla}\cdot\left(\rho\ \vec{v}[a]\,\right)}_{\text{Chemotactic ``Convection''}},\label{eq:KS1}
\end{align}
where $\rho$ is the local cell density, $a$ is the concentration of the attractant, $D_\rho(a)$ is the effective ``diffusion'' coefficient of cell motion (also known as the \textit{motility coefficient}), and $v[a]$ is the drift velocity due to chemotactic cell motion. Further, Keller \& Segel were able to show that analogous to Fourier's law of cooling, the drift velocity must be proportional to the chemical gradient (for sufficiently weak gradients, and ignoring threshold effects)~\cite{keller1974mathematical}. Thus, the drift velocity can be written as
\begin{align}
    v[a]=\chi(a)\cdot\vec{\nabla} a
\end{align}
where $\chi(a)$ is known as the \textit{chemotactic coefficient function}. These phenomenological parameters can be related to microscopic parameters such as the mean run time and the receptor kinetics. It must be reiterated that the above-defined ``diffusion'' and ``convection'' processes are not actual molecular diffusion and convection, but rather effective processes resulting from biased random individual movements that are analogous to their molecular counterparts. For a systematic derivation of the above-presented reaction-diffusion equation (along with an extensive review of the assumptions made in the derivation) in an arbitrary number of dimensions, the reader is directed to extensive existing literature reviews~\cite{alt1980biased,hillen2009user,horstmann20031970}).\par
\eqref{eq:KS1} can be coupled with reaction-diffusion equations for the attractant, to give rise to several experimentally observed spatial and temporal patterns in the cell density~\cite{brenner1998physical}. Keller and Segel attempted to employ \eqref{eq:KS1} to investigate one such pattern: the formation of traveling bands of chemotactic bacteria when placed in a stationary rich medium with a uniform attractant, the first extensive and modern treatment of which was performed by Julius Adler in 1966~\cite{adler1966effect,adler1966chemotaxis}. This pattern has subsequently been observed in capillary tubes~\cite{adler1966chemotaxis,adler1966effect,adler1966chemotaxis}, agar plates~\cite{croze2011migration}, and microfluidic chambers~\cite{salek2019bacterial,ahmed2010microfluidics,englert2009flow}. The traveling band observed in these experiments indicates a region of locally maximal bacterial density which appear to be formed by an ``accumulation'' of fast-moving bacteria. The existence of such a local maximum is in contrast to the resulting fronts from other models of front propagation into unstable states, such as the Fisher-Kolmogorov–Petrovsky–Piskunov Equation (F-KPP Equation), which feature a monotonic front with no local maxima, or periodic front~\cite{van2003front}.\par 

In their analysis, Keller and Segel assumed that $D_\rho$ is constant, and that the drift velocity due to chemotactic bacterial motion is determined by logarithmic sensing, i.e., $v[a(z)]=\frac{\chi_0}{a(z)}\vec{\nabla}a(z)$ where $\chi_0$ is a phenomenological proportionality constant known as the ``chemotactic coefficient'' and is a function of the bacterial strain and its internal state, the medium in which the experiment is conducted, and of the attractant being used~\cite{si2012pathway}. Such a form for the velocity is inspired by the Weber-Fechner law, which states that the sensitivity to a stimulus is inversely proportional to the background intensity of the stimulus. The Weber-Fechner law was first formulated in 1860 to describe human perception of physical magnitudes in the newly-created field of psychophysics~\cite{fechner1966elements,thompson1967foundations}, but it has been replicated in hundreds of studies across all sensory modalities and many animal species over the last two centuries~\cite{pardo2019mechanistic,gescheider2013psychophysics}. In particular, it has been shown that \coli cells sense the spatial gradient of the logarithmic ligand concentration for a range of concentrations~\cite{dahlquist1972quantitative,menolascina2017logarithmic,kalinin2009logarithmic}.\par 
The dynamics of the attractant field in chemotactic bacteria, are determined by molecular diffusion and uptake and secretion by the bacteria.
\begin{equation}
\frac{\partial a}{\partial t}=\underbrace{D_a\nabla^2 a}_{\text{Molecular Diffusion}}-\underbrace{\mu(a)\rho}_{\text{Uptake}}+\underbrace{\delta(a)\rho}_{\text{Secretion}}.\label{eq:KSa}
\end{equation}
In their analysis, Keller and Segel assumed that the molecular diffusion of the attractant is negligible compared to the motility coefficient and the chemotactic coefficient of bacteria, and that the rate of uptake by bacteria is linear in bacterial concentration (which is the case when attractant availability is saturated). Keller and Segel also only considered chemotactic systems in which there is little secretion of the attractant (secretion of attractant can lead to much more complex behaviors~\cite{budrene1991complex,brenner1998physical} and is not considered in this work). Thus, the one-dimensional form of the dynamical system analyzed by Keller and Segel is given by (where $x$ is the single spatial coordinate):
\begin{align}
    \frac{\partial \rho}{\partial t}&=D_\rho\frac{\partial^2\rho(x)}{\partial x^2}-\frac{\partial}{\partial x}\left(\frac{\chi_0}{a(x)}\frac{\partial a(x)}{\partial x}\right)\label{KS_simplrho}\\
    \frac{\partial a}{\partial t}&=-\mu\rho\label{KS_simpla}
\end{align}
They also assumed an initially localized population in a uniform attractant background with no finite size boundaries. Keller and Segel were able to solve the system exactly in one dimension and found that Eqs.~\textbf{\ref{KS_simplrho}-\ref{KS_simpla}} admit the following travelling wave solutions (where $z\equiv x-ct$ is the coordinate in the moving frame)
\begin{align}
    a(z)&=a_0\left[1+\exp\left(-\frac{cz}{D}\right)\right]^{-\frac{D}{\chi-D}} \label{KS_rho}\\
    \rho(z)&=\frac{N_0c}{\chi-D}\left[1+\exp\left(-\frac{cz}{D}\right)\right]^{-\frac{\chi}{\chi-D}}\exp(-cz/D)\label{KS_a}
\end{align}
where $N_0$ is the total number of cells in the inoculum. It must be noted that in this case, the total number of cells remains constant, and thus equal to $N_0$, as the net growth/death rate is assumed to be 0. The expansion speed (also referred to in literature as the \textit{traveling-wave velocity} or the \textit{linear spreading speed}), $c$, is given by $\mu N_0/a_0\equiv c_{KS}$.\par
\vspace{-8pt}
The KS model was extremely influential, but its results are highly sensitive to some of the assumptions made, many of which are biologically unrealistic. In particular, Keller and Segel identified that in order to generate traveling-wave solutions under their other assumptions, $v(a,\nabla a)$ must be singular or constant as $a\rightarrow 0$~\cite{Keller1971traveling}. This is unrealistic as cells cannot perform chemotaxis when concentrations fall below detectable values, which are determined by the kinetics of the enzymatic chemical reactions of the attractant. Novick-Cohen and Segel thus later analyzed a model in which $v\rightarrow 0$ for $a\rightarrow0$, by including a lower Weber cutoff in the form of the drift velocity~\cite{novick1984gradually}:
\begin{equation}
    \vec{v}(a,\nabla a)\equiv\chi_0\frac{\vec{\nabla}a}{(a+a_m)}.
\end{equation} 
In line with the original mathematical analysis, unstable wave-like solutions were obtained, with propagation slowing down noticeably during the time scale of the experiment~\cite{novick1984gradually} (see Fig.~S1) and the front gradually vanishing over time.\par

Besides being unable to describe the observed stable migration under biological realistic conditions, the KS model also fails to account for a number of important experimental observations such as the independence of the expansion speed on the initial inoculum size~\cite{Adler1978,adler1966chemotaxis,Cremer2019a}. Since the original formulation of the KS model, many additional aspects have been considered to explain a stable migrating population~\cite{horstmann20031970,keller1980assessing,wang2013mathematics,wong2018exploring}. Soon after the introduction of the KS model, bacterial growth was considered~\cite{lapidus1978model,kennedy1980traveling,lauffenburger1982effects,lauffenburger1984traveling,pedit2002quantitative,hilpert2005lattice,koster2012surface,ai2015traveling,fraebel2017environment} to recover the stability of the migrating population. However, the introduced models failed to account for key experimental observations such as the distinct migrating band or the rapid expansion speed~\cite{Cremer2019a}. A common feature of these models was that they took the attractant to be the same as the substrate for growth. However, by imposing a single substrate which plays both roles, these models unduly constrain the population dynamics and limit the expansion speed as recently pointed out~\cite{Cremer2019a}. Further, models often preserved the unrealistic form of the drift velocity without a Weber cutoff assumed in the original KS-model~\cite{lapidus1978model,kennedy1980traveling,lauffenburger1982effects,lauffenburger1984traveling,ai2015traveling,rugamba2018green}. 
More recently introduced models consider more complex attractant uptake and excretion dynamics observed for certain environmental conditions~\cite{horstmann20031970,funaki2006travelling,brenner1998physical,hillen2009user,brenner1998physical}. While these models describe fast and stable expansion, they are not able to describe population migration over several generations since growth is not explicitly included.
\vspace{-30pt}
\section{The Crucial Role of Growth}
The logarithmic sensitivity to attractant concentration results in a constant drift velocity even as $\nabla a(z)\rightarrow 0$ as long as $a(z)\rightarrow 0$ in the same limit. However, this is unreasonable as in the case of a vanishing attractant, the drift velocity would be expected to also vanish. In their analysis, Keller and Segel demonstrated that for constant per capita uptake of attractant by bacteria, traveling wave solutions to the system of equations require a singularity in the chemotactic coefficient function, $\chi(a)$ of order one or greater at $a=0$~\cite{Keller1971model}. However, relaxation of the constraint on the uptake by bacteria does not guarantee that the resulting solution would be stable. In fact, without the introduction of any new terms, a vanishing drift velocity would necessarily lead to a ``leakage'' of cells from the front of the wave. To demonstrate this, we operate in one dimension and assume that a travelling wave solution exists for the system defined by \eqref{eq:KS1} and \eqref{eq:KSa}. We define the population of the front, $N$, to be the total bacteria in a region right of a point, $x^*$, in the laboratory frame.
\begin{align}
    \frac{dN}{dt}&=\frac{d}{dt}\int_{x^*-ct}^{\infty} dx\  \rho(x,t)\\
    &=-c\rho(x^*,t)+\int_{x^*-ct}^{\infty} dx\  \frac{\partial\rho(x,t)}{\partial t}\\
    \intertext{As the boundary conditions, $\partial_x \rho,\rho\rightarrow0$ as $x\rightarrow\infty$, by performing integration by parts and plugging in \eqref{eq:KS1}, we obtain}
    &=-(c-v(x^*,t))\rho(x^*,t)-D_\rho[a(z)]\frac{\partial\rho(x^*,t)}{\partial x}
    \intertext{Going back to the moving frame with $\zd\equiv x^*-ct$}
    \frac{dN}{dt}&=-(c-v(z^\dagger))\rho(z^\dagger)-D_\rho[a(z)]\frac{\partial\rho(z^\dagger)}{\partial x}
\end{align}
As we require that $v(z)\rightarrow0$ as $z\rightarrow-\infty$, for a position sufficiently to the left, $(c-v(z^\dagger))\rho(z^\dagger)>0$ and $\frac{dN}{dt}$ must be negative if $\partial_x\rho(z^\dagger)>0$. But we must have that $\partial_x\rho(z^\dagger)>0$ for the boundary condition that $\rho(z)\rightarrow0$ as $z\rightarrow-\infty$. Thus, we immediately note that for a stable travelling wave solution with a vanishing velocity as $\partial_z a(z)\rightarrow0$ (without assuming anything of the velocity or the chemotactic coefficient function other than continuity), $dN/dt<0$. Thus, for a stable propagating wave, additional terms may be needed. In particular, the ``leakage'' due to the vanishing drift velocity must be counteracted by an additive term, such as growth.
\vspace{-24pt}
\vspace{3.5pt}
\section{The Growth-Expansion Model: General Form and Simplification}
Cremer and Honda et al.~\cite{Cremer2019a} demonstrated experimentally and numerically that the inclusion of the growth of the bacteria is sufficient to counteract the effect of the leakage due to the lower Weber cutoff and obtain stable migratory bands. They further demonstrated that such an expansion affords a novel physiological benefit to bacteria: guided range expansion which takes place well before the consumption of the nutrient at the inoculation site by the bacteria, and thus allows for rapid colonization. They introduced the generalised Growth Expansion (GE) model given by the following set of equations:\\
\begin{align}
\frac{\partial \rho}{\partial t}&=r(n,a)\rho-\nabla(\vec{v}\rho)+D_\rho\Delta\rho, \label{eq:rho_gen}\\
\vec{v}&=\chi_0\vec{\nabla} \log \left[\frac{1+a/a_-}{1+a/a_+}\right], \label{eq:v_gen}\\
\frac{\partial n}{\partial t}&=-\frac{r(n,a)}{Y}\rho+D_n\Delta n, \label{eq:n_gen}\\
\frac{\partial a}{\partial t}&=-\mu(r,a)\rho+D_a\Delta a, \label{eq:a_gen}
\end{align}
where $\rho$ is the bacterial density, $a$ is the concentration of the attractant, $v$ is the drift velocity of the bacterial population and $n$ is the concentration of the nutrient. All other symbols denote functions and parameters, both environmental and physiological as described below.
\begin{enumerate}
    \item $r(n,a)$ is the rate of growth of the bacteria. It is assumed to depend on only the local nutrient and attractant concentrations. For bacteria, the Monod equation provides an adequate relation to the nutrient concentrations~\cite{monod1949growth}. To simplify the system by eliminating the nutrient, the logistic growth equation  may be used to approximate the decrease in growth rate due to the consumption of nutrient~\cite{kargi2009re,charlebois2019modeling}. A further analysis of the effect of this simplification is explored below. The relation between growth rate and the attractant concentrations depends on the physiological effect of the attractant on the species of bacteria being considered, and the attractant may even hinder growth~\cite{brenner1998physical}. However, the effect of the attractant on growth is typically much smaller than the other limiting nutrients~\cite{Cremer2019a} and may be ignored.
    \item $D_\rho$ is the motility-induced diffusion of the bacteria. Bacteria are too large for Brownian motion to be significant in comparison to their size, however they engage in run-and-tumble motion which leads to a mean run length which is similar to the mean free path of a particle experiencing Brownian motion. Even when chemotaxis is biased in one direction, the movement of the bacteria can be viewed as a diffusion-convection process as described in Section \ref{hist}. For bacteria such as \coli in a 0.25\% agar gel, it is typically of the order of 50 \si{\mu m^2/s}~\cite{Cremer2019a}.
    \item $a_+$ is the upper Weber cut-off. It has been found empirically that the bacterial sensitivity saturates at high attractant concentration because at high attractant concentrations, the bacteria is chemoreceptor-limited in its ability to sense attractant concentrations. For bacteria such as \coli and a attractant such as aspartate, it is typically of the order of 30 \si{mM}~\cite{Cremer2019a}.
    \item $a_-$ is the lower Weber cut-off. Since the bacteria cannot be infinitely sensitive to attractant concentration, the lower Weber cut-off ensures that at very low attractant concentrations, the chemotaxis induced drift-velocity goes to 0. For bacteria such as \coli and a attractant such as aspartate, it is typically of the order of 1 \si{mM}~\cite{Cremer2019a}.\par
    It must be noted that an equivalent form for the drift velocity in one dimension is
    $$v=\chi_0\frac{a_+(a+-a_-)\nabla a}{a_-(a+a_+)^2}.$$
    This form, with appropriate substitution of constants, is more commonly found in literature. The case without $a_-$ can be studied by taking $a_-\rightarrow0$, and the case without $a_+$ can be studied by taking $a_+\rightarrow\infty$. In subsequent analysis, for visual clarity, we shall be using the symbol $a_m$ instead of $a_-$ which was used by Cremer and Honda et al.~\cite{Cremer2019a}.
    \item $Y$ is the biomass yield of the nutrient. It reflects a mass conversion factor from the nutrient to the bacterial density. For bacteria such as \coli and a nutrient such as glucose, it is typically of the order of 0.1 OD/mM~\cite{Cremer2019a}.
    \item $D_n$ is the diffusion constant for the nutrient. For a nutrient such as glucose in a 0.25\% agar gel, it is typically of the order of 800 \si{\mu m^2/s}~\cite{Cremer2019a}.
    \item $\mu(n,a)$ is rate of uptake of the attractant by the bacteria per unit bacteria. The nutrient dependence is to allow for a growth-dependent rate of uptake of the attractant. For the case of nutrient saturation, the rate of uptake of the attractant may be taken to be growth-rate independent. The dependence on the attractant is typically of the Michaelis-Menten form:
    $$\mu(a(z))=\mu_0 \frac{a(z)}{a(z)+a_k}$$
    This is contrasted to the constant form assumed by Keller and Segel, and others. The Michaelis-Menten form is crucial if growth is to be included as for low attractant concentrations the bacterial density may not be small as is the case without growth. Thus, $\mu(a)$ is required to be vanishing for low attractant concentrations and is roughly linear in attractant concentration. For relatively higher attractant concentrations, the constant form of attractant consumption is recovered.
    \item $D_a$ is the diffusion constant of the attractant. $D_a$ was typically taken to be negligible in the literature, as it was presumed that it is of a much smaller magnitude than the motility-induced diffusive and chemotactic movements of the bacteria. However, for small molecule attractants such as aspartate, serine and glucose, $D_a$ is typically larger than $D_\rho$ and $\chi_0$. Moreover, in porous media such as agar, $D_a$ can be significantly larger than $D_\rho$~\cite{licata2016diffusion} as bacteria are not able to complete their full run-and-tumble motions due to collisions with the polymer gel in agar. In their experiments, Cremer et al. found that in agar gels with 0.25\% final agar concentration, $D_\rho$ was 50.2 \si{\mu m^2/s}. In contrast, the diffusivity constant for a typical attractant is around 800 $um^2$/s~\cite{ribeiro2014binary}.
    \item $\chi_0$ is the phenomenological parameter known as the chemotactic sensitivity parameter. It is shown to be strain-dependent and is evolutionary selected by the location of the bacterium relative to other bacteria~\cite{Liu2019}, and for bacteria such as \coli in a 0.25\% agar gel, it is typically of the order of 300 \si{\mu m^2/s}~\cite{Cremer2019a}.
\end{enumerate} 
While providing excellent numerical agreement to the experimental results, the generalised model of Eqs.~\textbf{\ref{eq:rho_gen}-\ref{eq:a_gen}} is analytically intractable. The system can effectively be understood by making the non-crucial assumptions mentioned above, to recover a system of equations which is much closer to the simplified Keller-Segel equations:
\begin{align}
\frac{\partial \rho(z)}{\partial t}&=-\nabla(\vec{v}\rho)+D_\rho\Delta\rho(z)+r\rho(z)(1-\rho(z)/\rho_c),\\
\vec{v}&=\chi_0\vec{\nabla} \log [1+a(z)/a_m],\\
\frac{\partial a(z)}{\partial t}&=-\mu\frac{a(z)}{a(z)+a_k}\rho(z)+D_a\Delta a(z),
\end{align}
where $\rho_c$ is the carrying capacity of the system. Most notably, we have eliminated the nutrient field since in the GE model the effects of the nutrient and the attractant on the bacterial concentration are decoupled. The effect of limited availability of nutrient can be mimicked by limiting $\rho_c$. Thus, we reduce the equation to a system of two coupled partial differential equations. The system of equations has a degree of 4 and is accompanied by appropriate initial value and boundary conditions. In our analysis, we specify the initial conditions to be a localized profile of $\rho(z)$ (any localized profile leads to the same steady state solution; see Fig.~S3) and a constant profile of $a(z)$ at a value of $a_0$. With some reordering and working in one dimension, we obtain the following equations:
\begin{align}
\frac{\partial \rho}{\partial t}&=D\frac{\partial^2}{\partial x^2}\rho-\chi_0\frac{\partial}{\partial x}\left(\frac{\rho}{a(z)+a_m} \frac{\partial a}{\partial x}\right)+r\rho\left(1-\rho/\rho_c\right)\label{eq:simplrho}\\
\frac{\partial a}{\partial t}&=D_a\frac{\partial^2}{\partial x^2} a-\mu\frac{a(z)}{a(z)+a_k}\rho\label{eq:simpla}
\end{align}

We seek traveling-wave solutions of the forms $$\rho(x,t)=\rho(z),\ a(x,t)=a(z);\ \text{with } z=x-ct$$ where $c>0$ is the traveling wave speed (also referred to in previous literature as the expansion speed or the linear spreading speed). This converts the system from a system of partial differential equations to a system of ordinary differential equations as follows:
\begin{align}
-c\frac{\partial \rho}{\partial z}&=D\frac{\partial^2}{\partial z^2}\rho+r\rho(1-\rho)-\chi_0\frac{\partial}{\partial z}\left(\frac{\rho}{a(z)+a_m} \frac{\partial a}{\partial z}\right) \label{eq:travrho}\\
-c\frac{\partial a}{\partial z}&=D_a\frac{\partial^2}{\partial z^2} a-\mu\frac{a(z)}{a(z)+a_k}\rho. \label{eq:trava}
\end{align}

The boundary conditions of the system are specified such that the concentration of the attractant far to the left of the front is 0, and far to the right of the front is $a_0$, the initial concentration of the attractant; and the bacteria density is the carrying capacity of the system, $\rho_c$ far to the left of the front, and 0 far to the right of the front. The boundary conditions can be represented as follows:
\begin{align}
\rho(\infty)\rightarrow0,\ a(\infty)\rightarrow a_0,\ \rho(-\infty)\rightarrow\rho_c,\ a(-\infty)\rightarrow0\\
\partial_x\rho(\infty)\rightarrow0,\ \partial_x a(\infty)\rightarrow 0,\ \partial_x\rho(-\infty)\rightarrow0,\ \partial_x a(-\infty)\rightarrow0.
\end{align}\par

We compared the variation of the expansion speed with $\chi_0$ for both the full version of the GE model, and our simplified version, and found that the expansion speed remains roughly the same as can be seen in Fig.~S2.

\section{The exponential \textit{ansatz}}
In the case of no growth and $a_m,a_k\rightarrow0$ as in the simplified Keller-Segel model (Eqs.~\textbf{\ref{KS_simplrho}-\ref{KS_simpla}}), away from the right boundary such that $a_0\rightarrow\infty$, the solution to the system of differential equations is straightforward:

\begin{equation}
a(z)\propto\rho(z)\propto\exp(\lambda_{KS} z)\text{ where }\lambda_{KS}=0\text{ or }\lambda_{KS}=\frac{c}{\chi_0-D_\rho}    
\end{equation}

The non-trivial solution is asymptotically the solution obtained by Keller and Segel (Eqs.~\textbf{\ref{KS_rho}-\ref{KS_a}}) away from the right boundary (i.e., $z\rightarrow-\infty$). Since the inclusion of growth seeks to stabilize the dynamics of the front by counteracting the leakage due to a reduced drift velocity, we expect the results of the model including small growth to be qualitatively similar, and this expectation was confirmed using numerical simulations (see Fig.~1).\par
Assuming that $a(z) \sim a_1\exp(\lambda z)$, from \eqref{eq:simpla}, we obtain
\begin{align}
-c\lambda a&=D_a\lambda^2 a-\mu\frac{a(z)}{a(z)+a_k}\rho,\\
\implies \rho&=\underbrace{\frac{(D_a\lambda^2+c\lambda)}{\mu}}_{\equiv \beta} (a(z)+a_k).
\end{align}
Subsequently from \eqref{eq:simplrho}, in the case that $\rho_c\rightarrow\infty$, we have that
\begin{align}
% -c\lambda \rho&=D_\rho\lambda^2\rho-\chi_0\beta\frac{\partial}{\partial x}\left(\frac{a(z)+a_k}{a(z)+a_m}\frac{\partial}{\partial z}a\right)+r\rho\\
% -c\lambda \rho&=(D_\rho-\chi_0)\lambda^2\rho+r\rho+\chi_0\lambda^2\beta \left(\frac{(a_m-a_k)a_ma}{(a(z)+a_m)^2}\right)\\
% -c\lambda&=(D_\rho-\chi_0)\lambda^2+r+\chi_0\lambda^2 \left(\frac{(a_m-a_k)a_ma}{(a(z)+a_k)(a(z)+a_m)^2}\right)
% \intertext{Or alternatively, that}
-c\lambda a&=D_\rho\lambda^2a-\chi_0\beta\frac{\partial}{\partial x}\left(\frac{a(z)+a_k}{a(z)+a_m}\frac{\partial}{\partial z}a\right)+r(a(z)+a_k)\\
-c\lambda a&=(D_\rho-\chi_0)\lambda^2a+ra+ra_k+\chi_0\lambda^2\beta \left(\frac{(a_m-a_k)a_ma}{(a(z)+a_m)^2}\right)\\
-c\lambda&=(D_\rho-\chi_0)\lambda^2+r+\frac{ra_k}{a}+\chi_0\lambda^2 \left(\frac{(a_m-a_k)a_ma}{(a(z)+a_k)(a(z)+a_m)^2}\right)
\end{align}
In the regime that $a_k/a(z)\to 0$ and for the case that $a_k=a_m$, we obtain a quadratic equation in $\lambda$, and thus our \textit{ansatz} approximately holds. The case $a_k=a_m$ is biologically motivated since the pathways (such as the periplasmic binding proteins) that are relevant to sensing of the attractant in bacteria are similar to those that are relevant to the consumption of the attractant in the bacteria. This assumption also eliminates the final term. Thus, for $a\gg a_k$ we can solve for $\lambda$ and we obtain
\begin{align}
    \lambda_\pm &= \frac{c}{2(\chi_0-D_\rho)}\pm\frac{\sqrt{c^2+4r(\chi_0-D_\rho)}}{2(\chi_0-D_\rho)}
\end{align}
Thus, we require that $\chi_0>D_\rho$ for a solution where an exponentially increasing profile of $\rho$ is observed. Since we assume a finite (and large) $D_a$, we will later assume that $\chi_0\gg D_\rho$. Otherwise, we find that the Chemotaxis Regime is very narrow and thus the expansion speed is often set by the transitionary regimes between the different regimes. As we are only interested in the exponentially increasing solution, for $r\ll \lambda c$, we have that\\
\begin{equation}
    \lambda\approx\frac{c}{\chi_0-D_\rho}\label{lambda_c}
\end{equation}
\section{Calculation of Expansion Speed}
\label{exp_calc}

To find $c$, we must use another boundary condition. However, the exponential increase in $\rho$ does not continue do to the right boundary condition for $\rho$. Instead, $\rho\rightarrow0$ fast enough for the $\rho(z)$ to be integrable in the interval $[z^\dagger,\infty)$ such that $a_k\ll a(\zd)$ and $\partial_z\rho(z)\approx \lambda \rho(z)$. Thus, we integrate \eqref{eq:simplrho} and \eqref{eq:simpla} and employ suitable approximations:
\begin{align}
-ca\big|_{z^\dagger}^{+\infty}&=D_a\frac{\partial a}{\partial z}\bigg|_{z^\dagger}^{+\infty}-\mu \int_{z^\dagger}^{+\infty}\frac{a(z)}{a(z)+a_m}\rho dz\\
\implies c(a_0-a(z^\dagger))&=D_a \lambda a(z^\dagger)+\mu \underbrace{\int_{z^\dagger}^{+\infty}\rho dz}_{\equiv \N}- \underbrace{\mu a_m \int_{z^\dagger}^{+\infty}\frac{\rho(z)}{a(z)+a_m} dz}_{\sim O\left(\frac{\mu}{\lambda}a_m\beta\log(a_0/a(z^\dagger))\right)}\end{align}
To obtain the order of the last term, we note that $\rho(z)=\beta (a(z)+a_m)$ from $z^\dagger$ to a point, $z_a$ where $a(z)\sim a_0$. For $z>z_a$, as shown in the calculation for the Diffusion Regime in the main text, $\rho(z)\approx\rho_0\exp(-cz/D_\rho)$ and $\rho(z)/(a(z)+a_m)\approx \frac{\rho_0}{a_0}\exp(-cz/D_\rho)$ where $\rho_0$ is the value of $\rho(z)$ at the interface of the Chemotaxis and Diffusion Regimes. Thus,
\begin{align}
   \int_{z^\dagger}^{+\infty}\rho dz=& \int^{z_a}_{\zd}\frac{\rho(z)}{a(z)+a_m} dz+\int_{z_a}^{+\infty}\frac{\rho(z)}{a(z)+a_m} dz\\
   \approx&  \beta (z_a-\zd)+\frac{\rho_0 D_\rho}{a_0 c}\approx\frac{\beta}{\lambda}\ln\left(\frac{a_0}{a(\zd)}\right)+\frac{\beta D_\rho}{c}\\
   =&O\left(\frac{\mu}{\lambda}a_m\beta\log(a_0/a(z^\dagger))\right),
\end{align} where the last equality is because $\lambda=c/(\chi_0-D_\rho)$. We note that since $\frac{a_m}{a_0}\ll 1$, the correction term is of a sub-leading order. We will, however, carry it forward to determine the order of the sub-leading term in the final calculation.
\begin{align}
\implies \N&=\frac{c(a_0-a(z^\dagger))-D_a \lambda a(z^\dagger)}{\mu}+O\left(\frac{a_m\beta}{\lambda}\log\frac{a_0}{a(z^\dagger)}\right)\\
\intertext{Now, from the integral of \eqref{eq:simplrho}}
-c\rho\big|_{z^\dagger}^{+\infty}&=D_\rho\frac{\partial}{\partial z}\rho\bigg|_{z^\dagger}^{+\infty}-\chi_0\left(\frac{\rho}{a(z)+a_m} \frac{\partial a}{\partial z}\right)\bigg|_{z^\dagger}^{+\infty}+r\N+O\left(\frac{ra_m\beta}{\lambda}\log\frac{a_0}{a(z^\dagger)}\right)\\
\implies c\rho(z^\dagger)&=-D_\rho\frac{\partial \rho}{\partial z}(z^\dagger)+\chi_0\beta \frac{\partial a}{\partial z}(z^\dagger)+\frac{rc(a_0-a(z^\dagger))-rD_a\lambda a(z^\dagger)}{\mu}+O\left(\frac{ra_m\beta}{\lambda}\log\frac{a_0}{a(z^\dagger)}\right)\\
\implies c\beta (a(z^\dagger)+a_m)&=-D_\rho\lambda\beta a(z^\dagger)+\chi_0\lambda \beta a(z^\dagger)+\frac{rc(a_0-a(z^\dagger))-rD_a\lambda a(z^\dagger)}{\mu}+O\left(\frac{ra_m\beta}{\lambda}\log\frac{a_0}{a(z^\dagger)}\right)
\intertext{Collecting the coefficients of $a(z^\dagger)$, we obtain the following two equations:}
c&=(\chi_0-D_\rho)\lambda-\frac{r(c+D_a \lambda)}{c\mu\beta}=(\chi_0-D_\rho)\lambda-\frac{r}{\lambda} \label{c_main}
\end{align}
This is the same relationship between $c$ and $\lambda$ that we had obtained earlier by assuming that $\rho(z)=\beta (a(z)+a_m)$. But from comparing the constant terms, we obtain
\begin{align*}
\beta\left(1+ O\left(\frac{r}{\lambda c}\log\frac{a_0}{a(z^\dagger)}\right)\right)&=\frac{(D_a\lambda^2+c\lambda)}{\mu}\left(1+ O\left(\frac{r}{\lambda c}\log\frac{a_0}{a(z^\dagger)}\right)\right)= \frac{ra_0}{\mu a_m}\\
\intertext{We assume that $r/(\lambda c)\ll 1$ and the lower order term can be ignored.}
\implies \frac{ra_0}{a_m}&\approx\lambda^2(\chi_0-D_\rho+D_a)-r
\end{align*}
As $a_m/a_0\ll1$, we ignore the the second order term on the RHS. Thus, as a final solution, we obtain that:
\begin{align}
\lambda\approx&\sqrt{\frac{r(a_0/a_m)}{\chi_0-D_\rho+D_a}}\label{lambda}\\
\implies c\approx&(\chi_0-D_\rho)\sqrt{\frac{r(a_0/a_m)}{\chi_0-D_\rho+D_a}}-\sqrt{\frac{r(\chi_0-D_\rho+D_a)}{(a_0/a_m)}}\approx (\chi_0-D_\rho)\sqrt{\frac{r(a_0/a_m)}{\chi_0-D_\rho+D_a}}\label{c_main}
\end{align}
For $\chi_0\gg D_\rho$ and $a_m\ll a_0$, the second term can be ignored. This is self-consistent with the assumption that $r\ll \lambda c$. The error in these calculations is of the order of $a_m/a_0$. Further, this result does not hold for $\chi_0\sim D_\rho$ (equivalent to the case that $\phi\rightarrow0$ shown in Fig.~S4) and in that regime the exponentially increasing profile for $\rho$ disappears and our approximations fail.
\vspace{-24pt}
\section{Form of expansion speed for finite carrying capacity}
For $\rho_c\nrightarrow\infty$, we can use the same techniques as earlier, but we have an additional term corresponding to the effect of the carrying capacity. The integral equation now reads:
\begin{align*}
c\rho(z^\dagger)&=-D_\rho\frac{\partial \rho}{\partial z}(z^\dagger)+\chi_0\beta \frac{\partial a}{\partial z}(z^\dagger)+r\N-\frac{r}{\rho_c}\int_{z^\dagger}^{\infty}\rho^2(z)dz+O\left(\frac{ra_m\beta}{\lambda}\log\frac{a_0}{a(z^\dagger)}\right)
\end{align*}
%c\beta (a(z^\dagger)+a_m)&=-D_\rho\lambda\beta a(z^\dagger)+\chi_0\lambda \beta a(z^\dagger)+\frac{rc(a_0-a(z^\dagger))-rD_a\lambda a(z^\dagger)}{\mu}+O\left(\frac{ra_m\beta}{\lambda}\log\frac{a_0}{a(z^\dagger)}\right)
To progress, we must make some assumptions regarding the form of $\rho^2$ (or equivalently, of $\rho$). Based on the numerical results, we assume a piece-wise exponential form for $\rho$: $$\rho(z)=\begin{cases}\rho_{\text{max}} \exp(\lambda z),\ z<0\\ \rho_{\text{max}},\ 0<z<\kappa D_a/c\\\rho_{\text{max}}\exp(-cz/D_\rho)\exp(\kappa D_a/D_\rho),\ z>\kappa D_a/c\end{cases}$$
where $\rho_{\text{max}}$ is the highest value of $\rho(z)$ obtained. The region $0<z<\kappa D_a/c$ is intended to reflect that that there is a transition region between the Chemotaxis Regime and the Diffusion regime where $\rho(z)$ does not behave exponentially is only slowly varying, as was observed in numerical simulations. The width of this region has been observed numerically to be set by $D_a$ and $c$, with an unknown proportionality constant $\kappa$. As the only relevant variables for $\kappa$ are $D_\rho,D_a,\chi_0$, we suspect that it is a function of these variables. Thus, we find that
$$\int_{-\infty}^{\infty}\rho^2(z)dz=\frac{\rho_{\text{max}}^2(\chi_0+2D_a\kappa)}{2c}$$
But we also have that $\N=\frac{(\chi_0+D_a\kappa)\rho_{\text{max}}}{c}$. Thus,
$$\int_{-\infty}^{\infty}\rho^2(z)dz=\frac{\N^2 c (\chi_0 + 2 D_a \kappa) }{2(\chi_0+D_a\kappa)^2}$$
Going back to our previous calculations, we now have that
\begin{align*}
-c\rho\big|_{z^\dagger}^{+\infty}&\approx D_\rho\frac{\partial}{\partial z}\rho\bigg|_{z^\dagger}^{+\infty}-\chi_0\left(\frac{\rho}{a(z)+a_m} \frac{\partial a}{\partial z}\right)\bigg|_{z^\dagger}^{+\infty}+r\N-\frac{r\N^2 c (\chi_0 + 2 D_a \kappa) }{2\rho_c(\chi_0+D_a\kappa)^2}\\
\implies c\rho(z^\dagger)&\approx-D_\rho\frac{\partial \rho}{\partial z}(z^\dagger)+\chi_0\beta \frac{\partial a}{\partial z}(z^\dagger)+\frac{rc(a_0-a(z^\dagger))-rD_a\lambda a(z^\dagger)}{\mu}\\&-\frac{r c (\chi_0 + 2 D_a \kappa) }{2\rho_c(\chi_0+D_a\kappa)^2}\left(\frac{c(a_0-a(z^\dagger))-D_a \lambda a(z^\dagger)}{\mu}\right)^2\\
\intertext{We ignore terms of $O(a^2(z^\dagger)/a_0^2)$ as $a_0\gg a(z^\dagger)$}
\implies c\beta (a(z^\dagger)+a_m)&=-D_\rho\lambda\beta a(z^\dagger)+\chi_0\lambda \beta a(z^\dagger)+\frac{rc(a_0-a(z^\dagger))-rD_a\lambda a(z^\dagger)}{\mu}\\&-\frac{r c (\chi_0 + 2 D_a \kappa) }{2\rho_c(\chi_0+D_a\kappa)^2}\left(\frac{c^2a_0^2-2c^2a_0a(z^\dagger)-2ca_0D_a \lambda a(z^\dagger)}{\mu^2}\right)
\intertext{Collecting the coefficients of $a(z^\dagger)$, we obtain the following equation:}
c&=(\chi_0-D_\rho)\lambda-\frac{r(c+D_a \lambda)}{c\mu\beta}+\frac{r (\chi_0 + 2 D_a \kappa) }{\rho_c(\chi_0+D_a\kappa)^2}\left(\frac{c^2a_0+ca_0D_a \lambda }{(D_a\lambda^2+c\lambda)\mu}\right)\\
&=(\chi_0-D_\rho)\lambda-\frac{r}{\lambda}+\frac{r (\chi_0 + 2 D_a \kappa) }{\rho_c(\chi_0+D_a\kappa)^2}\left(\frac{ca_0 }{\lambda\mu}\right)
\end{align*}
From comparing the constant terms, we obtain
\begin{align*}
\beta\left(1+ O\left(\frac{r}{\lambda c}\log\frac{a_0}{a(z^\dagger)}\right)\right)&=\frac{(D_a\lambda^2+c\lambda)}{\mu}\left(1+ O\left(\frac{ra_m}{\lambda}\log\frac{a_0}{a(z^\dagger)}\right)\right)= \frac{ra_0}{\mu a_m}-\frac{r c^2a_0^2 (\chi_0 + 2 D_a \kappa) }{2a_m\mu^2\rho_c(\chi_0+D_a\kappa)^2}\\
\intertext{However, for small $r$ and $\chi_0\gg D_\rho$, $c\approx (\chi_0-D_\rho)\lambda$ and thus, $r/(\lambda c)\ll 1$ and the lower order term can be ignored.}
\implies \frac{ra_0}{a_m}&\approx\lambda^2(\chi_0-D_\rho+D_a)-r+\underbrace{\lambda^2\frac{r (\chi_0-D_\rho)^2a_0^2 (\chi_0 + 2 D_a \kappa) }{2a_m\mu\rho_c(\chi_0+D_a\kappa)^2}}_{\equiv \lambda^2\frac{ r a_0^2}{a_m\mu\rho_c}(\chi_0-D_\rho)\gamma}
\end{align*}
where $\gamma$ is a dimensionless function of $D_\rho,D_a,\chi_0$ and $\kappa$. As $\kappa$ itself is suspected to be a function of the three variables, $\gamma$ replaces $\kappa$ as an equivalent constant. As $a_m/a_0\ll1$, we ignore the the second order term on the RHS. Thus, as a final solution, we obtain that:
\begin{align}
\lambda\approx&\sqrt{\frac{r(a_0/a_m)}{\chi_0-D_\rho+D_a+\frac{ r a_0^2}{a_m\rho_c}\gamma}}=\sqrt{\frac{r}{\frac{a_m}{a_0}\left(\chi_0-D_\rho+D_a\right)+\frac{ r a_0}{\mu\rho_c}(\chi_0-D_\rho)\gamma}}\\
\implies c\approx&(\chi_0-D_\rho)\sqrt{\frac{r}{\frac{a_m}{a_0}\left(\chi_0-D_\rho+D_a\right)+\frac{ r a_0}{\mu\rho_c}(\chi_0-D_\rho)\gamma}}=\left. c_\infty \middle/ \sqrt{1+\frac{ r a_0}{\mu\rho_c}\frac{(\chi_0-D_\rho)\gamma}{\left((\chi_0-D_\rho)+D_a\right)}\frac{a_0}{a_m}}\right. .
\end{align}
The attractant concentration for the maximum expansion speed is found to be
\begin{equation}
%\frac{a_0^{\text{max}}}{a_0}=\sqrt{\frac{\rho_c}{\beta a_0\gamma}\left(1+
%\frac{D_a}{\chi_0-D_\rho}\right)}.
\frac{a_0^{\text{max}}}{a_m}=\sqrt{\frac{\mu\rho_c}{ra_m\gamma}\left(1+
\frac{D_a}{D_\rho\phi}\right)}
\label{a0-max}
\end{equation}
The dimensionless function $\gamma$ is expected to be a non-trivial function of $\chi_0,D_a$ and $D_\rho$. It is related to the width of the assumed plateau for the top of the peak, and is expected to decrease with higher $D_a$ as the ``smoothening'' of the attractant gradient due to $D_a$ results in a broadening of the density bulge and reduces the effect of the carrying capacity. However, as $\chi_0,D_a$ and $D_\rho$ have the same dimensions, $\gamma$ may be a non-trivial combination of $\chi_0,D_a$ and $D_\rho$ which themselves may be raised to powers of combinations of $\chi_0,D_a$ and $D_\rho$.% A naive estimate based on comparing $D_a$ with $(\chi_0-D_\rho+D_a)$, the other recurring diffusion-related parameter, is that $\gamma = 1-D_a/(\chi_0-D_\rho+D_a)\approx 0.24$ for our choice of parameters. An excellent fit was obtained for the numerical results for $\gamma\approx 0.26$.

\section{F-KPP Dynamics in the Growth and Diffusion Regimes}
\subsection{Diffusion Regime}
To understand the selection of the asymptotic steepness of the front of the wave as $z\to+\infty$, we operate in the static frame and assume that the initial population is described by a Dirac delta function, i.e., $\rho(x,0)=\delta(x)$. Then, the solution to the F-KPP equation is
\begin{equation}
\frac{\partial \rho(x,t)}{\partial t}=D_\rho\partial_x^2 \rho(x,t)+r\rho(x,t)
\end{equation}
Since it is a second order PDE, we need two boundary/initial conditions, one spatial and one temporal. Our initial condition is given by the Dirac delta function, and we take the boundary condition as a zero population at infinity ($\rho(x\rightarrow,t)=0$). The solution is obtained by using the Fourier transform. In our convention, we define the Fourier transform as
\begin{equation}
\hat{\rho}(k,t)=\int d r \rho(x,t) \exp(-ix\cdot k)
\end{equation}
and the inverse is
\begin{equation}
\rho(x,t)=\frac{1}{(2\pi)}\int d k \hat{\rho}( k,t) \exp(ix\cdot k)
\end{equation}
where  $ k$ is the Fourier space dual of the position, and can be understood as a form of spatial frequency. \par
Going back to our equations, by plugging in the equation for the inverse Fourier transform into the diffusion equation.% Firstly we restrict ourselves to the radially symmetric case (for the non-radially symmetric case, it might be worth it to do a spherical harmonic decomposition\footnote{https://en.wikipedia.org/wiki/Spherical_harmonics}.
\begin{equation}
\frac{\partial \rho(x,t)}{\partial t}=\frac{1}{(2\pi)}\frac{\partial}{\partial t}\int d k \hat{\rho}( k,t) \exp(ix\cdot k)=\frac{1}{(2\pi)}\int d k \frac{\partial \hat{\rho}( k,t)}{\partial t} \exp(ix\cdot k)
\end{equation}
\begin{equation}
\begin{split}
D_\rho\partial_x^2 \rho(x,t)=D_\rho\partial_x^2 \frac{1}{(2\pi)}\int d k \hat{\rho}( k,t) \exp(ix\cdot k)=\frac{D_\rho}{(2\pi)}\int d k \hat{\rho}( k,t) \partial_x^2 \exp(ix\cdot k)\\
=\frac{D_\rho}{(2\pi)}\int d k \hat{\rho}( k,t) (- k^2 \exp(ix\cdot k))\\
\end{split}
\end{equation}
Thus, plugging everything into the F-KPP equation, we get that
\begin{align}
\frac{\partial \hat{\rho}( k,t)}{\partial t}&=(-D_\rho k^2+r)\hat{\rho}( k,t)\\
\implies \hat{\rho}( k,t)&=\hat{\rho}( k,t=0)\exp(-D_\rho k^2t+rt)\\
\text{ where }\hat{\rho}( k,t=0)&\equiv \int d^n r \rho(x,t=0) \exp(-ix\cdot k)
\end{align}
Consider the initial condition where you start with $N_0$ at an infinitesimal volume, which can be taken as a delta function, i.e., $\rho(x,t=0)=N_0\delta^n(x)$
Then,
\begin{align*}
\hat{\rho}( k,t=0)&\equiv \int d r N_0\delta(x) \exp(-ix\cdot k)=N_0
\end{align*}
\begin{equation}
\implies \hat{\rho}( k,t)=N_0\exp(-D_\rho k^2t+rt)\\
\end{equation}
Now, we can get the inverse transform:
\begin{equation}
    \rho(x,t)=\frac{1}{(2\pi)}\int d k  \exp(-D_\rho k^2t+ix\cdot k+rt)=\frac{N_0}{2\sqrt{\pi D_\rho t}}\exp\left(\frac{-x^2}{4D_\rho t}\right)\exp(rt)
\end{equation}
For the long-term behavior, we take $x=z+ct$,
\begin{equation}
    \rho(z,t)=\frac{N_0}{2\sqrt{\pi D_\rho t}}\exp\left(\frac{-(z^2+c^2t^2+2zct)}{4D_\rho t}\right)\exp(rt)=\frac{N_0}{2\sqrt{\pi D_\rho t}}\exp\left(\frac{-z^2}{4D_\rho t}\right)\exp\left(rt-\frac{c^2t}{4D_\rho}\right)\exp\left(\frac{-cz}{2D_\rho}\right)
\end{equation}
The first term just signals the transition from $t=0$ to later time, and can be ignored for long times. The second term demonstrates that the speed of the front should be at least $2\sqrt{D_\rho r}=c_F$ otherwise the solution doesn't satisfy the boundary conditions. If the speed were actually lower than $c_F$, a new front with speed $c_F$ would emerge and ``pull'' the front, thus creating a ``pulled wave-front''. Any front traveling faster would die over time, unless it is supported by the bulk of the wave.\par And the final term indicates that the steepness of the front must be at least $c/2D_\rho$, which would correspond to $\lambda_F=\sqrt{r/D_\rho}$ for the case of pulled waves. However, for speeds propagated by the bulk at speeds faster than $c_F$, the front is effectively pushed by the bulk. In such a case, the steepness must be at least $\lambda_F$ as otherwise the asymptotic steepness would be less than the steepness for the Fisher speed, and since a solution with the Fisher speed is always permitted, it would emerge and dominate the front before being overtaken by the bulk. The only stable solution is for the steepness to be $\lambda_D^+=\frac{c+\sqrt{c^2-4r D_\rho}}{2D_\rho}$, which is the observed steepness. This is the case of a ``pushed wave-front'' and occurs in the Diffusion Regime of our system.
%\section{Range Expansion in Ecology}
 %Relation to ecological literature on range expansion and places where our model may be more appropriate (repeat key points of Hastings et al, and how FKPP doesn't describe many cases of range expansion)
 
 \subsection{Growth regime}
The analysis of the Chemotaxis Regime results in relations for expansion speed as discussed above. To understand the entire traveling wave, we next consider the growth regime, which is defined being left of the density trough and characterized by $a\lesssim a_m$ (Fig.~2). As the drift velocity vanishes with falling attractant concentrations (as $\frac{d}{dz}a/a_m\rightarrow 0$ in this regime), convection does not counteract the effects of back-diffusion and cells leave the moving frame. This ``leakage'' is explored further in Sec.~S2 of the \textit{Supplemental Text}. While the total number of bacteria in the moving front is conserved as growth counteracts this ``loss'', cells in the trailing region cannot catch up the fast chemotactic migration and thus ``stay behind''. However, cells still grow as long as $\rho<\rho_c$ and move diffusively. In the growth regime the governing equation for the population is thus given as:
 \begin{align}
-c\frac{d\rho}{dz}&=D_\rho\frac{d^2}{dz^2}\rho+r \rho(1-\rho/\rho_c). \label{FK}
\end{align}
This is the well-known F-KPP equation. However, in contrast to the standard scenario canonically used to describe range expansion, the right boundary condition is specified by the traveling-wave dynamics of the Chemotaxis Regime.
%removed because this was already introduced above well-known F-KPP equation describing the expansion of a growing population due to diffusion alone~\cite{fisher1937wave,tikhomirov1991study,kolmogorov1937etude,ablowitz1979explicit}, but with a different right boundary condition which is specified by the traveling-wave dynamics of the Chemotaxis Regime.
The expansion speed of the population can be obtained in this case by analyzing the ``growth front'' (the boundary of Growth and Chemotaxis Regime in this case), for which $\rho\ll\rho_c$ and the nonlinear term in \eqref{FK} can be neglected. The remaining linear equation yields the solution
 \begin{equation}
     \rho\propto \exp(-\lambda_G^\pm z)\text{, with } \lambda_G^\pm=\frac{c\pm \sqrt{c^2-4rD_\rho}}{2D_\rho}, \label{lambda_F}
 \end{equation}
 which relates the expansion speed of the growth front, $c_G$, in term of the decay parameter $\lambda_G$ of the density profile. For each value of the allowed expansion speed $c \geq c_F\equiv 2\sqrt{rD_\rho}$, the solution is degenerate with two possible values of $\lambda_G$, except when $c=c_F$.\par 
 A seminal result in the theory of the F-KPP equation is that the marginally stable density profile, with $\lambda_F = c_F/(2D_\rho)=\sqrt{r/D_\rho}$ is selected among all the allowed solutions, for sufficiently compact initial conditions~\cite{kolmogorov1937etude,ablowitz1979explicit}. 
 However, as we found for the Chemotaxis Regime the population moves with an expansion speed given by \eqref{c_main}, with $c> c_{F}$ as long as $\phi$ is not too close to $0$. Traveling speeds with $c>c_{F}$ are permitted as solutions of the F-KPP equation, but they correspond to $\lambda_G \neq \lambda_F$ and are not marginally stable, thus typically not selected~\cite{van2003front}. Thus, we may ask, how do the bacteria in this case beat ``marginal stability''? Or in other words, how is the propagation speed $c$ ``passed on'' from the Chemotaxis Regime to the trailing growth regime which is governed by the F-KPP equation?\par
This may be understood through another well-known result in the theory of the F-KPP equation that, independent of the precise non-linearities, if the front of the wave is maintained to be shallower than $\lambda_F$, then the front travels with a speed given by $c>c_F$~\cite{van2003front}. In our case, the shallower slope is $\lambda_G^-\approx r/c$. This shallower slope is maintained by the ``leakage'' of cells from the Chemotaxis Regime into the front of the growth regime which are deposited behind the Chemotaxis Regime along a boundary moving at an expansion speed $c$. This can be understood through our \textit{ansatz} ($\rho(z)=\beta(a+a_m)$) as for $a(z)\to 0$, $\rho(z)\to \beta a_m$ and the resulting constant boundary condition for $\rho(z)$ at the tail of the migrating band. Thus, in this way, the propagation speed $c$ is ``passed on'' from the Chemotaxis Regime to the trailing growth regime.\par
The dynamics of the growth regime reveal how the migrating band may leave behind a small number of ``settling'' cells, that can then grow exponentially. By continually leaving behind the small number of cells, the back formed by the ``settling'' cells also keeps up with the migrating band. However, if migrating bands are composed of cells with different motilities starting at the same inoculum spot, fewer cells will be deposited by the faster-moving than by the slower-moving cells. Thus, the slower-moving cells will quickly increase to a higher density and out-compete the cells left behind by the faster-moving band. This leads to the ecological principle that more motile cells have a higher fitness in regions far from the inoculating site, and the less motile cells have a higher fitness in regions close to the inoculating site. This was validated experimentally by Liu et al., who repeatedly selected cells at different distances from the inoculating site, finding that over time the cells close to the starting point formed populations with lower expansion speeds and the cells far from the starting point formed populations with higher expansion speeds~\cite{Liu2019}.

\section*{Analytical Solutions to the Growth Regime, $D_\rho\to0$}
Beyond the results obtained for the Growth Regime using the modified ansatz, exact mathematical statements can be made regarding the Growth Regime in the case that $D_\rho=0$. We note that the solutions should not qualitatively change for $D_\rho>0$ as it can only smoothen the density profile (and subsequently the attractant profile). Numerical simulations performed with $D_\rho=0$ confirm that all results hold in the limit $D_\rho\to0$ and all qualitative features are preserved. For simplicity, we will take the limit of large $\rho_c$ since at the front, $\rho(z)/\rho_c\to 0$.\par 
A magnified view of the transition region between the Chemotaxis and Growth Regimes is shown in Fig.~5A, with the location of the density minimum being at $\zmin$. Applying the \ansatz  to $z=z_m$ (where \emph{ansatz} Eq. 7 holds as per the condition 11, $a(z)\gg (r/\lambda c) a_m$), we find the flux of cells due to chemotaxis and diffusion, $J(z_m)\equiv -v(z)\rho(z)$, to be given by
\begin{equation}
    J(z_m)=-\chi_0\beta\frac{da(z)}{dz}\bigg|_{z=z_m}=-c\beta a_m, \label{Jz_m}
\end{equation}
where we used $v(z)\rho(z) = \chi_0 \beta da/dz$ based on our \ansatz Eq. 7, and $a(z)\propto \exp(\lambda z)$ with $\lambda$ given by \eqref{lambda_c}. Here, a negative value indicates a net flux to the left at $z_m$, i.e., out of the Chemotaxis Regime. For the wave-front to be at steady state, the loss of cells at $z_m$ due to chemotaxis and diffusion must be replenished by growth. Recall that in the solution by Novick-Cohen and Segel \cite{novick1984gradually} that also incorporated the lower Weber cut-off but maintained a constant total population size, the population ``left behind'' the front was the reason that the migrating wave-front slowed down. Incorporating growth, even at very low rates, allows the migrating wave-front to ``replenish'' itself and maintain stability. This is discussed in more detail in \textit{Supplemental Text S2}.

Eq. 19 and \eqref{c_main} also allow us to use Eq. 18 to find $\N(\zd)$:
\begin{equation}
    \N(\zd) \equiv \int_{\zd}^\infty \rho(z)dz \approx N_\text{band} \cdot \left(1- \frac{a(\zd)}{a_0}\frac{\chi_0-D_\rho+D_a}{\chi_0-D_\rho} \right),  \label{Nzd-full}
\end{equation}
with
\begin{equation}
    N_\text{band} \approx c a_0/\mu. \label{NKS}
\end{equation}
Thus, if $\zd$ is sufficiently to the left (such that $a(\zd)\ll {a_m}/\left({r/ \lambda c}\right)$, a regime with a significant overlap with the Chemotaxis Regime), the total population to the right of the position $\zd$ approaches a constant $N_\text{band}$, which is interpreted as the population size of the migratory band. Thus, from the expression for $\beta$, Eq. 19, and the result \eqref{Nzd-full} and \eqref{NKS}, we can rewrite \eqref{Jz_m} as 
\begin{equation}
    J(z_m)\approx-rN(z_m), \label{J_rN}
\end{equation}
where $N(z_m)$ is the population size of the wave-front integrated over the range $z_m < z < \infty$. Thus, \eqref{J_rN} explicitly relates the ``leakage'' of cells out of the front at $z=z_m$ to the growth of cells in the front. \par 
To connect the Growth Regime to the Chemotaxis Regime, we integrate the ODE describing the density $\rho(z)$ in the moving frame, i.e., \eqref{eq:simplrho} (with $\rho_c \to \infty$), from a position $z<z_m$ in the Growth Regime, to the position $z = z_m$ in the Chemotaxis Regime. This results in the exact relation
\begin{equation}
    -c\cdot [\rho(z_m)-\rho(z)] = J(z_m) - J(z) + r \int_{z}^{z_m}\rho(z')dz' \label{rho-growth}
\end{equation}
Thus, using EQ. 7,
\begin{equation}
    \rho(z) = \beta a_m  +\frac{\chi_0}{c}\frac{\rho(z)}{a(z)+a_m}\frac{da(z)}{dz} +\frac{r}{c} \int_{z}^{z_m}\rho(z')dz' \label{rho-growth}
\end{equation}
We note that as $a(z)\to 0$, ${da(z)}/{dz}\to0$ as $z\to-\infty$. Now, as $a(+\infty)=a_0>0$ and $a(z)\geq0$, for $a(z)\in {C}^2$, ${da(z)}/{dz}>0$ for at least part of the domain $(-\infty,+\infty)$. Suppose ${da(z)}/{dz}\leq0$ in the domain $(z_p,z_p+\epsilon)$. Then, by continuity as $a(z)\in {C}^2$
\begin{align}
\implies\frac{da(z)}{dz}\bigg|_{z_p}=0,\  \frac{da(z)}{dz}\bigg|_{z_p-\epsilon}>0,\  \frac{da(z)}{dz}\bigg|_{z_p+\epsilon}\leq0\\
\implies \frac{d^2a(z)}{dz^2}\bigg|_{z_p}<0.
\end{align}
However, from \eqref{eq:simpla}, $\frac{d a(z)}{d z}\big|_{z_p}>0$ as $\mu\frac{a(z)}{a(z)+a_m}\rho(z)>0\ \forall z$, which contradicts our supposition. Thus, ${da(z)}/{dz}>0$ in the domain $(-\infty,+\infty)$ and $a(z)$ is always monotonically increasing (even in the Chemotaxis and Diffusion regimes). Since $\frac{da(z)}{dz}>0$, we have that $\rho(z)>\beta a_m$ from \eqref{rho-growth}. Further, since $\frac{d\rho(z)}{dz}|_{z=z_m}>0$, we know that $\rho(\zmin)<\rho(z_m-\epsilon)<2\beta a_m$ as $\zmin<z_m$. We declare $\zstar<\zmin<z_m$ such that $\rho(\zstar)=2\beta a_m$ and $\beta a_m<\rho(z)<2\beta a_m$ for $z\in(\zstar,z_m)$.\\
%\an{Below, I consider $a(z)\gg (r/\lambda c)a_m$ and $a(z)\ll a_m$. Both show the bound, but I don't know if I'm missing any intermediate values.}\\
Now, for $a(z)\gg(r/\lambda c)a_m$, the \textit{ansatz} holds, and thus \eqref{rho-growth} can be written as
\begin{align}
    \rho(z)>\beta a_m+\frac{\chi_0\lambda}{c}\beta a(z)=\beta (a(z)+a_m)
\end{align}
%For $a(z)\sim(r/\lambda c)a_m,\ a(z)\ll a_m$ in our parameter regime, and thus $\beta a_m\approx\beta (a(z)+a_m)\implies \rho(z)>\beta (a(z)+a_m)$. Thus, $\rho(z)>\beta (a(z)+a_m)$ for $z<z_m$.\\
For $a(z)\ll a_m\implies a(z)\sim \frac{r}{\lambda c} a_m$,
\begin{align}
    \frac{r}{c} \int_{z}^{z_m}\rho(z')dz'>\frac{r}{c}\beta a_m(z_m-z)
    \end{align}
    As we will show below, $a(z)<a_m\exp(\lambda (z-z_m))$ for $z<z_m$
\begin{align}
    \implies \frac{r}{c} \int_{z}^{z_m}\rho(z')dz' >\beta\frac{r}{c\lambda} a_m\log (a_m/a(z))>\beta a(z)
\end{align}
And for the chemotactic drift, we have, \begin{align}
    \frac{\chi_0}{c}\frac{\rho(z)}{a(z)+a_m}\frac{da}{dz}>\frac{\chi_0}{c}\frac{\beta a_m}{a(z)+a_m}a(z)\lambda\approx\beta a(z)
\end{align}
Thus, from \eqref{rho-growth},
\begin{align}
    \rho(z)>\beta a_m+\beta a(z)=\beta (a(z)+a_m)
\end{align}
Thus, $\rho(z)>\beta (a(z)+a_m)$ for $z<z_m$.
From \eqref{eq:simpla} and this lower bound on $\rho(z)$, %\an{I realised a mistake here. I need $\rho(z)> \beta(a+a_m)$ while I just have $\rho(z)> \beta a_m$. I know the former to be true empirically and can show that the assumption is self-consistent, but I am yet to show that it is necessarily true from Eq.49}
\begin{align}
    \frac{da(z)}{dz} &=-\frac{D_a}{c}\frac{d^2a(z)}{dz^2}+\frac{\mu}{c}\frac{\rho(z)}{a(z)+a_m}a(z)\label{dadz}
    \\&>-\frac{D_a}{c}\frac{d^2a(z)}{dz^2}+\frac{\mu\beta}{c}a(z)\\
    &=-\frac{D_a}{c}\frac{d^2a(z)}{dz^2}+\frac{(\chi_0+D_a)}{\chi_0}\lambda a(z)
\end{align}
If $a(z)$ can locally be described as a slowly varying exponential such that $a(z)=\exp(\lambda_a(z) z)$, then for $(z_m-\zmin)\ll 1/(\ln(\lambda_a(z)))'$, we may take $a(z)=\exp(\lambda_A z)$ where $\lambda_A=\lambda_a(\zmin)$,% \an{This is the one assumption I need to make. It would be nice to have bounds on $1/(\ln(\lambda_a(z)))'$ such that this holds},
\begin{align}
\lambda_A a(z)>\lambda a(z)+\frac{D_a}{c}a(z)(\lambda^2-\lambda_A^2)\\
\implies \lambda_A-\lambda>\frac{D_a}{c}(\lambda^2-\lambda_A^2)
\end{align}
This is only possible if $\lambda_A>\lambda$. Thus, 
\begin{align}
a(z)=a_m\exp(\lambda_A (z-z_m))<a_m\exp(\lambda (z-z_m))\text{ for }z<z_m. \label{lambdaAlower}
\end{align} 
Further, for $\zstar<z<z_m$, from \eqref{dadz}% \an{If I can prove that $\rho(z)>\beta(a(z)+a_m)$, I can get rid of the factor of two introduced in the 2nd inequality below.}
\begin{align}
\frac{da(z)}{dz}<\frac{\mu \rho(z_m)}{c a_m}a(z)<\frac{2\beta \mu}{c }a(z)=\frac{2(\chi_0+D_a)\lambda}{\chi_0}a(z)\\
\implies \lambda_A<\frac{2(\chi_0+D_a)\lambda}{\chi_0} \label{lambdaAhigher}
\end{align}
This provides bounds on $\frac{da(z)}{dz}$:
\begin{align}
    \lambda a(z)< \frac{da(z)}{dz}<\frac{2(\chi_0+D_a)\lambda}{\chi_0}a(z)\label{da_bounds}
\end{align}
This also gives us bounds on $v(z)=\frac{\chi_0}{a(z)+a_m}\frac{da(z)}{dz}$:
\begin{align}
\frac{\chi_0\lambda a(z)}{2a_m}< v(z) <\frac{2\chi_0(\chi_0+D_a)\lambda}{(\chi_0)}\frac{a(z)}{a_m}\\
\implies v(z)<\frac{2\chi_0(\chi_0+D_a)\lambda}{\chi_0}\exp(\lambda(z-\zmin))
\end{align}
From \eqref{eq:simplrho}, $\frac{d\rho(z)}{dz}=0$ if 
\begin{align}
r\rho(z)=\frac{d(v(z)\rho(z))}{dz}=\frac{dv(z)}{dz}\rho(z)
\end{align}
\begin{align}
\implies r=\frac{dv(z)}{dz}=\frac{\chi_0}{a(z)+a_m}\left(\frac{d^2a(z)}{dz^2}-\frac{(da/dz)^2}{a(z)+a_m}\right)\\
=\frac{\chi_0}{a(z)+a_m}\lambda_A^2\left(a(z)-\frac{a(z)^2}{a(z)+a_m}\right)
\end{align}
From Eq. 11, we know that for growth to be comparable to the chemotactic drift, $a(z)\ll a_m$, and thus
\begin{align}
r \approx\frac{\chi_0}{a_m}\lambda_A^2a(\zmin)\implies a(\zmin)\approx \frac{ra_m}{\chi_0\lambda_A^2}
\end{align}
From the bounds on $\lambda_A$ from \eqref{lambdaAhigher} and \eqref{lambdaAlower},
\begin{align}
%\implies \frac{(\chi_0-D_\rho+D_a)}{\chi_0}\frac{a_m}{a_0}>\frac{a(\zmin)}{a_m}>\frac{(\chi_0-D_\rho)^2}{4(\chi_0-D_\rho+D_a)\chi_0}\frac{a_m}{a_0}
%\implies \frac{r}{\lambda c}\frac{(\chi_0-D_\rho)}{\chi_0}>\frac{a(\zmin)}{a_m}>\frac{r}{\lambda c}\frac{(\chi_0-D_\rho)^3}{4(\chi_0-D_\rho+D_a)^2\chi_0}
\implies \frac{r}{\lambda c}>\frac{a(\zmin)}{a_m}>\frac{r}{\lambda c}\frac{\chi_0^2}{4(\chi_0+D_a)^2}
\end{align}
Numerically, we find that $a(\zmin)\approx ra_m/\lambda c$ (see Fig.~5B), which means that $\lambda_A^2\approx \lambda c/\chi_0\approx \lambda^2$ where the final equality holds if $D_\rho\to0$. Thus, we find that $\lambda_A\lesssim\lambda=\lambda_a(z_m)$ and $\lambda_a(\zmin)\approx\lambda_a(z_m)$, numerically verifying our assumption above that $\lambda_A\approx \lambda_a(z)$ for $\zstar<z<\zmin$.\par
Now, at $z=\zhat$ such that $\rho(\zhat)=2e\beta a_m$, assuming that $\rho(z)\propto\exp(-\lambda_Gz)$ for $\zstar>z>\zhat$:
\begin{align}
J(\zhat) &=-v(\zhat)\rho(\zhat)\\
&=-\frac{r}{\lambda_A}\exp\left(-\frac{\lambda_A}{\lambda_G}+\lambda_A(\zhat-\zmin)\right)\rho(\zhat)\\
&<\frac{r}{\lambda_A}\exp\left(-\frac{\lambda_A}{\lambda_G}\right)2e\beta a_m
\end{align}
As we will show below, $\lambda_A\gg\lambda_G\implies \exp(\lambda_A/\lambda_G)\gg \lambda_A/\lambda_G$, and thus the final term is much smaller than the net growth in time $(z_m-\zhat)/c$:
\begin{equation}
%\begin{split}
-J(\zhat)\ll \frac{r}{\lambda_G}2(e-1)\beta a_m< r\int_{\zhat}^{z_m}\rho(z')dz'\label{v-condition}
%\end{split}
\end{equation}
Thus, using \eqref{Jz_m} and $\rho(z_m)=2\beta a_m$ (since the \ansatz is valid at $z=z_m$), we can rewrite \eqref{rho-growth} for $z<\zhat$ as
\begin{align}
&c\rho(z) 
     \approx c \rhohat  +  r \int_{z}^{\zhat}\rho(z')dz', \label{rho-growth2}\\
     &\rhohat = \beta a_m + \frac{r}{c} \int_{\zhat}^{z_m}\rho(z')dz', \label{rhohat}
\end{align}
for $z\le \zhat$, with the only approximation coming from the condition \eqref{v-condition}.
\par
%
%\eqref{rho-growth2} is the first integral of the F-KPP equation, with the boundary condition that there is an influx given by $c\rhohat$ at $z=\zhat$. 
Solving \eqref{rho-growth2} yields 
\begin{equation}
     \rho(z\le \zhat) =\rhohat \exp[\lambda_G (\zhat-z)], \label{rho-growth3}
\end{equation}
with 
\begin{equation}
    \lambda_G = -r/c,   \label{lambda_G}
\end{equation}
We note that if $D_\rho>0$, we would have another solution such that $\lambda_G\approx -c/D_\rho$. However, this solution would be rejected as it corresponds to a solution dominated by diffusion and is an unstable solution at the front of the wave (since growth is greater than or comparable to chemotactic drift, there is no term to balance diffusion in the diffusion-dominated solution and thus lead to the transition to the Chemotaxis regime. Thus, diffusion can only dominate when the right BCs are asymptotic, i.e., $\rho(z)\to0$ as in the Diffusion regime) \cite{van2003front}.
Thus, as cited earlier, we can see that $\lambda_A/\lambda_G=\lambda c/r\gg 1$ in our parameter regime.
%This solution \eqref{rho-growth2} recapitulates the exponential density profile seen in Fig.~1B, with a gentle rise $\lambda_G$ compared to the rise of the density bulge, i.e.,  $\lambda_G \ll \lambda$ as long as $r \ll \lambda c$. The rising form of $\rho(z)$ for $z$ deep into the Growth regime indicates that the last term in the ODE for $a(z)$, \eqref{eq:simpla}, is of the form $a(z)\cdot \exp[-\lambda_G z]$. This leads to a very steep drop in $a(z)$ and a corresponding steep drop in $v(z)$; see SIxxx. This self-consistently justifies the approximation \textbf{\ref{v-condition}}, hence, \eqref{rho-growth2} and its solution. \par
%
To determine the width of the transition zone between the exponentially decreasing and exponentially increasing profiles of $\rho(z)$ in the Growth and Chemotaxis Regimes respectively, we note that for $\zhat<z<z_m$, $\beta a_m<\rho(z)<2e\beta a_m$. Thus, \eqref{rhohat} can be written as
$$2e\beta a_m=\beta a_m+\frac{r}{c}(z_m-\zhat)\alpha\beta a_m$$
where $2e>\alpha>1$. Thus, $$(z_m-\zhat)=\frac{c(2e-1)}{r\alpha}\implies \frac{(2e-1)}{2e\lambda_G}<(z_m-\zhat)< \frac{(2e-1)}{\lambda_G}$$
%
%The above results are represented geometrically by the sketches in Figs.~5D,~5E. At a given instance (time $t_0$), the wave-front is shown as the solid green line in the lab frame. The front region, comprised of $N$ cells, grow at a rate $rN$. This growth is balanced by cells which leave the front (i.e., across the black dashed line indicating $x_0=z_m + c t_0$), with flux $J = -c\beta a_m$. At some time $\delta t$ later, the front has traversed a distance $\delta x = c \cdot \delta t$. The total amount of cells leaving the front during this time is $\delta N = J\delta t$. The corresponding density of the cells left behind the propagating front is $\rhomin = \delta N/\delta x = \beta a_m$ (shown as the brown region in Fig.~5E). The cells left behind will continue to grow at the rate $r$. For $\delta t$ much smaller than the doubling time, the density behind the front will not have grown much and thus remain at $\sim \beta a_m$ (Fig.~5D). After time $\Delta t$ large compared to the doubling time, the population size at the back will become $\rho(x_0, t) = \rho(x_0,t_0)\, e^{r\Delta t}=\rho(x_0,t_0)\, e^{r(t-t_0)}$ (Fig.~5E). Given that $t_0 = (x_0-z_m)/c$, we have 
%\begin{equation}
% \rho(x_0, t) \approx \rhomin \exp\left[-\frac{r}{c}(x_0-ct)\right]. \label{trail}
% \end{equation}
% Thus, the trailing exponential density profile \eqref{trail}, while looking like a moving front, is merely a result of the exponential growth of a stationary population, which is \textit{seeded} by the traveling wave-front at density $\rhomin$ and speed $c$. 
% \par 
 
\section{Relation to Literature on Chemotactic Pattern Formation}
%Where did inclusion of logistic growth start?\\
%What has been said about travelling wave solutions?\\
%How has the connection Fisher waves been made?\\
%General comment on the literature focusing on pattern formation rather than travelling wave solution. Inclusion of consumption and degradation, and the question for Keller-Segel remains unanswered.
Soon after the introduction of the KS model, other models that considered the growth of the bacteria were proposed. Following a misconception that chemotaxis is driven by a search for nutrients, most such models coupled the growth rate and the concentration of the attractant~\cite{lapidus1978model,kennedy1980traveling,lauffenburger1982effects,lauffenburger1984traveling,pedit2002quantitative,hilpert2005lattice}. However, such a coupling cannot account for the large expansion speeds observed~\cite{Cremer2019a}. Cremer et al. note that a small amount of an attractant in the presence of a nutrient can lead to significantly higher expansion speeds than if the attractant is the only nutrient source. Thus, though some of these models resulted in traveling waves~\cite{kennedy1980traveling,lauffenburger1984traveling}, they treat an unnatural case.\par Other models require production of attractant by the bacteria. There has been very extensive mathematical literature~\cite{horstmann20031970,funaki2006travelling,bonami2001singular,aida2006lower,efendiev2005continuous,hu2017boundedness,aida2006lower,woodward1995spatio} (for a comprehensive review, refer to \cite{arumugam2021keller}) focusing on the chemotactic models with attractant production due to crucial experiments that demonstrated that in motile bacterial cells aggregate in response to gradients of attractant which they excrete themselves, and form complex spatial patterns~\cite{budrene1991complex}. Much work has also been done on traveling-wave solutions and their connection the F-KPP equation~\cite{salako2019can,salako2019traveling,salako2017existence,mansour2008traveling,calvez2017traveling}. However, this work has been primarily motivated by the formation of complex patterns, rather than on the relatively simple migratory bands discussed in this paper. Further, the work has been mostly mathematical in nature, focusing on existence and uniqueness proofs, and scaling results and concise approximations with phenomenological parameters that can be utilized by experimentalists are lacking.\par 
In light of recent experimental evidence for physiological roles of chemotaxis such as range expansion, we hope that tools developed for complex chemotactic pattern formation may be adapted to understanding the relatively simple migratory bands, and build on our simplified model. The minimal nature of our simplified model allows for further analysis upon perturbation by inclusion of more terms such as an upper Weber cutoff, or production of attractant by the bacteria. Further, by relating the system of chemotaxis to other well-studied systems such as the F-KPP equation, we hope that the understanding of the other systems can be drawn to our system, and that physical, biological and experimental insights related to our system can be utilized for the other systems. We hope that the confluence of the these three systems, which have historically been pursued by different scholastic communities, spurs insightful exchange.

 \newpage
\section*{Supplemental References}
\addcontentsline{toc}{section}{Supplemental References}
\bibliography{pnas-sample}